\documentclass{aa}
 \usepackage{graphicx}
  \begin{document}


 \title{Polarization behavior of periodic optical outbursts in blazar OJ287}
 \author{S.J.~Qian\inst{1}}
 \institute{
   National Astronomical Observatories,
    Chinese Academy of Sciences, Beijing 100012, China} 
   \date{Complied by using A\&A latex}
  \abstract{As a characteristic feature of generic blazars the polarization
    behavior of the quasi-periodic optical outbursts  observed in OJ287 
   is investigated. The optical light-curves of the December/2015 outburst
   are also simulated  in terms of the precessing jet nozzle model previously 
   proposed.}
   {The polarization behavior of three primary quasi-periodic optical 
   outbursts peaking in $\sim$1983.0, $\sim$2007.8 and $\sim$2015.8
    are analyzed in order to understand the nature of their optical radiation.}
   {A two-component model has been applied, showing that the variations in 
   flux density, polarization degree and polarization position angle can 
    be consistently interpreted with two polarized components: one 
   steady-component with constant  polarization and one burst-component 
   with varying polarization (e.g., relativistic shock propagating along 
   the jet-beam axis).} {The flux light curves of the December/2015 outburst
   (including its first flare and second flare) are well model-simulated 
   in terms of 14 elementary synchrotron sub-flares, each having a
   symmetric profile. The model-simulations of polarization behavior for 
   the three major outbursts (in 1983.0, 2007.8 and 
   2015.8) demonstrate that they all exhibit rapid and large rotations in
    polarization position angle, implying
   that they are synchrotron flares produced in the jet.}{Combining with 
  the results previously obtained  for interpreting the optical
 light curves in terms of lighthouse effect for both quasi-periodic and 
  non-periodic  outbursts, we suggest that relativistic jet models 
  may be the most appropriate models for understanding the nature of the 
  optical flaring radiation in blazar OJ287: its optical outbursts may comprise
   a number of  blended ``elementary synchrotron flares'', each produced by 
   the helical motion of individual superluminal optical knots via 
  lighthouse effect.}
  \maketitle
  \section{Introduction}
    Blazars are active galactic nuclei with their relativistic jets pointing at
   small angles with respect to the line of sight. They emit radiation across 
   the entire electromagnetic spectrum from radio, IR/optical, UV, X-ray
    through to high energy (TeV) $\gamma$-rays. Their emissions are highly
    variable with a wide range of time-scales from minutes/hours to years
   (e.g., Angel \& Stockman \cite{An80}, Aller et al. \cite{Al10}, \cite{Al14},
    Ackermann et al. \cite{Ac11}).\\
       OJ287 (z=0.306) is a unique blazar: its optical variability 
     not only has the characteristic properties of radiation in generic
    blazars, but also reveals a $\sim$12-yr
   quasi-periodicity with a double-peaked structure  at time-intervals
    of $\sim$1-2 years (in its optical
    light curve recorded during more than one hundred years since $\sim$1890;
   e.g., Sillanp\"a\"a et al. \cite{Si88}, Lehto \& Valtonen \cite{Le96},
    Sundelius et al. \cite{Su97},  Valtonen et al. \cite{Va11}).\\
     Thus, the investigation (both observational and theoretical) of the 
    OJ287 phenomena mainly includes four subjects: (1)  mechanisms
    which cause the $\sim$12-yr quasi-periodicity and double-flare structure
    of the optical outbursts with time-intervals of $\sim$1-2\,yr;
     (2) characteristics of the optical radiation and its associated
     radiation in other wavebands (from radio to $\gamma$-rays).Especially, 
     correlation between the optical outbursts and the ejection of superluminal
     radio knots; (3) connection between the characteristics  of the optical
     outbursts and the mechanism for its quasi-periodicity; (4) properties
     of the putative supermassive black hole binary (precessing 
     orbital motion and accretion processes) and possible tests of 
     general-relativity effects (gravitational radiation, precession of binary 
     orbit and no-hair theorem, etc.).\\
      Various  models have been proposed to interpret the quasi-periodicity
      and double-peaked structure in its  optical light curve 
      (e.g, Sillanp\"a\"a et al. \cite{Si88}, 
      Lehto \& Valtonen \cite{Le96}, Sundelius et al. \cite{Su97},
      Valtonen et al.\cite{Va19}, Dey et al. \cite{De19}, Valtaoja et al.
      \cite{Val20}, Katz \cite{Ka97}, Villata et al. \cite{Villa98}, 
      Britzen et al. \cite{Br18}, Tanaka \cite{Tan13}) .\\ 
       Many authors investigate the nature and properties of the optical/radio
       emission, especially, polarization behavior and helical motion of the
      emitting components (e.g., Usher \cite{Us79},  Aller et al. \cite{Al81},
      Holmes et al. \cite{Holm84}, Kikuchi et al. \cite{Ki88}, Gabuzda et al. 
      \cite{Ga01}, \cite{Ga04}, \cite{Ga99}, D'Arcangelo et al. \cite{Da09},
      Villforth et al. \cite{Vil10}, Aller et al. \cite{Al10}, \cite{Al14},
      \cite{Al16}, Hodgson et al. \cite{Hod17},
     Kushwaha et al. \cite{Ku18a}, \cite{Ku18b}, 
     Qian \cite{Qi18b}, \cite{Qi19a}, \cite{Qi19b}, Cohen et al. \cite{Co17}, 
     \cite{Co18},  Myserlis  et al. \cite{My18}).\\
      The $\sim$12\,yr quasi-periodicity is widely suggested to be related 
     to the orbital motion of the putative black-hole binary in the nucleus.
     The precessing jet model (or binary black hole-impact model)
     originally proposed by Lehto \& Valtonen 
     (\cite{Le96}) and its improved versions (e.g., Valtonen et al. 
     \cite{Va19}, Dey et al. \cite{De19}) suggest that the quasi-periodicity
     is due to the precessing orbital motion of the binary due to 
      the gravitational interaction between the supermassive secondary and 
     primary holes.  The periodic outbursts are 
     caused by the secondary black hole's penetrating
     into the disk of the primary black hole. The double-flare structure 
     is assumed to be due to the two impacts occurring per one orbital cycle
     near pericenter and apocenter passages. Each secondary hole's
     crossing of the primary disk will  produce strong thermal optical outburst 
     from the gas-bubble torn off the primary disk. In addition, the 
     penetrations lead to enhanced accretion onto the primary hole and
     optical knot ejections from the jet, producing the follow-up synchrotron 
     flares (tidal flares). At present, the disk-impact model can well explain
     the quasi-periodicity and the double-flare structure, and have successfully
     predicted the flare times of a few double-outbursts (including the
     2015/2019 pair-flares, Laine et al. \cite{La20}). The optical light curve
      has been modeled in terms of the combination of impact flares and 
     tidal flares (e.g., Dey et al. \cite{De18}).\\
       Recently, Britzen et al. (\cite{Br18}) have made a detailed 
     analysis of the kinematic properties of the radio jet on pc-scales
     and suggested that the radio jet precession is related to the orbital 
     motion of the black-hole binary, and the optical variability could 
     be  related to the precession and nutation of the radio jet.\\
       In order to investigate the nature of optical emission for both 
     quasi-periodic and non-periodic outbursts in OJ287, Qian (\cite{Qi19a})
     has  made model simulations of their optical light curves under
     the precessing jet nozzle scenario, which has been  previously used to
     study the the kinematics of superluminal radio knots (including their
     helical motion and variable Doppler boosting effects), jet-beam
     precession,  connection between optical and radio variability (including 
    simultaneous  radio/optical variations, evolutional 
    relation between  optical and radio
     knots, etc.)  in a number of blazars (3C345: Qian et al. \cite{Qi91a},
     \cite{Qi09}; 3C454.3: Qian et al. \cite{Qi14}; NRAO 150: Qian \cite{Qi16};
      B 1308+326: Qian et al. \cite{Qi17}; PG 1302-102:
     Qian et al. \cite{Qi18a}; 3C279: Qian et al. \cite{Qi19c}, \cite{Qi13};
     OJ287: Qian \cite{Qi18b}, \cite{Qi15}, \cite{Qi19a}, \cite{Qi19b}).\\
      The model simulation of the optical light curves are based on two basic
     assumptions: (1) the optical outbursts are decomposed into a number of
     elementary synchrotron flares (defined as a single flare with smallest 
     time-scales of $\sim$10 days); (2) each elementary synchrotron
     flare is produced by an individual superluminal optical knot moving 
     along helical trajectory via lighthouse effect.\\
      The optical light curves of the periodic outbursts
     observed in $\sim$1983.0, 1984.1, 1994.6, 2005.7, 2007.8 
     and 2015.8 and a few non-periodic outbursts were well simulated
     in terms of the precessing jet nozzle model. The light curve of the 
     periodic outburst in 1995.8 was also model-simulated  and its simultaneous 
     optical and radio variations was  explained (Qian \cite{Qi19b}).\\
      In our  precessing nozzle model we use the following terms to
      describe the jet phenomenon: jet, jet-beam, jet-nozzle, beam-axis
      and jet axis ($\equiv$precession axis). They describe the picture of jet
     phenomenon: the plasma/magnetic-field and superluminal knots ejected 
     from the jet-nozzle form the jet-beam and the jet-beam precesses 
     around the precession-axis (i.e. jet-axis) with a period of 
     $\sim$12\,yr, producing the jet. Since the superluminal knots (both 
     radio and optical) and  magnetized plasma move along helical trajectory 
     around the precessing beam-axis, the term "jet" actually represents the
     whole jet which is made up of all the magnetized plasma and 
     superluminal knots ejected from the precessing nozzle. The 
     plasma/magnetic-field within the jet should be rapidly swirling and 
     the jet-axis may be also 
     precessing in space, which has not been taken into account in the 
     current precessing nozzle model.\\ 
      In the precessing jet-nozzle scenario, the optical light curves are 
     assumed to comprise a number of  synchrotron subflares for both periodic 
     and non-periodic outbursts. That is, we assume that the double-peaked
      outbursts observed in the optical light curve are synchrotron flares
      produced in the relativistic jet
      and related to the ejection of  superluminal optical knots from the core,
      which are closely 
     related to the mass-accretion  onto the primary hole, the rotation 
     of its magnetized disk, the spin of the primary hole and the magnetic 
     acceleration mechanism through strong toroidal fields
     in its magnetosphere (e.g., Blandford \& Znajek \cite{Bl77}, Blandford \&
     Payne \cite{Bl82}, Li et al.\cite{Lizy92}, Camenzind \cite{Cam90}, Meier 
     \cite{Mei13}, \cite{Mei01}, Beskin \cite{Be10}, Vlahakis \& K\"onigl
     \cite{Vl04}) . The precessing jet nozzle model does not treat the 
     quasi-periodicity in the optical variability and is not able to predict 
     the flare-times of the periodic outbursts. It requires additional model(s)
      for explaining the quasi-periodicity and double-peaked structure, which
      can match its interpretation of the quasi-periodic optical 
     outbursts being  synchrotron in origin within the jet.
     \footnote{Based on the analysis of the kinematics of the
      superluminal radio knots, there might have some evidence for the
      existence of two jets in OJ287 (Qian \cite{Qi18b}).}\\ 
     In brief, if the quasi-periodicity with double-flare structure observed in
     the optical light curve are not involved temporarily, the precessing
      nozzle model is different from the binary
     black-hole impact model mainly in two aspects:
     \begin{itemize}
     \item  The disk-impact model assumes that the first flares 
      of the double-peaked 
     outbursts are thermal flares produced  by the impacts of the
     secondary black hole penetrating the disk of the
     primary hole, when gas-bubbles are torn off the primary disk and
     emit thermal emission. At the same time, the impacts (near pericenter and
     apocenter passages) result in strong disturbances to the primary disk.
     The disturbances and tidal effects lead to enhanced accretion onto
     the primary hole, producing the follow-up flares ("tidal flares")
     which are synchrotron flares produced in the relativistic jet of the 
     primary hole. The difference between the precessing nozzle model
     and the disk-impact model is only in the interpretation of the nature of 
    radiation from the first flares of the double-peaked outbursts: 
    the former claims their origin from synchrotron process, while
    the latter claims their origin from bremsstrahlung process;
      \item The impact models involve both thermal and nonthermal flares,
      it needs dual-energetics for understanding the optical outbursts (thermal
     and nonthermal). The energy source of the nonthermal flares mainly 
     comes from the spin energy of the primary hole and the angular momentum 
     of its disk. The synchrotron outbursts are strongly Doppler-boosted 
    \footnote{Doppler-amplification factor could be $\sim{10^5-10^6}$.} and 
     their timescales are shortened by Doppler-beaming effect. 
      But the  energy source of the impact (thermal) flares mainly comes 
    from the kinematic energy of the orbiting secondary hole.
     The flux density of the thermal flares  is not Doppler-boosted,
       and their time-scales are not shortened either. Thus the energy 
     source and energetics of the thermal and nonthermal outbursts are 
     completely different. It seems difficult to unify the two kinds of flares
     (both in flux density scale and time scale) only by scaling the
      impact energy and impact-induced accretion rate onto the 
      primary hole (e.g. Dey et al. \cite{De18}, \cite{De19}). In other words, 
      synchrotron flares gain energy-input from the spin of the 
      primary hole and angular momentum of its disk, which is completely
      different from the energy source for the thermal outbursts. Thus it seems
      difficult to understand the observational fact: the non-thermal and 
     thermal outbursts observed in OJ287 could have very similar behaviors 
     in variations of flux density and polarization with similar time-scales
      (e.g. Valtaoja et al. \cite{Val20}).
      In contrast, the precessing nozzle model has no problem
     in the unification of the energetics. Moreover, the impact-disk 
     model seems not able to explain the structure of the optical outbursts 
      which comprise several spike-like flares on timescales
     of $\sim$10 days  with symmetric profiles, rapid polarization variation
     on time scales of $\sim$a day, and  the simultaneous variations in optical
     and radio regimes (Valtaoja et al. \cite{Val20}, Qian \cite{Qi19a},
     \cite{Qi19b}). Recently, We have applied the precessing nozzle model with 
     helical motion of the superluminal optical knots to well simulate
     the R- and V-band light curves of the December/2015 outburst 
     (Qian \cite{Qi19a}). In the following we will investigate
      the polarization behavior (especially
      the rotations in polarization position angle observed in the periodic
     outbursts  in 1983.0, 2007.8 and 2015.8, providing further 
    evidence for their origin in synchrotron process.
      \end{itemize}      
       Recently, Myserlis et al. (\cite{My18}) observed the fast
      rotations of polarization position angle in OJ287 at V-band and radio
     wavelengths (10.5, 8.4 and 4.8\,GHz) during December/2015-January/2017
    ($\sim$JD2457300-800), showing a time-delay of the radio PA rotation
    with respect to the optical PA rotation.  They suggested that these
     position angle rotations (on time-scales of $\sim$10\,day in optical and 
     on time-scales of about a month in radio) are due to the
    helical motion of the superluminal optical knots and radio knots, 
    respectively.\footnote{Unfortunately, Myserlis et al.'s observation 
   did not obtain any polarization data for the first flare of the 
   December/2015 outburst (peaking at $\sim$JD2457361.5), which was
   claimed as the "impact (thermal)" flare in the impact-disk 
   scenario.}   Rotations in polarization position angle were 
    also observed by Kushwaha et al. (\cite{Ku18a}) in the December/2015
     outburst. Cohen et al. (\cite{Co18}) have reported observational
    evidence for polarization angle rotations in the cm-light
    curves observed in the Michigan Monitoring projects.\\
    These polarization observations seem strongly supporting the
    precessing nozzle model. According to our precessing nozzle 
    scenario for explaining  the optical light curves of the
    quasi-periodic and non-periodic outbursts
    (Qian \cite{Qi19b}), superluminal optical knots move along helical
     trajectories, which would result in fast rotations of polarization
      position angle. This phenomenon has already been observed in 
   a few blazars in earlier years: for example, in BL Lacertae 
    and 0727-115 (Aller et al. \cite{Al81},
       Sillanp\"a\"a \cite{Si93}, 
         Marscher et al. \cite{Ma08}).\\ 
     In this paper we shall show the rapid variations in polarization 
         position angle during the periodic optical outbursts (in 1983.0,
         2007.8 and 2015.8) and discuss the interpretation of their 
      light curves of flux density,
         polarization degree and position angle as a whole. We would like to
         note that low polarization degrees alone  do not necessarily imply
         the outbursts being thermal, because thermal flares can greatly
          reduce the source  polarization degree, but can not cause
          rapid changes in polarization position angle.\\
           The cavity-accretion flare model proposed by Tanaka (\cite{Tan13}) 
   is another type of disk-impact model, which assumes that the primary hole
   and the secondary hole having comparable masses and are in near-coplanar
   orbital motion. Hydrodynamic/magneto-hydrodynamic (HD/MHD) 
   simulations for such binaries surrounded by circumbinary disks
    have shown that cavity-accretion processes would create 
   two gas-flow streams impacting onto the disks of the black holes per
    pericenter passages, possibly causing the double-peak structure of the 
   quasi-periodic outbursts. This model suggests that the gas-flow impacts
   produce thermal outbursts, but it is not able to make
    accurate timing of the quasi-periodic outbursts. 
   \begin{figure*}
   \centering
   \includegraphics[width=5.5cm,angle=-90]{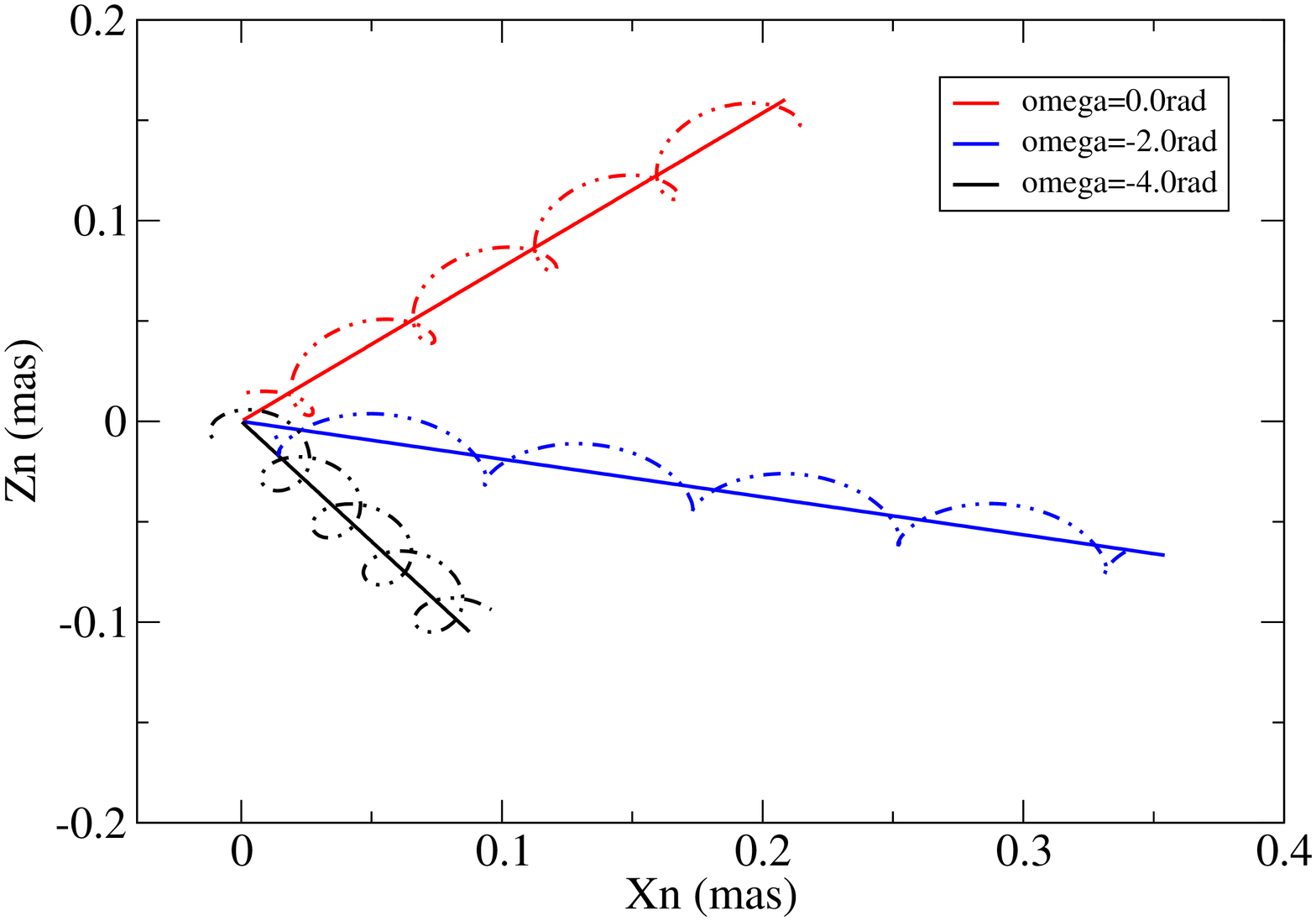}
   \includegraphics[width=5.5cm,angle=-90]{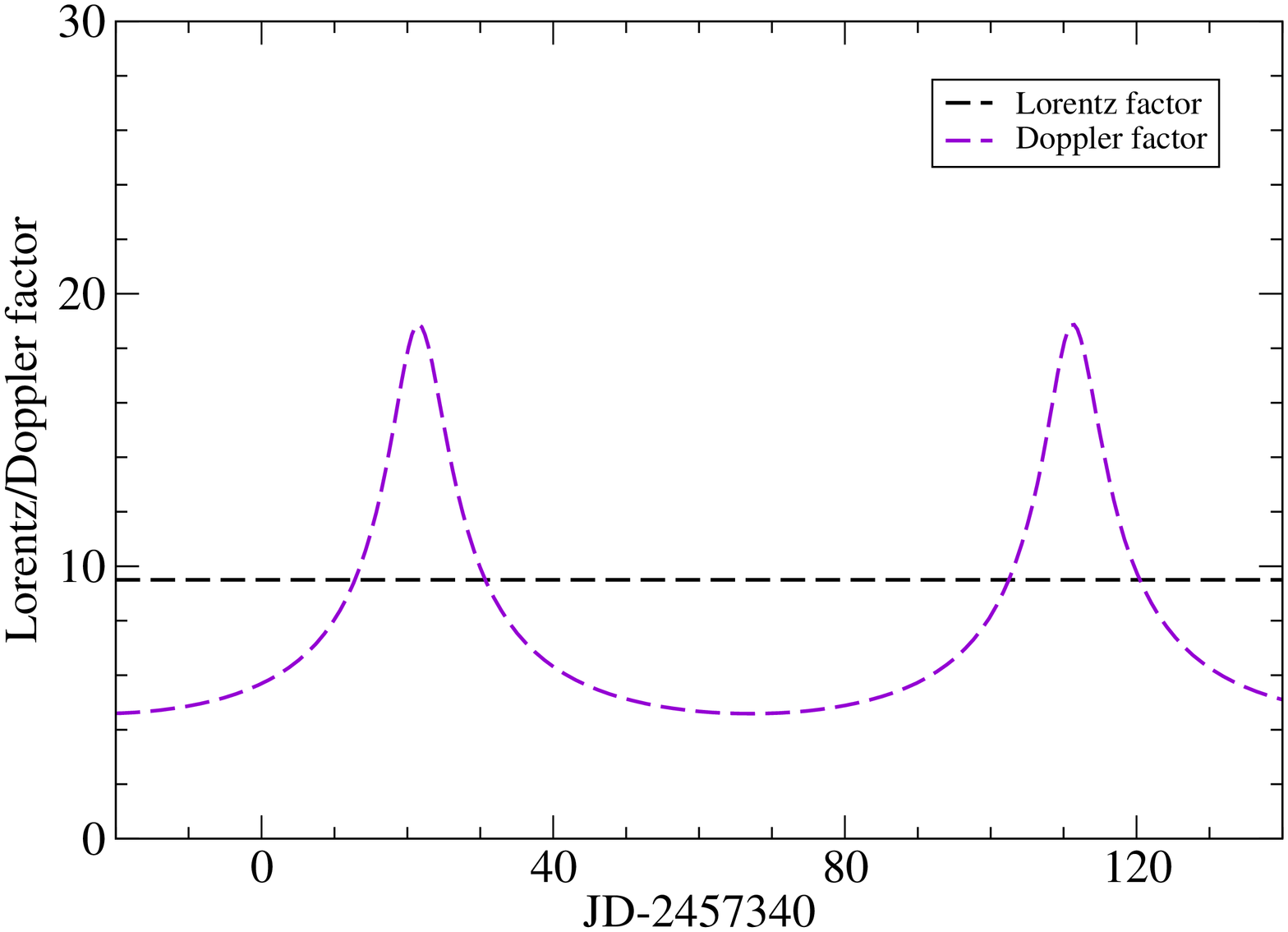}
   \includegraphics[width=5.5cm,angle=-90]{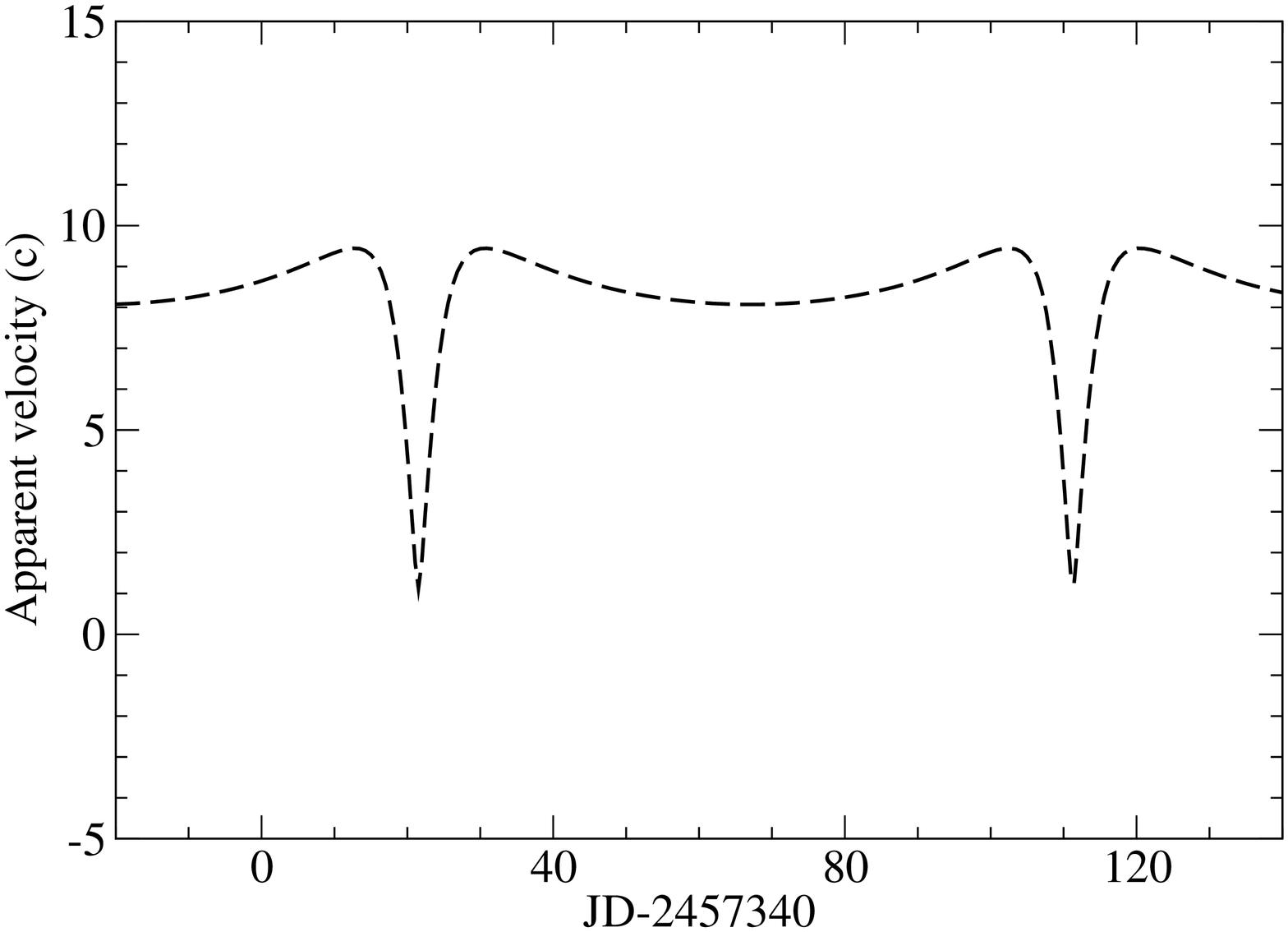}
   \includegraphics[width=5.5cm,angle=-90]{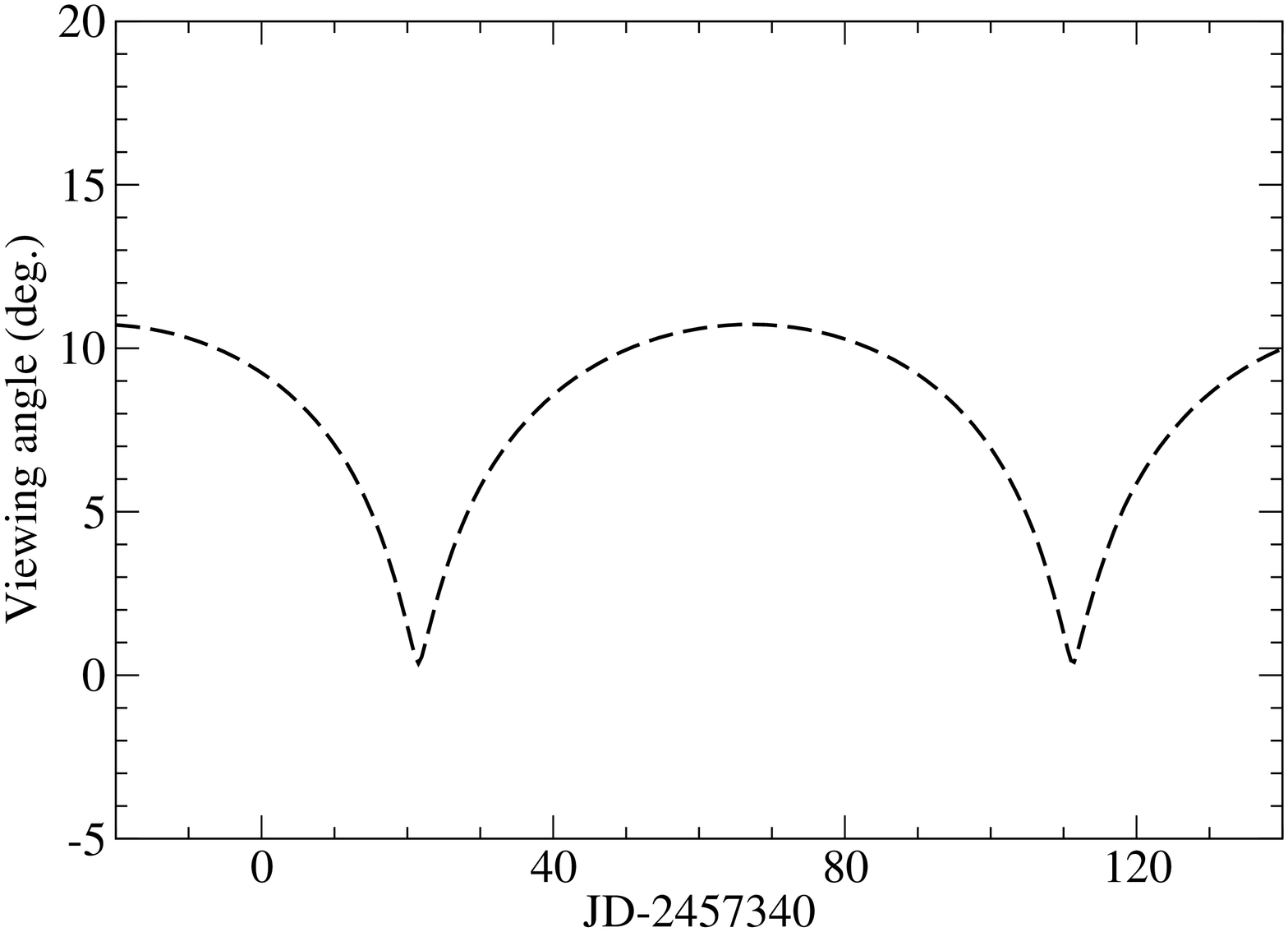}
   \caption{A sketch of the precessing nozzle scenario with helical motion
    proposed in a previous work: superluminal optical knots move along 
    helical trajectory around the beam-axis which precesses with a period of
    12\,yr (left-upper panel). The Lorentz/Doppler factor, apparent velocity 
   and viewing angle versus time are shown for the precession phase 
    $\omega$=-2\,rad (helical trajectory in blue). Periodic optical 
    variations with a period of $\sim$90 days are obtained (details referring 
    to Qian 2019a; $\Gamma$=9.5).}
   \end{figure*} 
     \section{The precessing jet-nozzle model}
    \subsection{Introduction}
   It is widely suggested that blazars are extragalactic sources with
   relativistic jets pointing close to our line of sight. 
   Recently, Qian (\cite{Qi19a}) has
   tentatively proposed an alternative jet model to understand the 
   phenomena observed in OJ287, which is based on the optical  multi-wavelength 
    observation and $\gamma$-ray observation performed for OJ287 
   (e.g., Kushwaha et el. \cite{Ku18a}, \cite{Ku18b}), combining  with 
   the distinct features previously found in the optical and radio 
   (flux and polarization) variations (e.g, Sillanp\"a\"a et al. \cite{Si96a},
   Valtaoja et al. \cite{Val20}, Usher \cite{Us79}, Holmes et al. 
    \cite{Holm84}, Kikuchi et al. \cite{Ki88}, D'Arcangelo et al. \cite{Da09},
    Kushwaha et al. \cite{Ku18a}, Britzen et al. \cite{Br18}, 
   Qian \cite{Qi19b}).\\
      The precessing jet-nozzle model was originally proposed in Qian et al.
    (\cite{Qi91a}) to study the VLBI-kinematics of the superluminal radio
    components observed in QSO 3C345, which were suggested to move along
    helical trajectories. This scenario has been further applied to investigate 
    the VLBI-kinematics of several blazars, including 3C345, 3C454.3, OJ287,
     NRAO 150, B 1308+326 and PG 1302-102, 3C279. These studies have not only 
    made good model-fits to the observed 
    kinematic properties (trajectory, core-separation and apparent 
    velocity versus time) of their superluminal radio components, but also
    obtained some new results, e.g.,:\\
     (1) Some convincing evidence for the existence of two radio jets
        in QSO 279 (Qian et al. \cite{Qi19c})\\
     (2) Some evidence for the possible existence of two radio jets
        in BLO OJ287 (Qian \cite{Qi18b})\\
     (3) Derivation of the precession period of jet-beams (or jet-nozzles)
       and investigation of  the 
        jet-nozzle precession mechanisms (Newtonian-driven precession,
         Lense-Thirring effect and spin-orbit coupling;
         Qian et al. \cite{Qi17}, \cite{Qi18a}) \\
     (4) Derivation of  the kinematic Lorentz (and Doppler) factor and 
       investigation of  the intrinsic evolution of superluminal radio 
       components (Qian et al. \cite{Qi96})\\
     (5) Tentative derivation of the mass and spin of black holes in the
         nuclei of blazars through  precession mechanisms (Qian et al.
         \cite{Qi19c}, \cite{Qi17}, \cite{Qi18b})\\
     (6) Interpretation of the simultaneous optical and radio outbursts
         observed in OJ287, helping to clarify the nature of quasi-periodic
         optical outbursts (Qian \cite{Qi19b}).
    \subsection{Formulation of the precessing jet nozzle scenario}
     we have formulated the scenario to calculate the VLBI-kinematics of 
     superluminal components (optical or radio knots), details referring to
     Qian et al. (\cite{Qi91a}, \cite{Qi19a}, \cite{Qi18b}, \cite{Qi17}).
     We assume that the jet-nozzle ejects magnetized plasma and  superluminal
     knots (blobs or shocks) which form the collimated jet-beam and move along
       helical trajectories around the beam-axis. The jet-nozzle and the
     beam-axis precess around an axis fixed in space (designated as the 
     precession axis) with a certain precession  period. The precession of the
     jet-beam produces the overall jet  which occupies the whole 
     region where superluminal components and magnetized plasma ejected at
     different times move outward in different directions. 
     \footnote{Here we would like to use 
      three  terms (jet-nozzle, jet-beam and  jet) to 
     describe the three physically distinct entities of the jet phenomenon.} 
     We chose the following set of parameters to define the precessing nozzle
    scenario which are the same as used in Qian (\cite{Qi19a}, \cite{Qi18b}):\\
      (1) Precession axis is defined by parameters $\epsilon$=$3^{\circ}$ 
        and  $\psi$= $0^{\circ}$ .\\
      (2) The axis of the jet-beam is assumed to be rectilinear described by
          the parameters $\it{x}$=1 and $\it{a}$=0.0402 as in Qian 
         (\cite{Qi18b}). Thus the beam-axis precesses on a cone with an opening 
         angle of $2.3^{\circ}$ .\\
      (3) The parameters describing the helical motion of a superluminal
        optical knot around the beam-axis are: amplitude $A_0$=0.0138\,mas
        and rotation rate d$\phi$/d$z_0$=-7.04\,rad/mas.\\
      (4) The precession phase is chosen to be: $\omega$=-2.0\,rad which 
      corresponds to the blue line shown in Figure 1.\\
      (5)  The spectral index $\alpha_{R,V}$ is assumed to be 1.5  
      (${S_{\nu}}{\propto}{{\nu}^{-\alpha}}$)\\
      (6) The base-level flux density at R- and V-bands are assumed to be
      $S_{\nu,0}$=3.5\,mJy and 3.0\,mJy, respectively\\
      (7) The concordant cosmology model ($\Lambda$CDM model) is assumed with
      $\Omega_m$=0.27 and $\Omega_{\Lambda}$=0.73 and Hubble constant $H_0$=
      71 km $s^{-1} Mpc^{-1}$ (Spergel et al. \cite{Sp03}, Komatsu et al.
      \cite{Ko09}). Thus for OJ287, z=0.306, its luminosity distance is
      $D_L$=1.58\,Gpc (Hogg \cite{Ho99}, Pen \cite{Pe99}) and angular distance
      $D_A$=0.9257\,Gpc. Angular scale 1\,mas= 4.487\,pc and proper motion 
      1\,mas/yr is equivalent to an apparent velocity =19.1\,c (c--speed of
      the  light).\\
      A sketch describing the precessing jet nozzle scenario is shown 
      in Figure 1, which
      was proposed for interpreting the $\sim$90\,day periodicity in the
      light curve of the December/2015 outburst (between the flares in 
      $\sim$JD2457360 and  $\sim$JD2457450, Qian \cite{Qi19a}).\\   
      In the following we shall make model-simulation for the 
      flux density light curve of the 
      2015 outburst in Section 6, decomposing it into 14 elementary synchrotron
      flares.
       \section{Two-component model for polarization analysis}
    Blazars are highly variable in their optical continuum and especially in
    their polarization. Generally, polarization properties are the main 
    characteristics revealing their nature of synchrotron  radiation originated
    from the relativistic  jets. For blazar OJ287, there may be a major issue 
    to be solved: whether the  first flares of the quasi-periodic optical 
    outbursts are  thermal flares originated from the 
    secondary hole  penetrating into the disc of the primary hole  or 
    they are nonthermal synchrotron flares originated in the jet.\\
     To solve this issue analysis of the polarization properties of the first
     flares of the quasi-periodic outbursts observed in OJ287 are important.
     According to Qian (\cite{Qi19a}) the  periodic optical outbursts in OJ287 
     can be decomposed into a number of 
    elementary flares, each of which is produced by a superluminal optical
     knot moving along a helical trajectory in helical magnetic fields 
     due to lighthouse effect. 
    This explanation naturally predicts
    that their polarization position angle should rotate during the helical 
    motion. PA rotations have been observed in a few blazars on timescales 
    of $\sim$10\,day in optical bands (Aller et al. \cite{Al81}, 
    Sillanp\"a\"a et al. \cite{Si93}, Marscher et al. \cite{Ma08})
    or about a month in radio bands (Aller et al. \cite{Al81}, \cite{Al14}, 
    K\"onigl \& Choudhuli \cite{Ko85a}). Mechanisms 
    for explaining these large-amplitude PA rotations
    have been suggested (e.g., Blandford \& K\"onigl \cite{Bl79}, 
    K\"onigl \& Choudhuli \cite{Ko85b}, Bj\"onrson \cite{Bj82}, Qian \& Zhang 
    \cite{Qi03}, Qian \cite{Qi93}, \cite{Qi92}). Most of the studies
    prefer the mechanisms in 
   which PA rotations are caused by relativistic shocks moving through helical
    magnetic fields or the composition of two polarized components.\\
    In order to analyze the polarization properties, especially the 
    rotation of the polarization position angle of the quasi-periodic outbursts
    observed in blazar OJ287, we will apply a simple method to investigate
     the polarization behavior
    of three periodic optical outbursts in 1983.0, 2007.8 and 2015.8
     in terms of two
    component model (Qian et al. \cite{Qi91b}, Qian \cite{Qi93}): 
    one steady component (or underlying quiescent jet component) 
    and one variable component (burst component). 
     Both are polarized synchrotron components.
    Assuming ($I$, $p$, $\theta$), ($I_1$, $p_1$, $\theta_1$) and 
    ($I_2$, $p_2$, $\theta_2)$ being the intensity, polarization degree (\%) 
    and polarization position angle (deg.) of the integrated
    outburst, component-1  and component-2, respectively, then we have 
       \begin{equation}
         {p_2}={\frac{I_1}{I_2}}{p_1}{\frac{\tan{2\theta\cos{2\theta_1}}
                  -\sin{2\theta_1}}
                         {\sin{2\theta_2}-\tan{2\theta}\cos{2\theta_2}}}
       \end{equation}
       \begin{equation}
         {p^2}={\frac{(I_1p_1)^2+(I_2p_2)^2+
             2I_1p_1I_2p_2\cos(2\theta_1-2\theta_2)}{I^2}}
       \end{equation}
       Stokes parameters ($Q$, $U$) are
       \begin{equation}
          {Q}={Ip}\cos{2\theta}
       \end{equation}
       \begin{equation}
           {U}={Ip}\sin{2\theta}
       \end{equation}
       Similar equations are  for ($Q_1$, $U_1$) and ($Q_2$, $U_2$).
    In our case ($I$, $p$, $\theta$) are known, ($I_1, p_1, \theta_1$) 
    will be  appropriately chosen, then $I_2$=$I-I_1$ is known. Only
     two parameters $p_2$ and  $\theta_2$ need to be determined. And equations 
    (1) and (2) are just sufficient to solve the two parameters.
   \section{Model simulation of polarization behavior for 1983.0 outburst}
    As shown above the polarization behavior of
   the periodic optical flares is a  key ingredient to determine the nature of 
   optical outbursts, distinguishing the relativistic models from the 
   disk-impact models. Therefore  we will make model simulation of the light
   curves of flux density, polarization degree and polarization position
    angle (as a whole) for three periodic optical outbursts  in 1983.0, 2007.8
    and 2105.8.\\
    For the outbursts in 1983.0 and 2007.8 very low polarization degrees 
   ($\sim$0.4\%-2.4\%) were observed, and both outbursts were claimed
   to be thermal flares  produced by the bubbles
   torn off the disk of the primary hole when the secondary hole penetrates
   into the primary disk. However, alternative interpretations are also 
   possible. For example,  a non-thermal (synchrotron) flare can also
   cause very low polarization degree if the non-thermal flare component has 
   its  direction of polarization nearly perpendicular to that of the 
   preexisting polarized component with similar polarized flux. But in this
    case rapid variations in  polarization position angle of the source
   would be observed.  Thus changes in polarization position angle may be 
   particularly important  for distinguishing non-thermal flares from 
   thermal flares.
      \subsection{Introduction}
      The 1983.0 optical outburst is an instructive event for understanding
      the nature of the periodic outbursts and distinguishing different models.
      The multi-waveband observations from IR (JHK) to optical (RVU) carried
       out by Holmes et al. (\cite{Holm84}, also Smith et al. \cite{Sm87})
     have provided full information about the flux density, polarization degree
       and polarization position angle during a 4-day period (7, 8, 9 and 10
      January, 1983), which
      was just coincided with the peaking period of the outburst. A model
      simulation of its flux density light curve (V-band) is shown in 
      Figure 2 (left panel), and the right panel shows the fits to the
      data-points obtained during the 4-day peaking period (at V- and R-bands).
      Some distinct features can be recapitulated : (1) The outburst peaked on 
      8 January, with very low polarization degrees of 0.4\% and 0.8\% at
      R- and V-bands, respectively; (2) The minimum polarization degrees at R- 
      and V-band are concurrent with the peaks of flux density (see
      left panels of Figure 3);
      (3) position angles at R- and V-band
      rotated, especially at V-band: during the 3\,day period (7, 8 
      and  9 Jan.) its polarization position angle rotated clockwise by
      $\sim{50^{\circ}}$ and then rotated counter-clockwise by
       $\sim{110^{\circ}}$ (see left/bottom panel of Figure 3); (4)
      The multi-frequency observations showed that the outburst had convex
       spectra  with a change in spectral index $\Delta{\alpha}{\sim}$0.4
       from infrared ($\alpha{\sim}$0.9) to optical ($\alpha{\sim}$1.3).\\
        Interestingly, the observed features-1 and -2 are just the
       characteristics of a thermal outburst required by the impact-disk
      model. But feature-3 (polarization position angle swing) does not support
      the model, because a thermal outburst can not cause large position angle
      swing. Feature-4  indicates that the observed spectra 
      are more complex than that of a thermal outburst as predicted by 
      the impact-disk model and very much like that of a synchrotron source
       with high-frequency steepening due to radiation losses.\\
        Therefore, as a whole, the multi-wavelength light curves observed
        for the 1983.0 outburst (including 
       flux density, polarization degree and polarization position angle)
       can not be explained 
      in terms of the appearance of a strong thermal outburst.\\
       \begin{figure*}
       \centering
       \includegraphics[width=5.5cm,angle=-90]{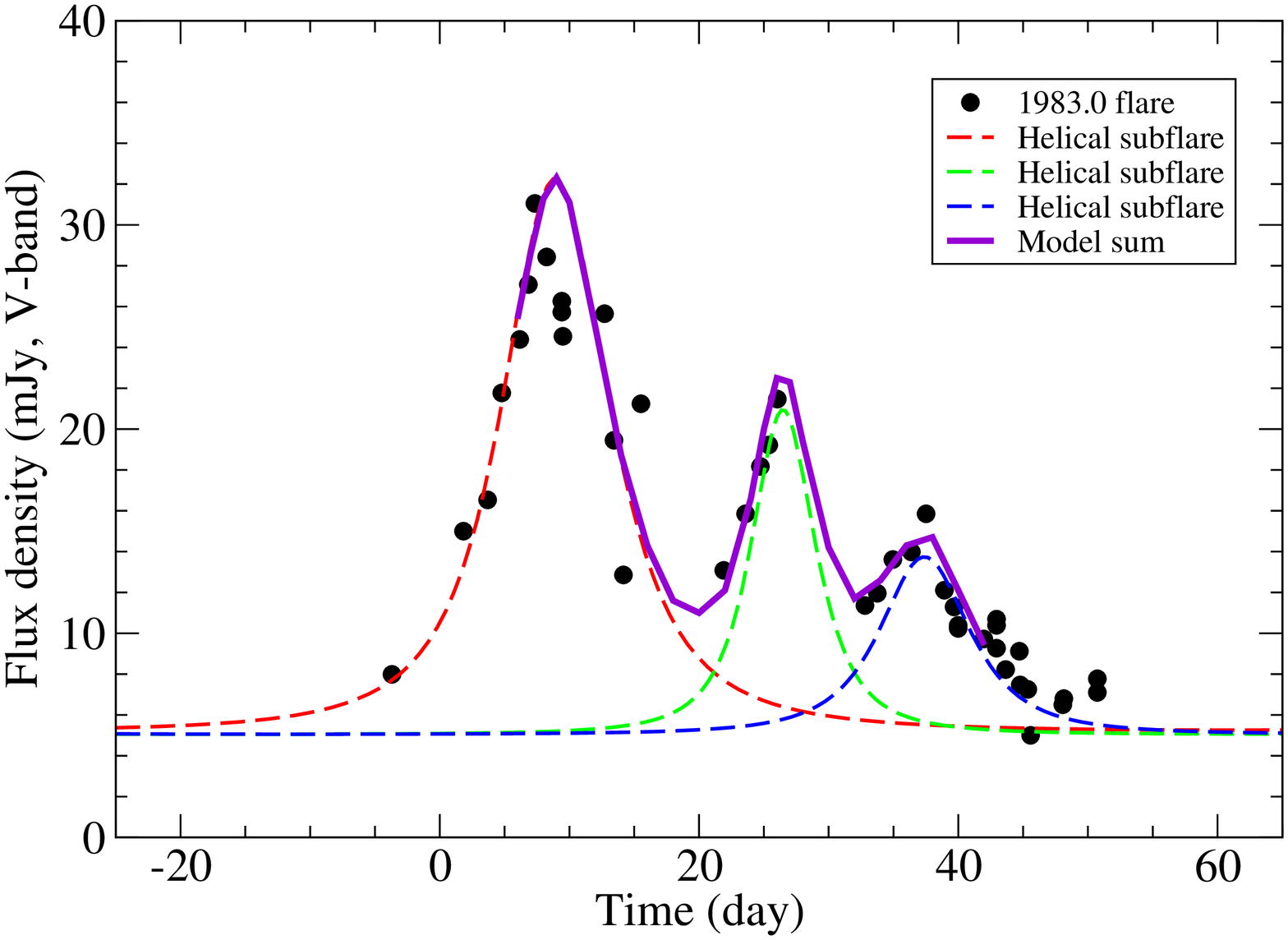}
       \includegraphics[width=5.5cm,angle=-90]{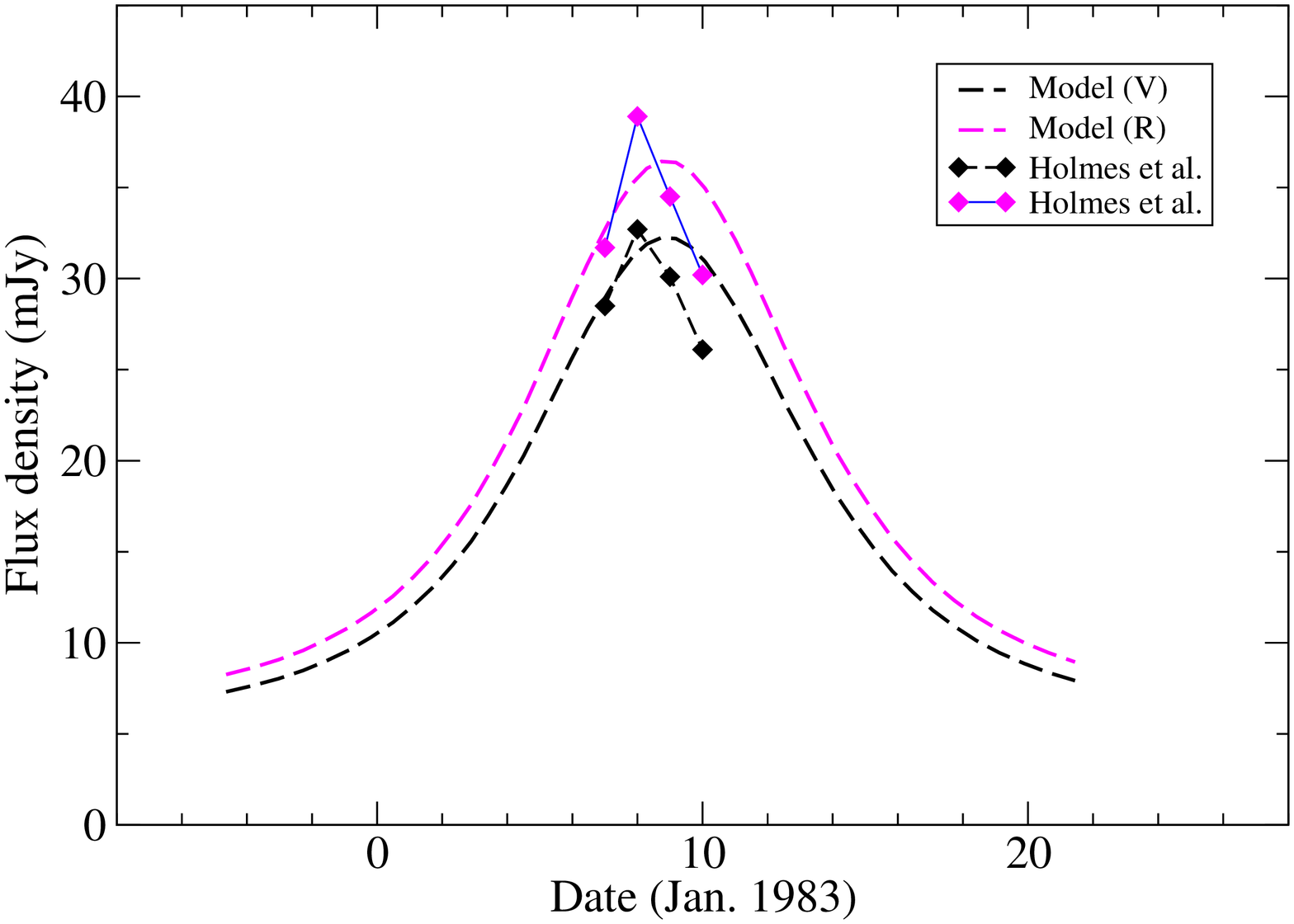}
       \caption{Left: Model-simulation of the 1983.0 outburst in terms
       of the helical-motion model under the precessing jet-nozzle scenario, 
       consisting of  three subbursts (reproduced from Qian, \cite{Qi19a}; 
       $\Gamma$=8.0 for the first subburst). 
       Right: The 4-day data-points obtained by Holmes et al. (\cite{Holm84})
       are well fitted by the model light curves at R- and V-bands for the 
       first flare.}
       \end{figure*}
       \begin{figure*}
       \centering
       \includegraphics[width=5.5cm,angle=-90]{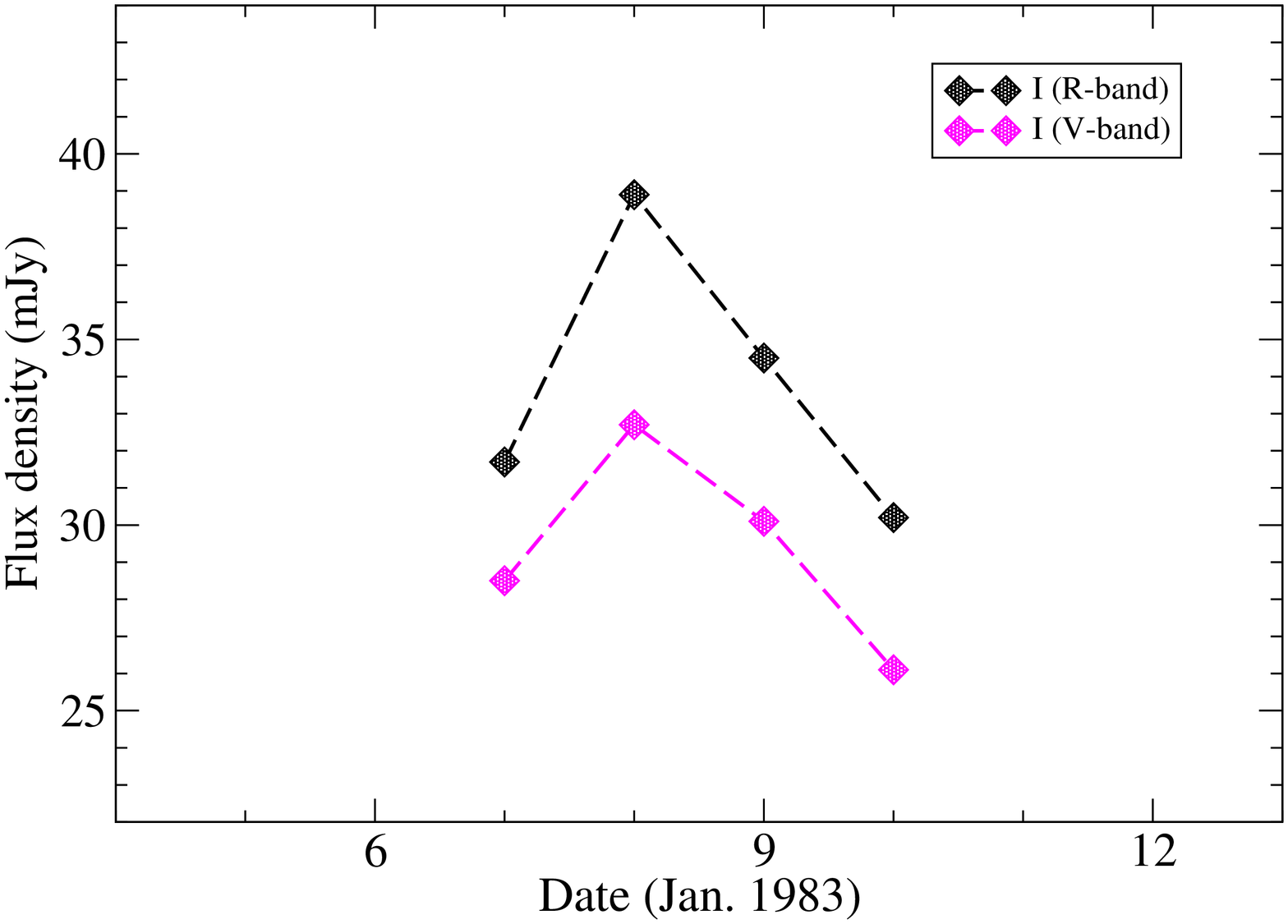}
       \includegraphics[width=5.5cm,angle=-90]{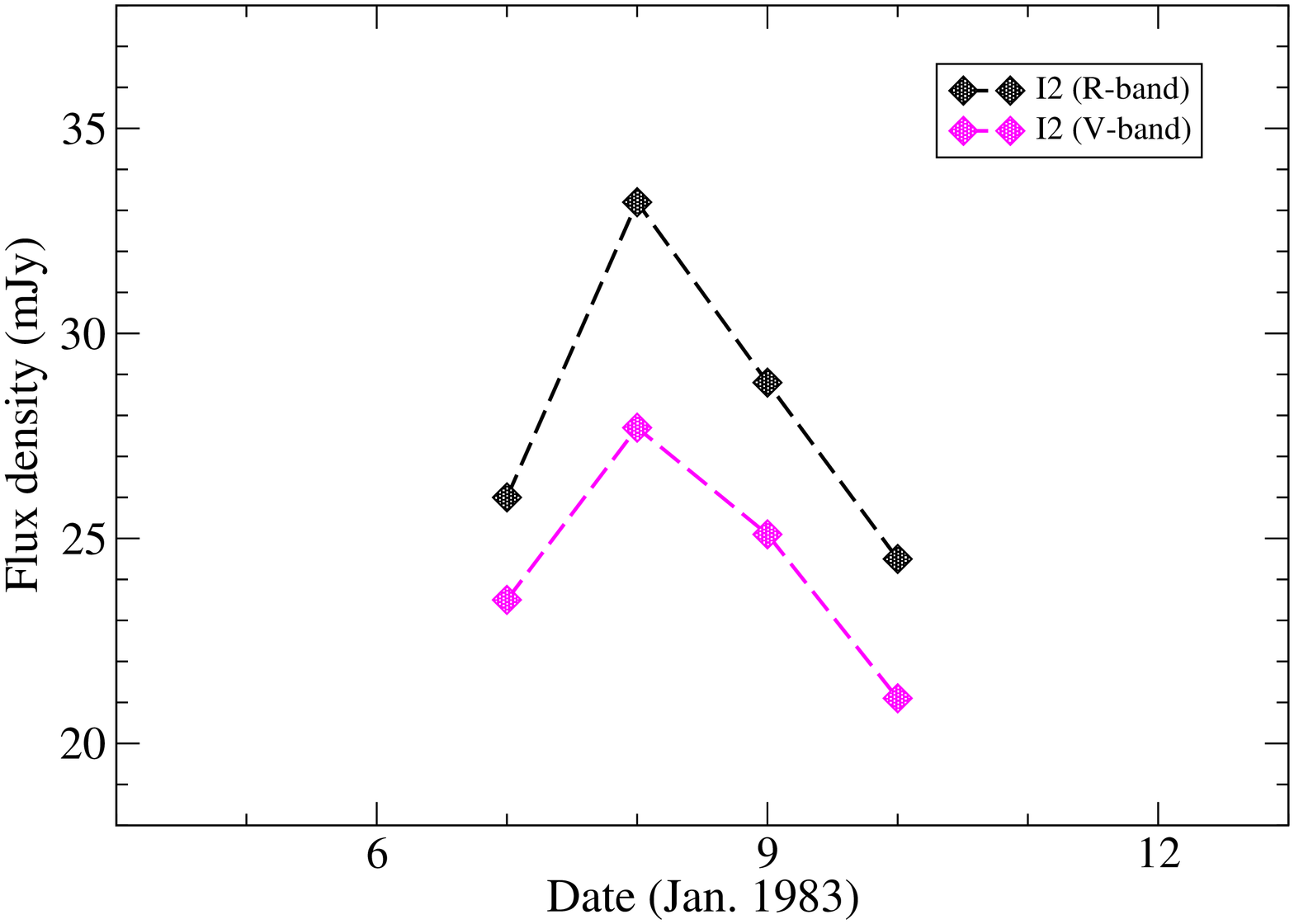}
       \includegraphics[width=5.5cm,angle=-90]{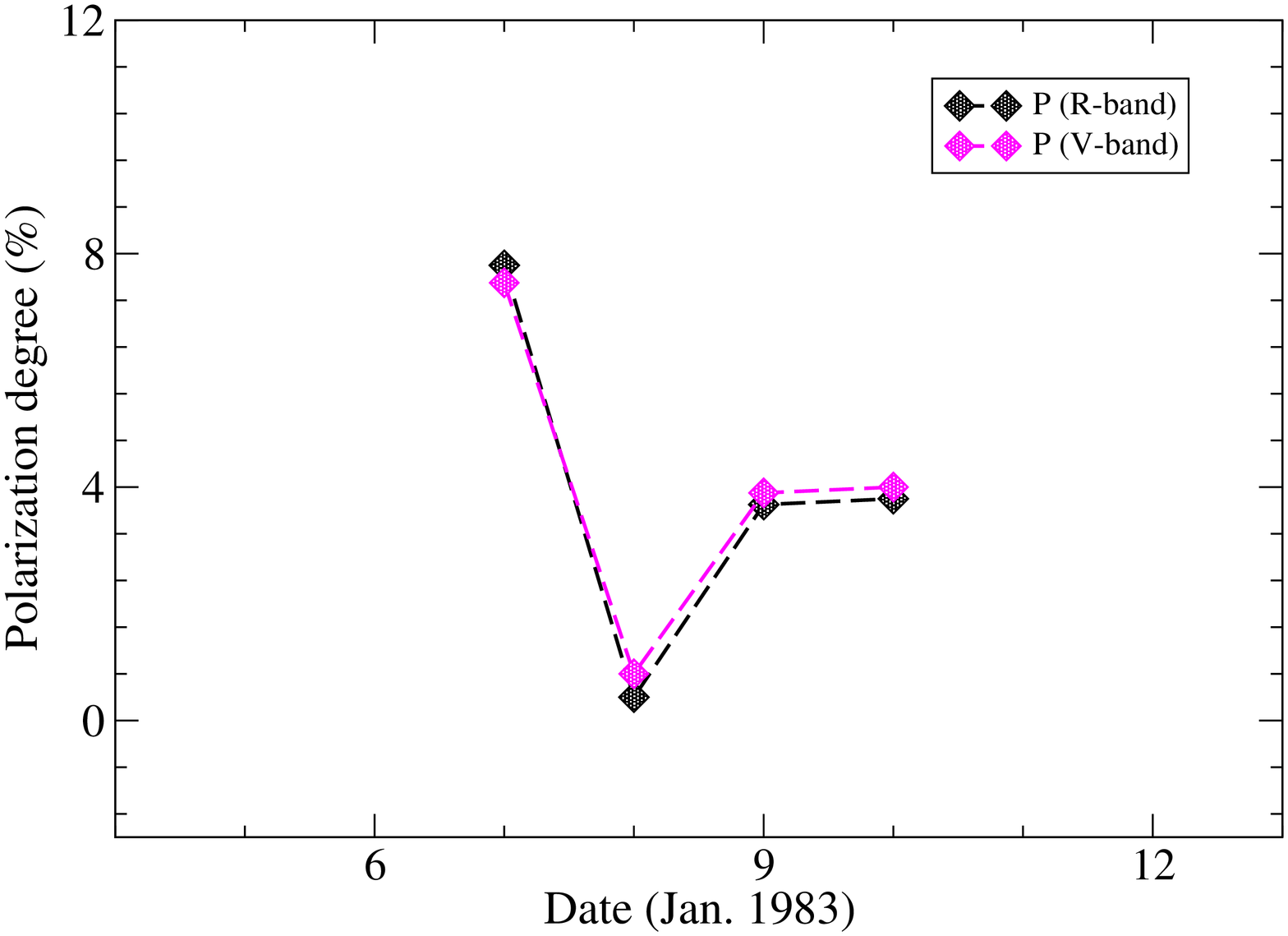}
       \includegraphics[width=5.5cm,angle=-90]{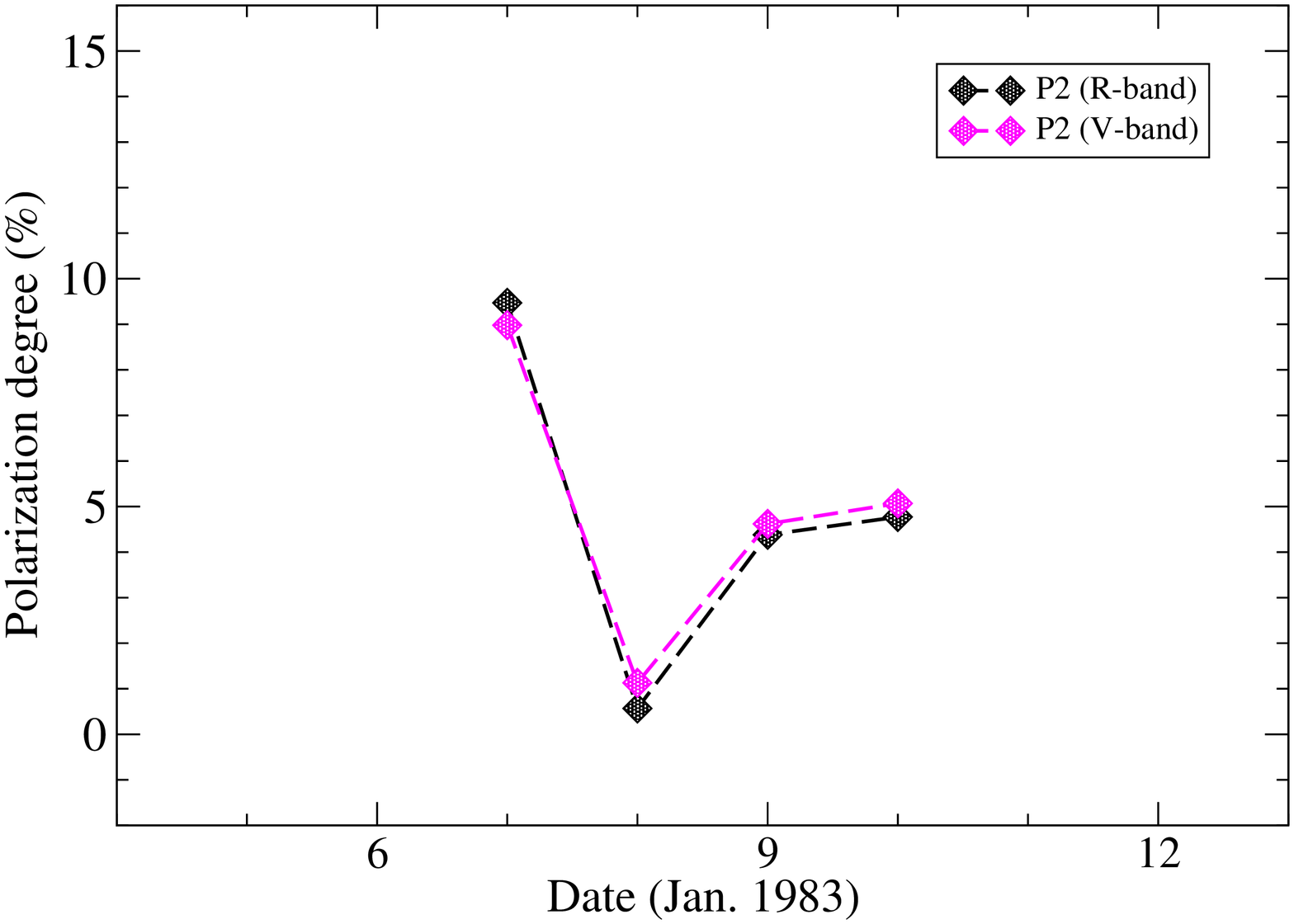}
       \includegraphics[width=5.5cm,angle=-90]{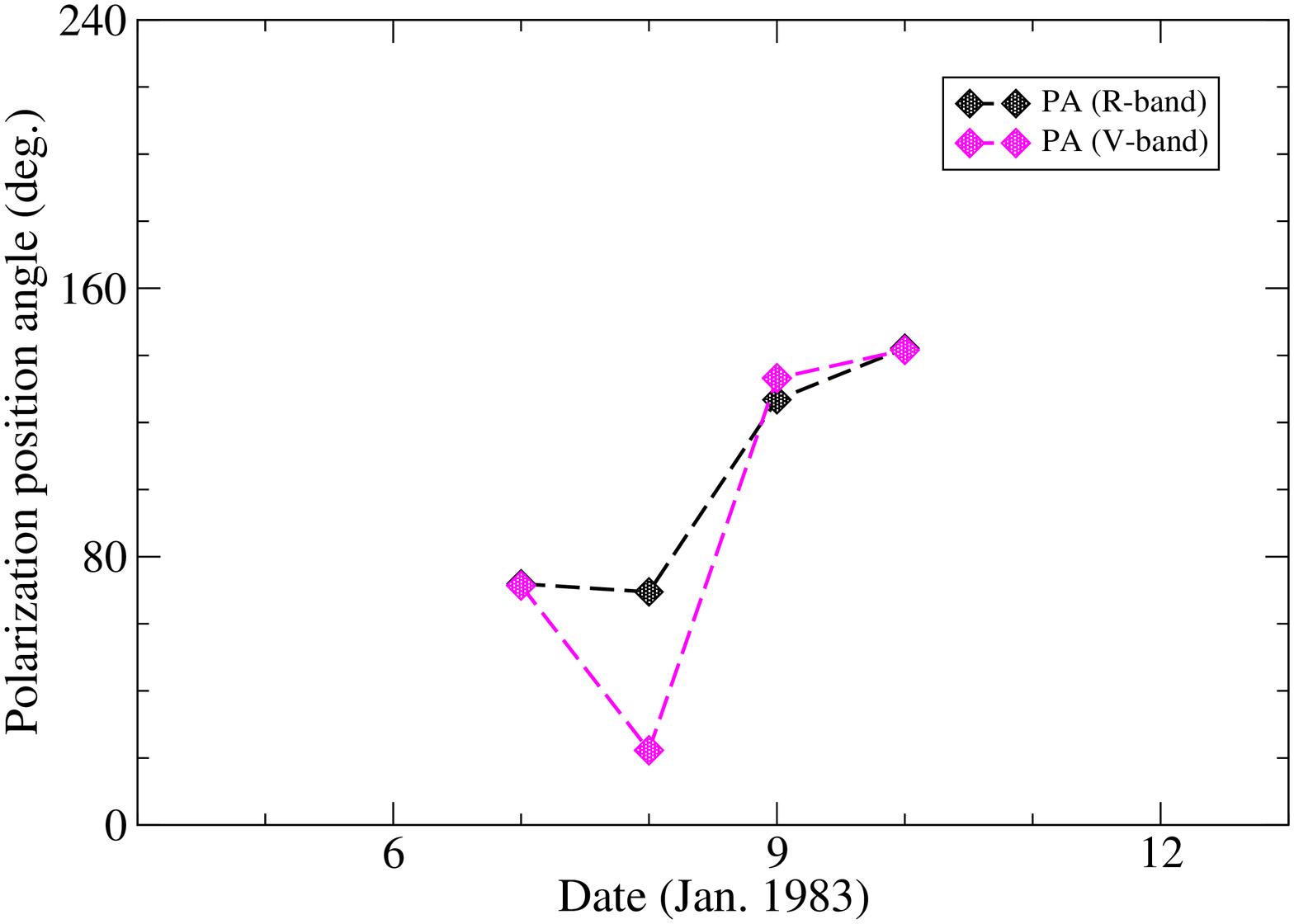}
       \includegraphics[width=5.5cm,angle=-90]{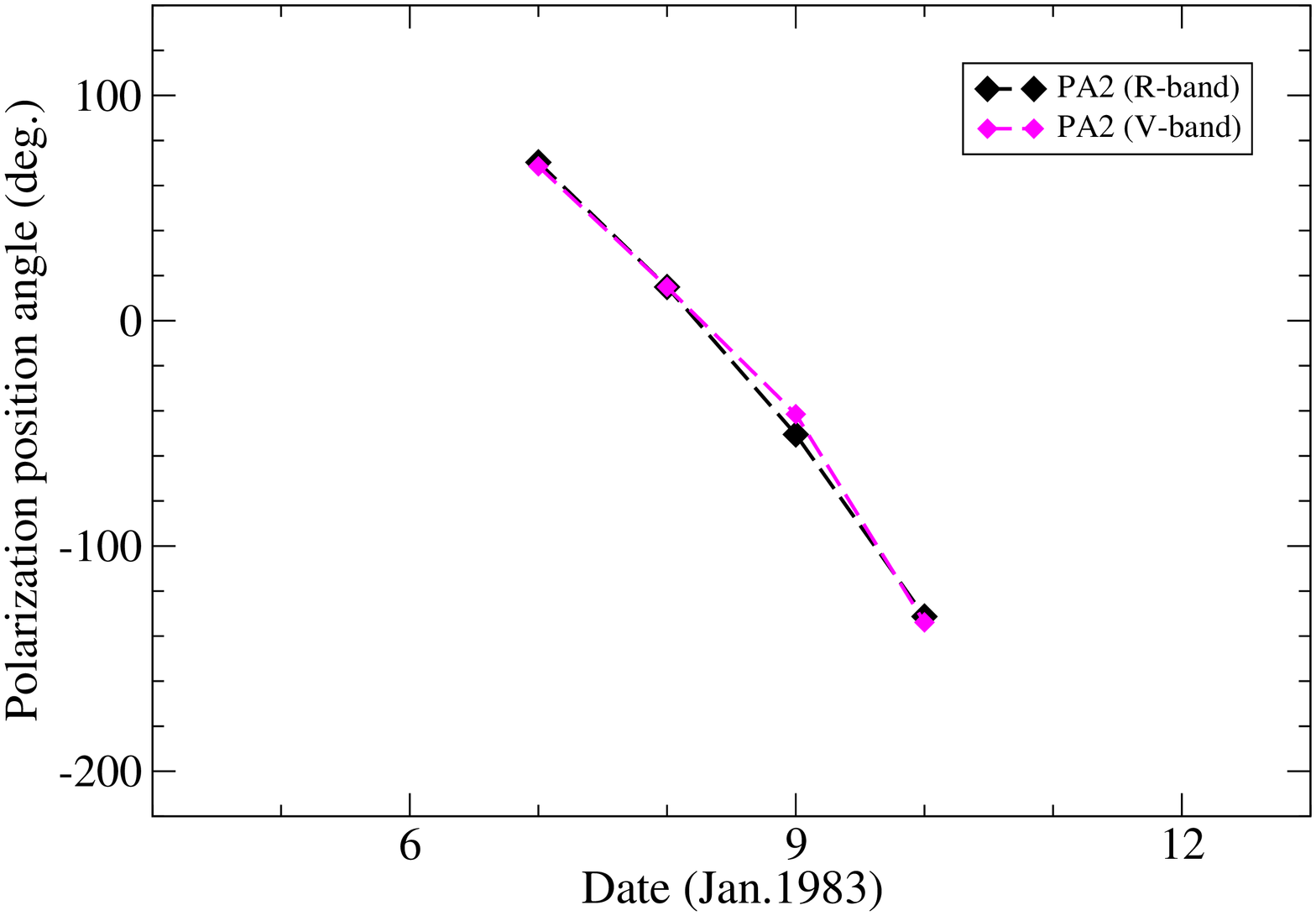}
       \caption{Simulation for the 1983.0 outburst. Left column: the observed
       light curves at V- and R-bands for the  integrated 
       flux, polarization degree and position angle. The position angle swings
       were very large between 8 January and 9 January:
       $\sim{110^{\circ}}$/day (V-band) and $\sim{50^{\circ}}$/day (R-band).
       Right column: the  light curves derived  from the model simulation
       for the flare component: continuous clockwise position angle rotations
        of $\sim{-65^{\circ}}$/day during the 4-day period at both R- and 
       V-bands are shown in right/bottom panel.}
       \end{figure*}
       \begin{figure*}
       \centering
       \includegraphics[width=5.5cm,angle=-90]{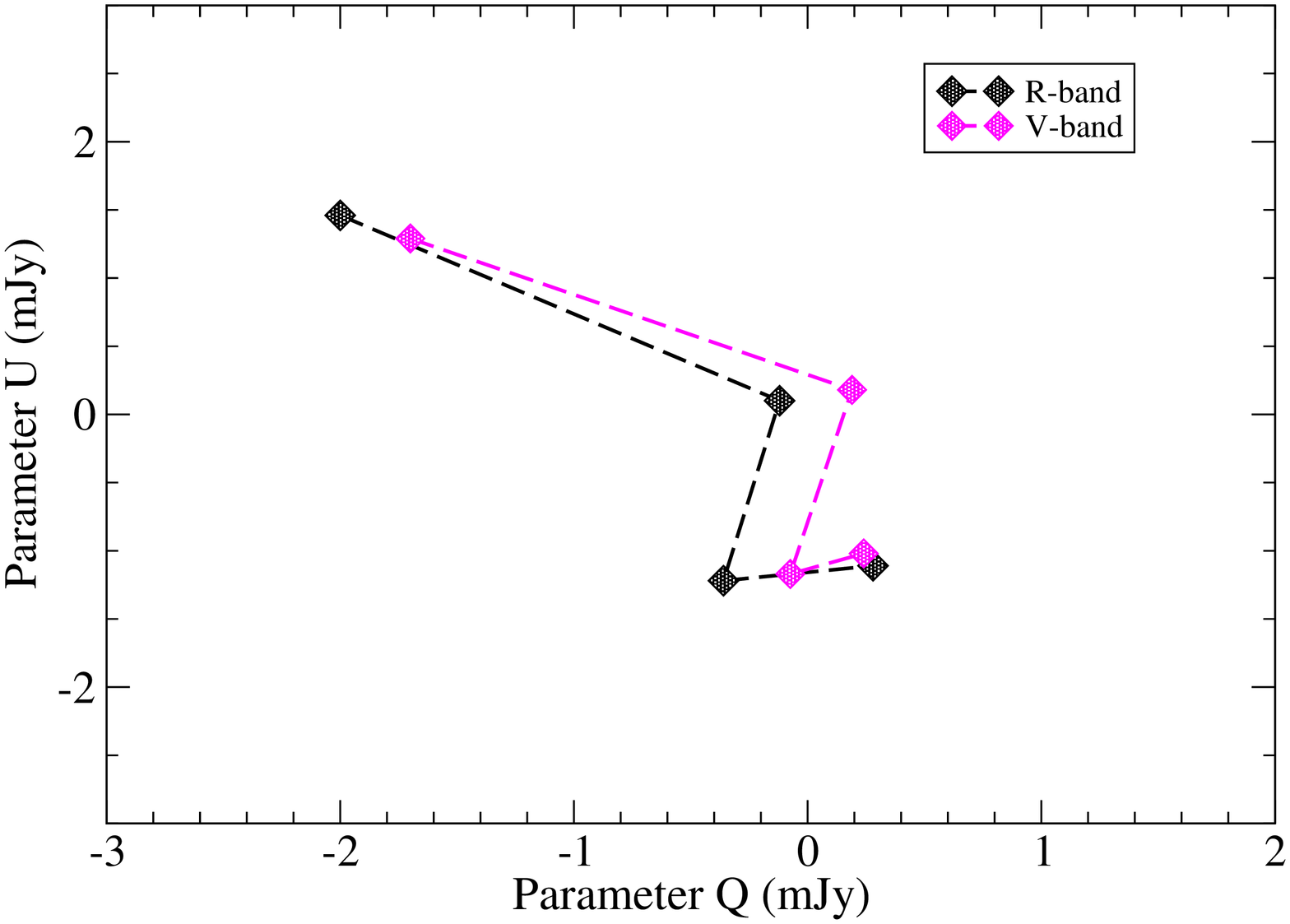}
       \includegraphics[width=5.5cm,angle=-90]{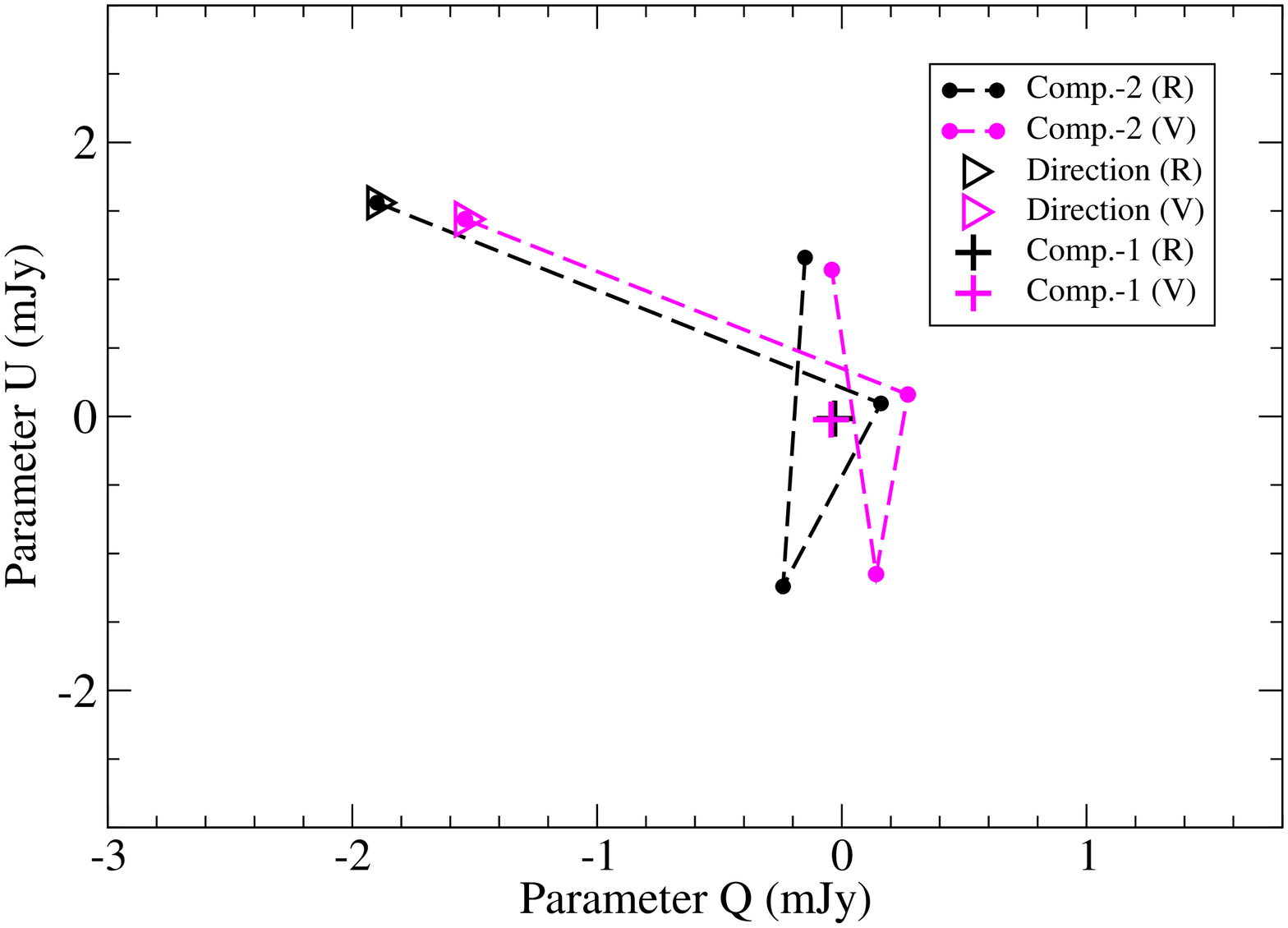}
       \caption{Modeled QU-plots for the integrated outburst (left panel) 
       and for the flare component (right panel). The flare component more
       clearly exhibits the position angle rotation. Triangles 
       indicate the start of the QU-tracks.}
       \end{figure*}
      Holmes et al. (\cite{Holm84}) proposed a two component model to explain
      the observed multi-wavelength light curves of flux density, 
      polarization degree and position angle as a whole.  
       They suggested that the two variable synchrotron (polarized)
       components are optically thin, having similar fluxes and polarization
       degrees  (or similar polarized fluxes), 
      but their position angles are different by near $\sim{90^{\circ}}$, 
      thus the superposition of the two components  leads to the resultant
      very low polarization degrees. More interestingly, they found that
       the two components
      are physically connected in the jet: one had  a stable position angle
     while the other exhibited a gradual rotation in position angle.
      They argued that the observed behavior of position angle rotation 
      may be due to the  physical rotation of the magnetic field.\\
       The detailed modeling of the multi-wavelength light curves (especially
     the wavelength-dependent polarization) performed by Holmes et al. 
      demonstrates that the 1983.0 periodic optical outburst
     is synchrotron in origin, consistent with the assumptions of our
     precessing nozzle scenario (Qian \cite{Qi19a}, \cite{Qi18b}).\\
       Although Holmes et al. used a two-component model to simulate the 
     integrated light curves of flux density, polarization degree and 
     position angle, they did not give the modeled light curves specifically
     for the variable polarized component (or the flare component). In 
     addition, the rotation rate obtained for the component with variable 
     position angle  was less than $\sim{30^{\circ}}$/day, which is much
     smaller than the values observed for the integrated polarization position
     angle. This result seems inappropriately describing the polarization
     behavior of the component with variable position angle.\\
       In the following we will propose an
      alternative two-component model to explain the observed features of 
      the 1983.0 outburst, concretely showing the position angle rotation
      and the Stokes Q-U plot of the flare component.
      \subsection{An alternative model}
       We  suggest an alternative two-component model as follows.
      We assume that the outburst consists of two polarized components defined 
      by ($I_1,p_1,\theta_1$) and ($I_2,p_2,\theta_2$) respectively (see 
      Section 3). The vector combination of the two components forms the 
      observed (integrated)
      polarized outburst described by ($I,p,\theta$). Since polarization
      observations in blazars usually show that just before the beginning 
      or/and just after the end of outbursts the position angle keeps to be 
      very near their preferred directions (e.g. in BL Lacertae,
       Sillanp\"a\"a et al. \cite{Si93}), we will choose the steady component
       to be the underlying quiescent jet component before the
      outburst. According to Holmes et al. component-1
      has a stable position angle $\sim{100^{\circ}}$. Its polarization
      degree is constrained by the very low polarization 
      degrees observed at R-band (0.4\%) and V-band (0.8\%). Its intensity
       is chosen according to the model simulation
      of the V-band light curve and the spectral index ${\alpha}_{RV}$=0.8
       (Qian \cite{Qi19a}). Thus we take:\\
            R-band:  $I_1$=5.7mJy, $p_1$=0.58\%, $\theta_1=104^{\circ}$ \\
            V-band:  $I_1$=5.0mJy, $p_1$=1.0 \%, $\theta_1=104^{\circ}$ \\
       Having these values for the component-1, the values of 
       ($I_2,p_2,{\theta}_2$) 
      for the component-2 (the outburst component) can be  determined
      from solving the equations (1) and (2) given in Section 3.  
      The left panels of Figure 3 show the light curves of integrated flux
       density, polarization degree  and polarization position angle. The
     modeled  light curves for the flare component are shown in the right
     panels.  Note that the flare component has similar rotation rates
     ($\sim{60^{\circ}}$/day clockwise) at R- and V-bands during 
     the 4\,day period.\\
     It is instructive that the flare
     component  has minimal polarization degrees corresponding to its flux
      peaks in R- and V-bands (Figure 3, right/middle panel),
     having large position angle rotations during the peaking period. This
     behavior is similar to that of a superluminal knot moving along a
      force-free helical field (m=1 mode) proposed by  K\"onigl \& Choudhuri 
     (e.g. for the PA swing in radio
     wavebands observed in 0727-115,\cite{Ko85a}, \cite{Ko85b}; Aller et al. 
     \cite{Al81}; also Qian \& Zhang \cite{Qi03}, Qian \cite{Qi92}),
      in which relativistic aberration leads to position angle swings 
     (of $\sim{180^{\circ}}$ and more) for a single flaring component 
     (a relativistic shock). In addition, in the present case, there
     should exist a dominated random field component and the observer's
     direction in the shock frame should be nearly perpendicular to the
      shock front to  produce its very low polarization degrees 
      at R- and V-bands (Laing \cite{Lai80}).\\
       For comparison we note that similar  polarization behaviors 
      in optical regime have been
      observed in BL Lac: Sillanp\"a\"a et al. (\cite{Si93}) observed 
      an event  of large polarization position angle swing during 
      26 September - 1 October 1989, having a minimal polarization degree of
      $\sim$3.6\% (V-band), while  Marscher et al. (\cite{Ma08}) 
      observed a large position angle swing of $\sim{240^{\circ}}$
      during 2005.81-2005.83,
       having  minimal polarization degrees of $\sim$2-3\%. The very low 
      polarization degrees revealed in both events combined with the 
      position angle swings were interpreted in terms of relativistic 
      shocks propagating along helical magnetic fields. In particular,
      Marscher et al. found that the
      large position angle swing occurred when a superluminal knot moving 
      through the VLBI core. This association indisputably demonstrates that
      the low polarization degree and position angle swing are produced in
      the jet.\\
      Finally, we would like to point out that the assumed and resultant 
      values obtained in the model simulation are by no means unique and more
     elaborate models are still required for  explaining the entire event, 
     especially taking its multi-wavelength spectral properties into 
     consideration.
     \subsection{Stokes QU-plots}
      Tracks of the Stokes parameters Q and U for the 1983.0  outburst
      are shown in Figure 4. Q-U tracks for the integrated polarization
      and the flare component are  shown in the left and  right
      panels, respectively. It can be seen that after reducing the background 
      component, the Q-U track of the flare component reveals its position 
      angle rotation more clearly. \\
      \section{Simulation of polarization behavior for 2007.8 outburst}
     \subsection{Introduction}
      Villforth et al. (\cite{Vil10}) have made optical observations to study
     the polarization behavior of the periodic optical outburst in 2007.8.
     This outburst showed a very low polarization degree ($\sim$2.4\%, R-band
     at $\sim$JD2454356.75, near the peak of the outburst) and  was thus
      claimed to be  the impact thermal flare predicted by the disk-impact 
     model.\\
     We have collected the data given in Villforth et al. (\cite{Vil10}) 
     to investigate its polarization behavior.  A model-fit to its
     light curve in terms of the precessing nozzle model is shown in Figure 5
     (left panel), which is reproduced from Qian (\cite{Qi19a}). The right 
     panel represents part of the observed light curve where only the 
     flux densities corresponding to the polarization measurements are shown.\\
     The observed (integrated) light curves (during $\sim$JD2454353.5-366) of 
     flux density, polarization degree and polarization
     position angle are presented in the left panels of Figure 6. It can be
     seen that during the peaking stage of the outburst 
     ($\sim$JD2454356.5-361.5) 
      a large position angle swing was  clearly revealed: first a clockwise
     rotation of $\sim{180^{\circ}}$ and then a counter-clockwise rotation
     of $\sim{180^{\circ}}$, forming a "trough" in the light curve of 
     position angle. Obviously, this large position angle swing is a distinct
     feature of the 2007.8 outburst as for the 1983.0 outburst, which should
      not be neglected and  certainly demonstrate the outburst being 
     synchrotron in origin.\\
     It worths noting that the very low polarization degree (2.4\%, at
      $\sim$JD2454356.75) occurred  earlier than the flux density peak
     ($\sim$JD2454357.5) by a day, where the position angle  only slightly 
     changed. Thus the relations between the variation in flux density and 
     polarization degree or position angle are quite complex, which can not 
      be explained in terms of the disk-impact model.\\ 
      Here we suggest a two-component model to  simulate the observed 
     light curves of flux density, polarization degree and position angle
     as a whole.
     \subsection{A two-component model for 2007.8 outburst}
      We assume that the outburst in 2007.7 consists of two polarized 
     components: one steady component-1 defined by the underlying quiescent 
     jet background before the outburst and a variable flare
      component-2, which are described by parameters ($I_1,p_1,\theta_1$) and 
     ($I_2,p_2,\theta_2$), respectively. the values for ($I_1,p_1,\theta_1$)
     are chosen as\\
      $I_1$=7.5\,mJy, $p_1$=8.5\%, $\theta_1=163.3^{\circ}$\\
     The value of $p_1$ is constrained by the low polarization degree (2.4\%)
     observed at $\sim$JD2454356.75, when the position angle only 
    slightly changed. The value of $I_1$ is chosen
      from the model simulation of the light curve (Qian \cite{Qi19a}).\\
     Having given the values of $(I_p,p_1,\theta_1)$, we can derive the modeled 
     light curves of flux density, polarization degree and position angle for 
     the flaring component ($I_2(t),p_2(t),\theta_2(t)$) from solving the 
     equations (1) and (2) given in Section 3.\\
     The model simulation results are shown in Figure 6 (right column). It can
      be seen that the model-derived polarization degree of the flare component
     (component-2) is quite moderlate during the peaking stage 
    ($\sim$JD2454356-2454361), reaching $\sim$10\% (middle panel). \\
      The rotation of the polarization position angle 
     during this period is not uniform  with a mean rate of 
     $\sim{25^{\circ}}$/day. The maximum rate of $\sim{50^{\circ}}$/day 
     occurred at the beginning of the outburst (bottom panel). \\
       \begin{figure*}
       \centering
       \includegraphics[width=5.5cm,angle=-90]{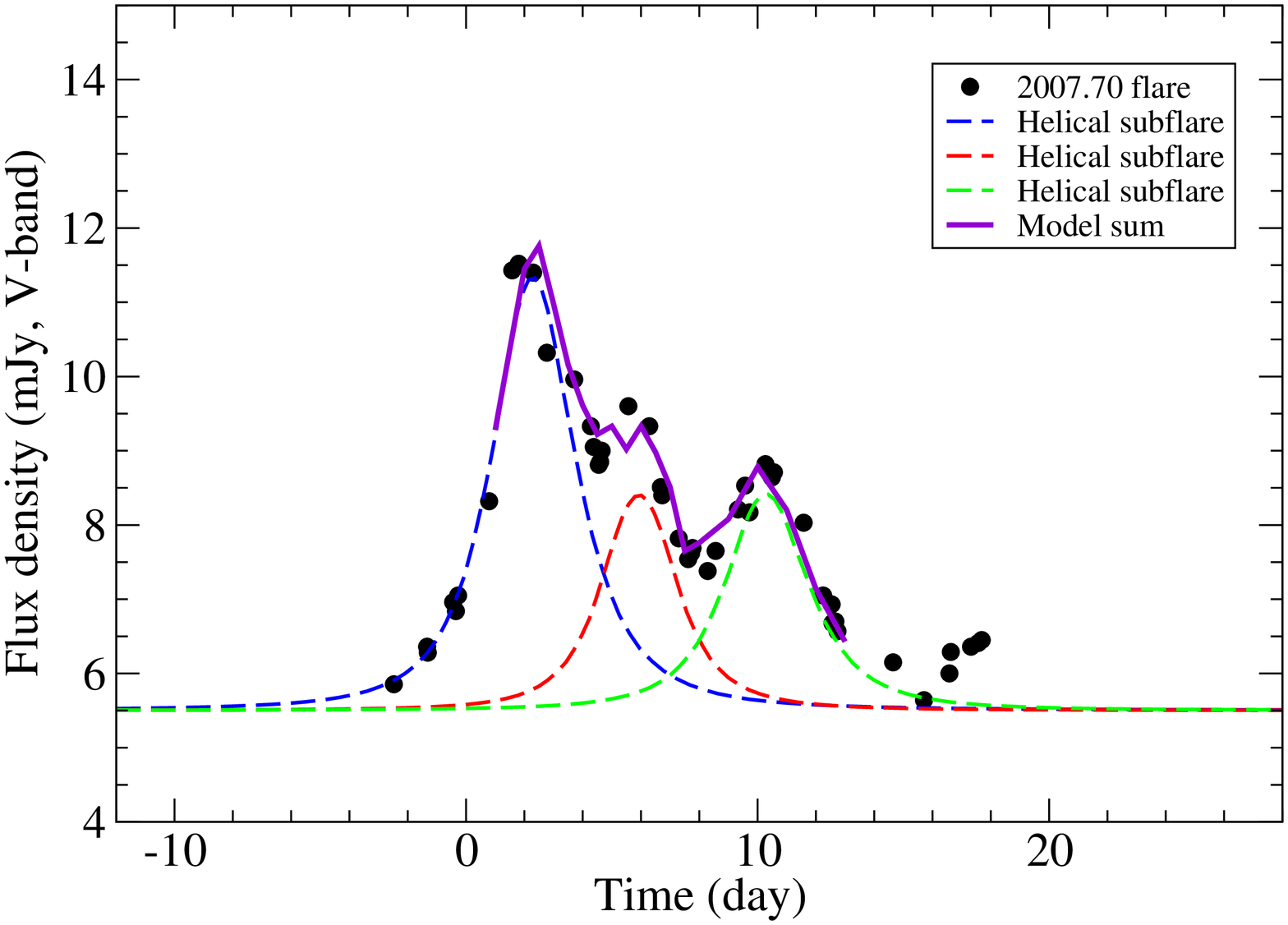}
       \includegraphics[width=5.5cm,angle=-90]{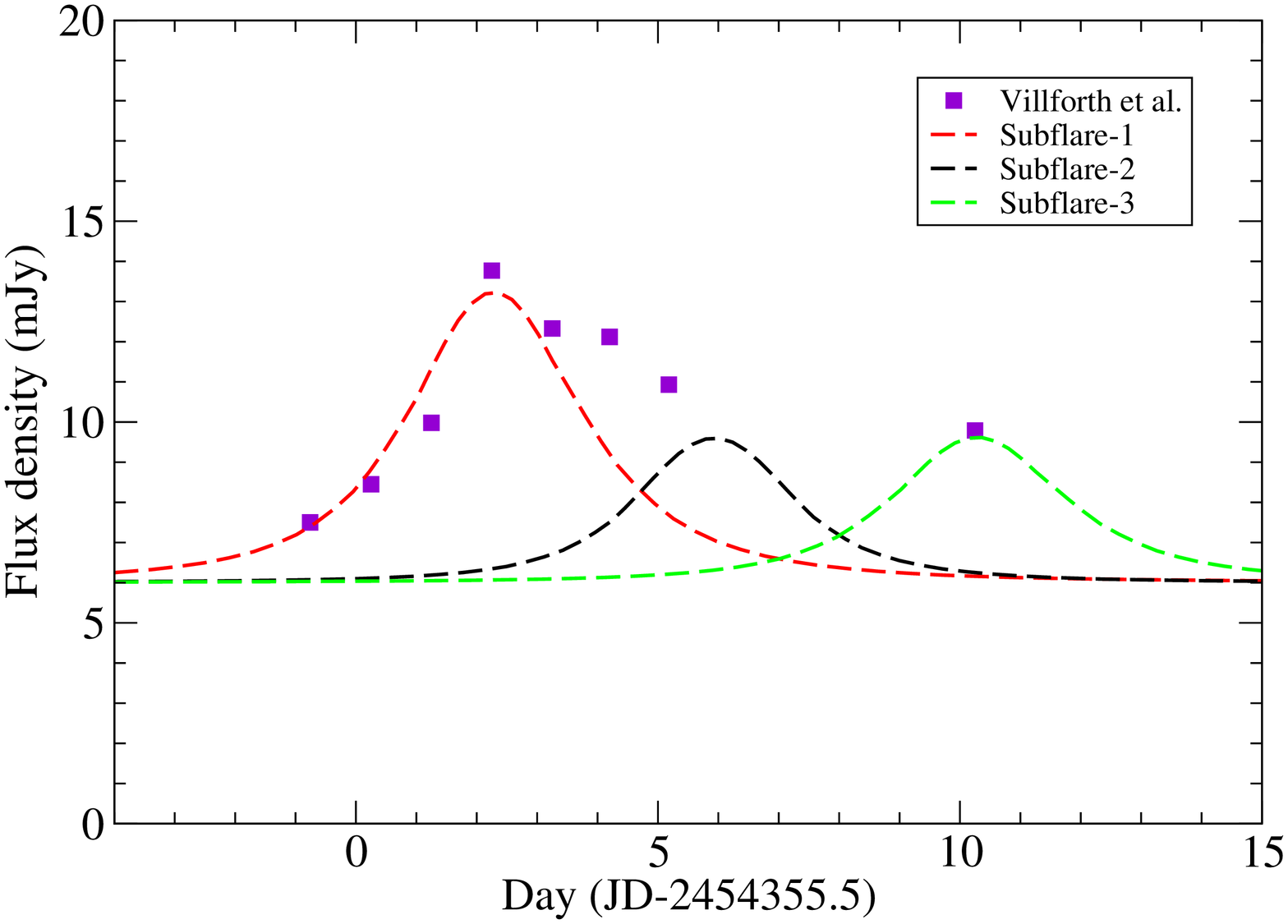}
      \caption{Left panel: A model-fit to the V-band light curve for 
      the 2007.8 outburst reproduced from Qian (\cite{Qi19a}): three elementary 
      subbursts are used ($\Gamma$=11.4 for the first subburst).
        Right panel shows the model-fit of the R-band data
       points collected from Villforth et al. (\cite{Vil10}).}
       \end{figure*}             
     \begin{figure*}
     \centering
     \includegraphics[width=5.5cm,angle=-90]{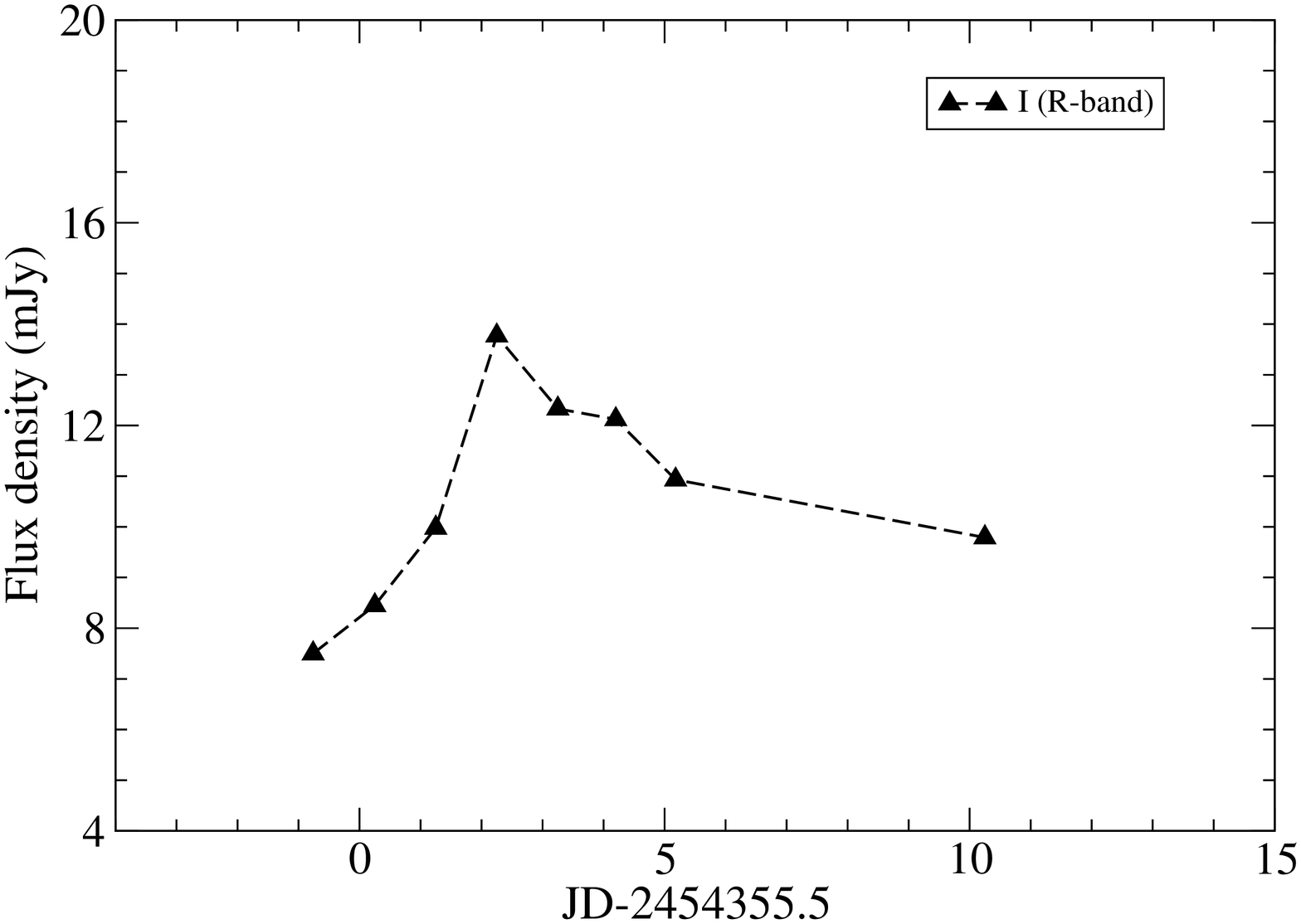}
     \includegraphics[width=5.5cm,angle=-90]{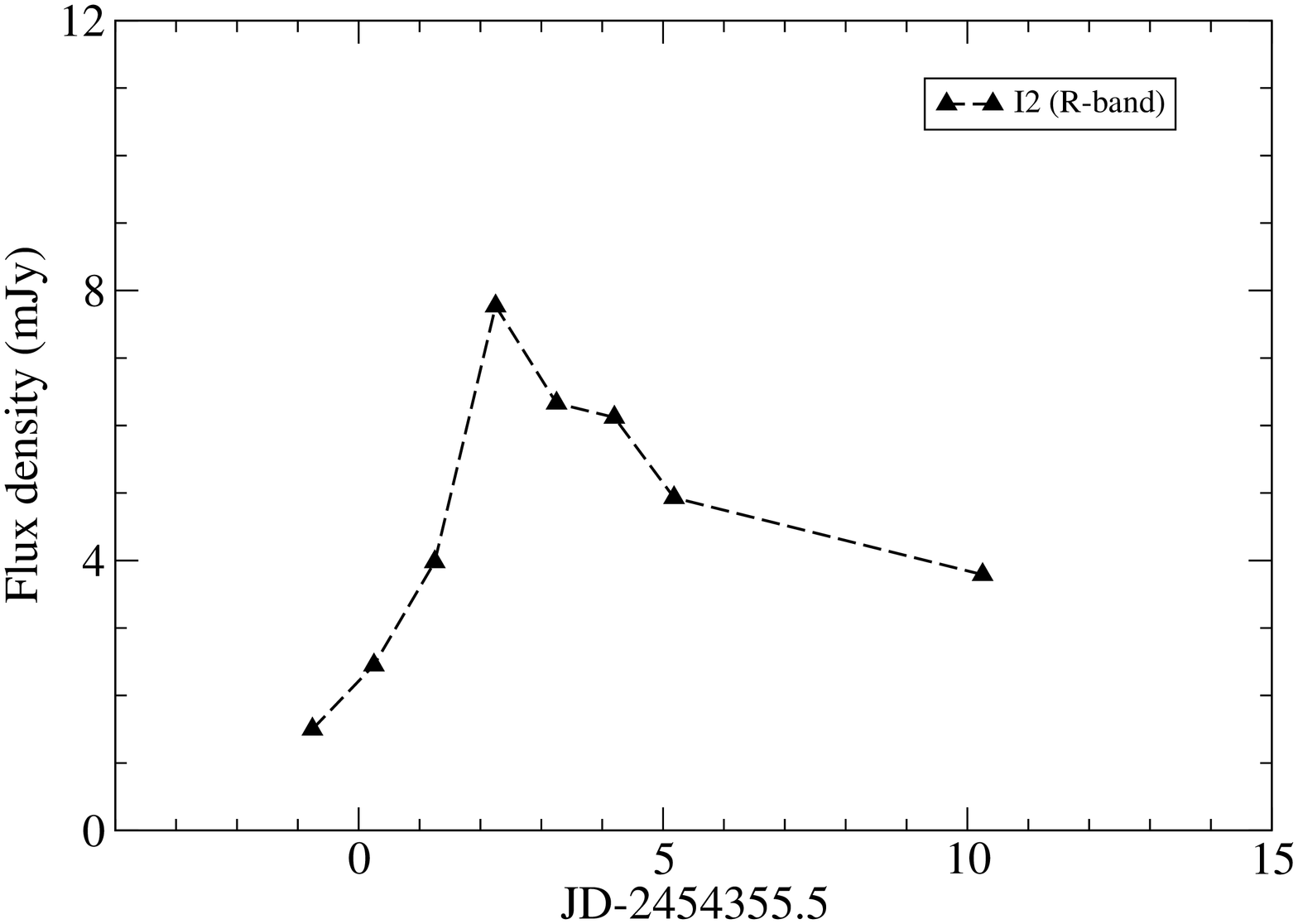}
     \includegraphics[width=5.5cm,angle=-90]{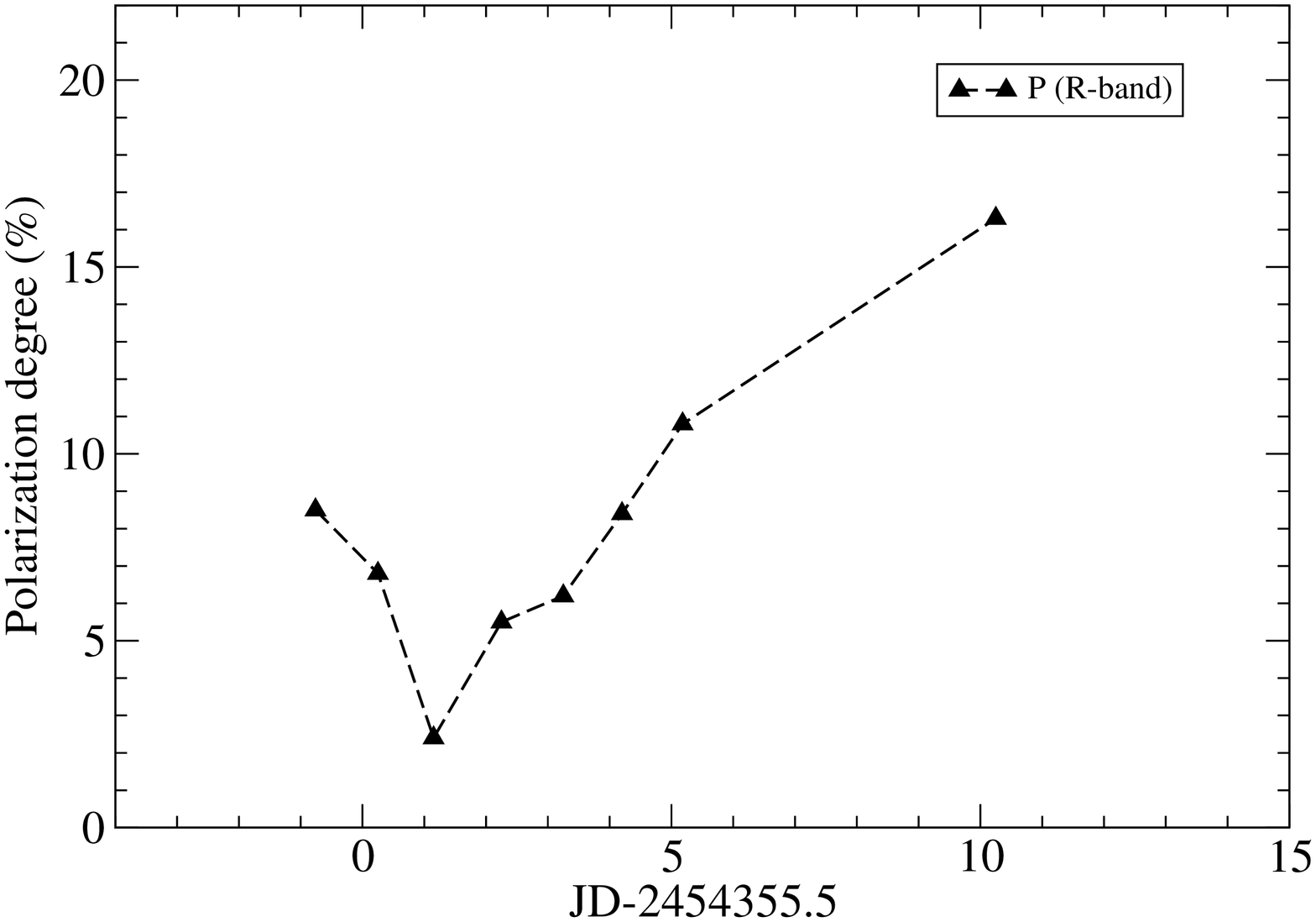}
     \includegraphics[width=5.5cm,angle=-90]{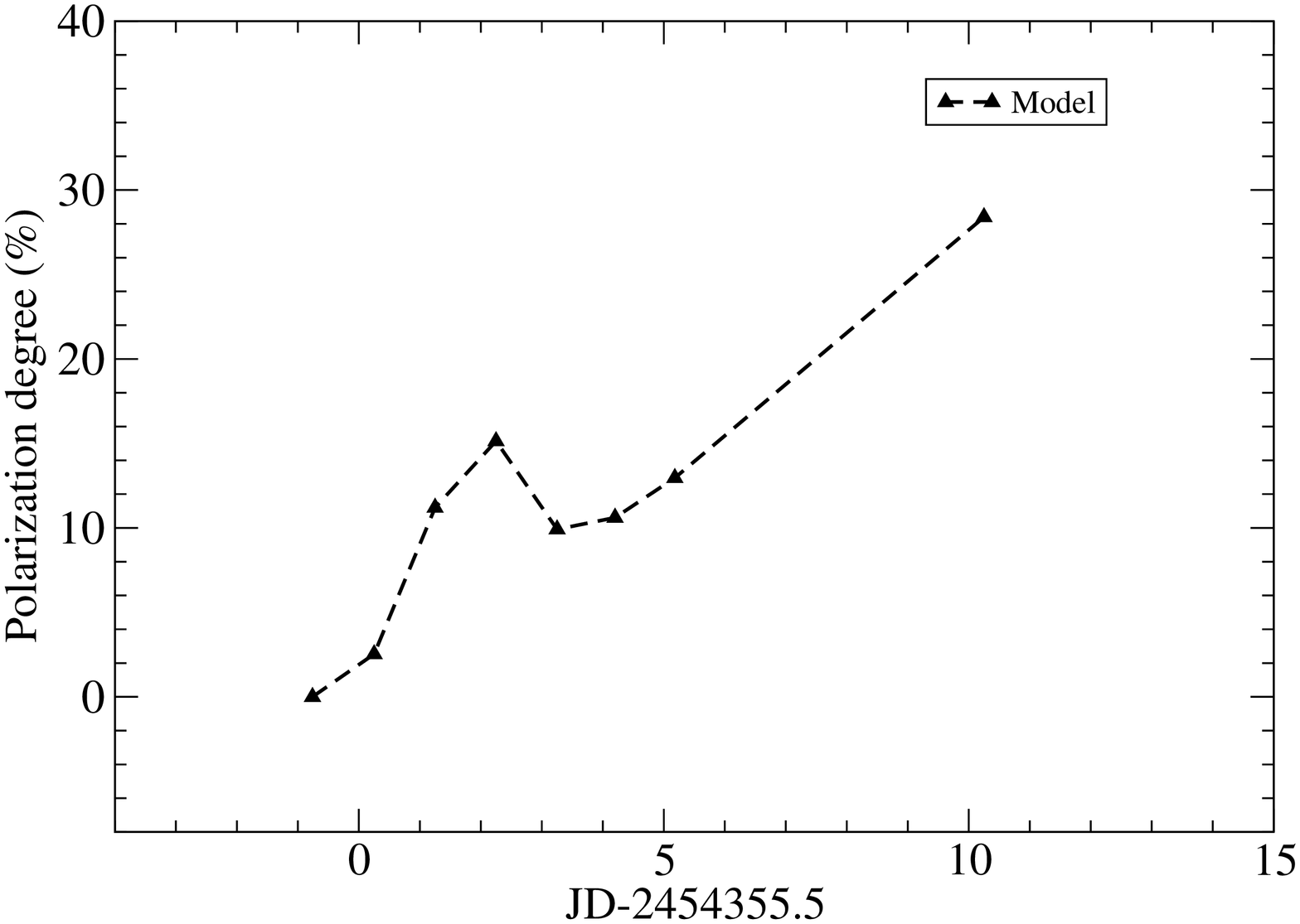}
     \includegraphics[width=5.5cm,angle=-90]{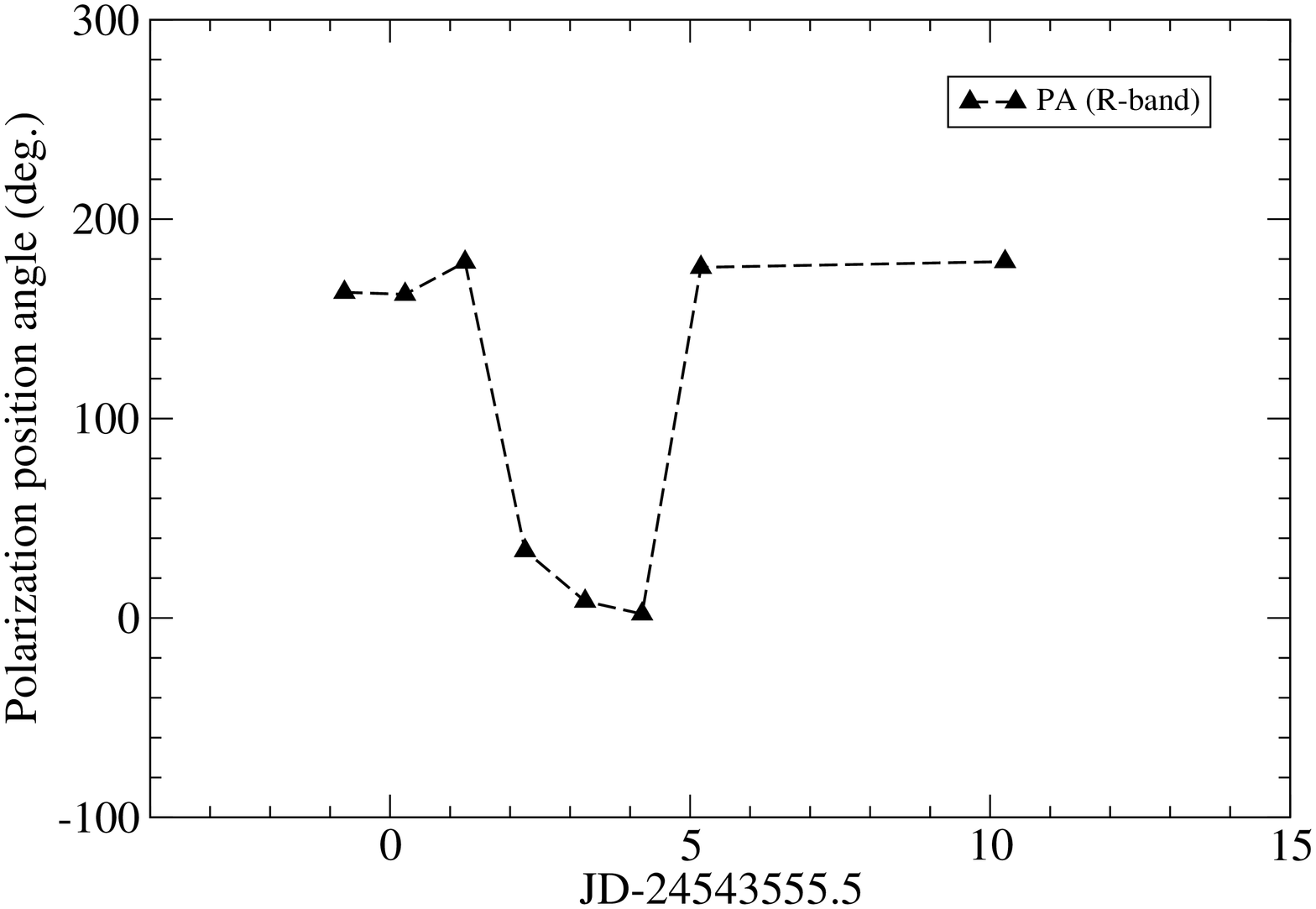}
     \includegraphics[width=5.5cm,angle=-90]{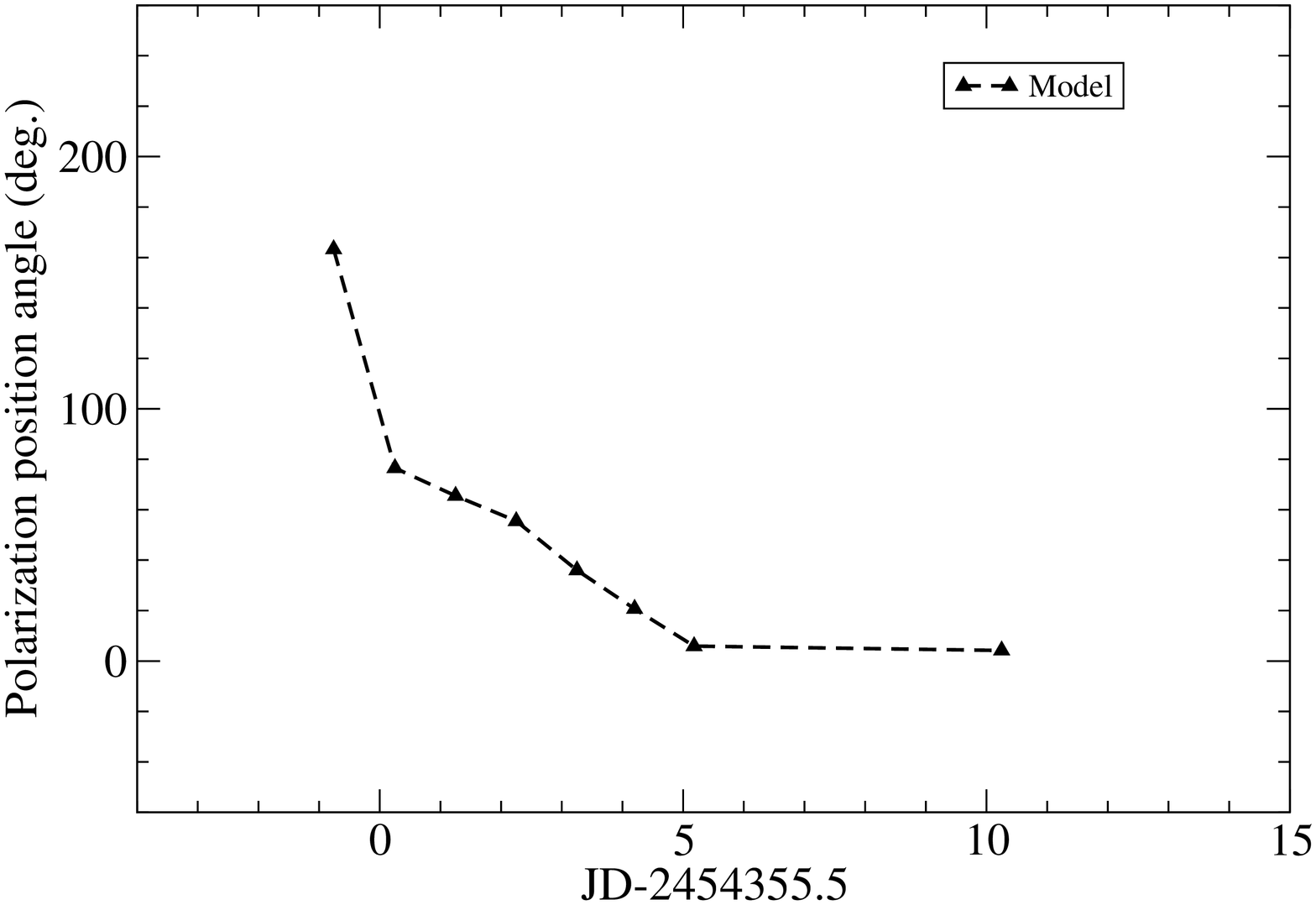}
     \caption{Left column: light curves of the integrated flux density,
     polarization degree and position angle ($I(t),p(t),\theta(t)$) for 
     the 2007.8 outburst, clearly showing the large position angle swings
      during a 5\,day period ($\sim$JD2454356.5-361.5: 
     a clockwise rotation of $\sim{180^{\circ}}$ and  then a counter-clockwise 
     rotation of $\sim{180^{\circ}}$. Right column: modeled light curves 
    ($I_2(t),p_2(t),\theta_2(t)$) during $\sim$JD2454354-366 for the 
    flare component.  A continuous clockwise 
     rotation is clearly revealed during $\sim$JD2454354-361: 
     an average rotation rate of about ${-15}^{\circ}$/day during 
     $\sim$JD2454355.75-360.68 and the maximum rate during 
      $\sim$JD2454354.74-355.75 is about ${-85}^{\circ}$/day.}
     \end{figure*}
     \begin{figure*}
     \centering
      \includegraphics[width=5.5cm,angle=-90]{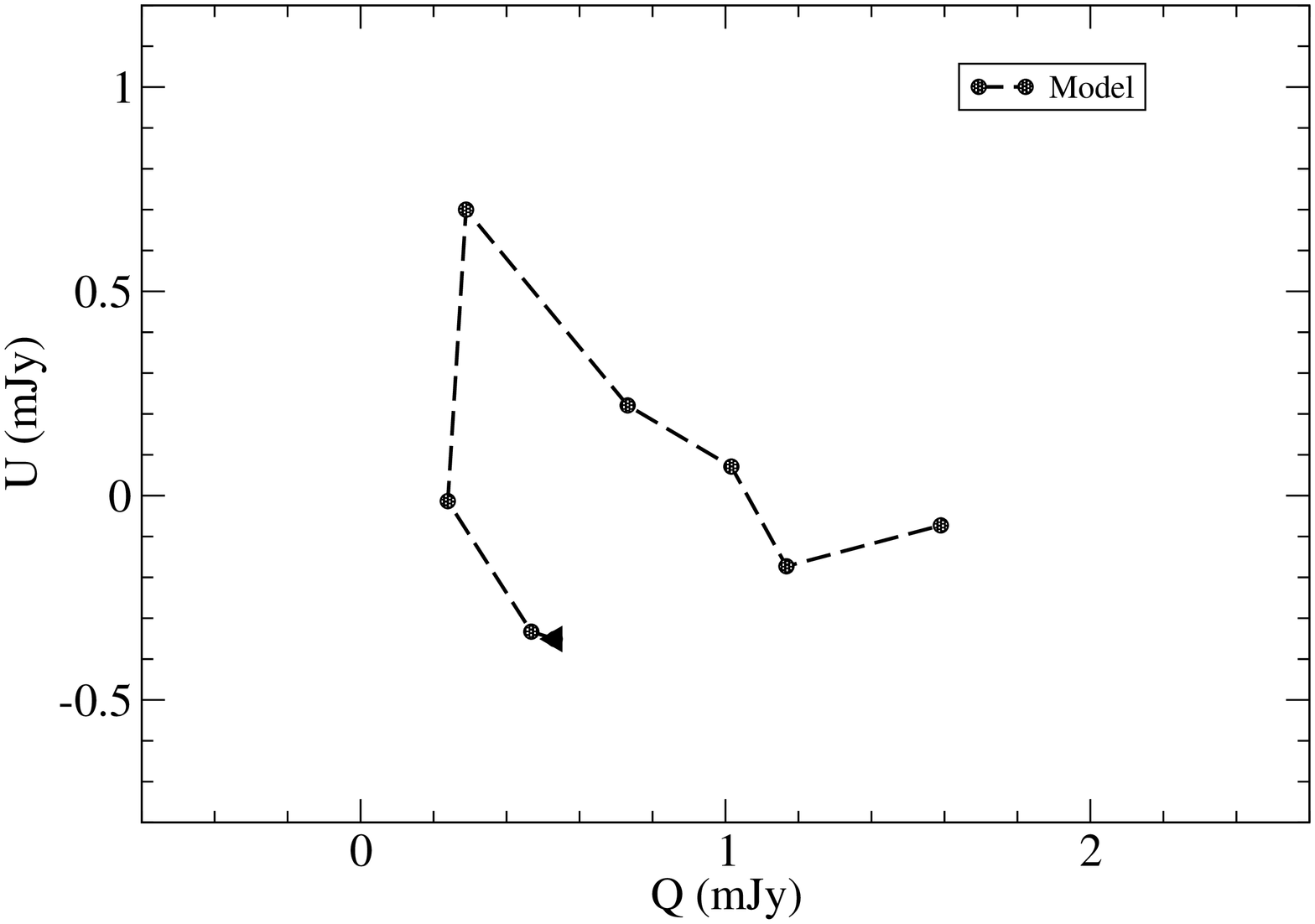}
     \includegraphics[width=5.5cm,angle=-90]{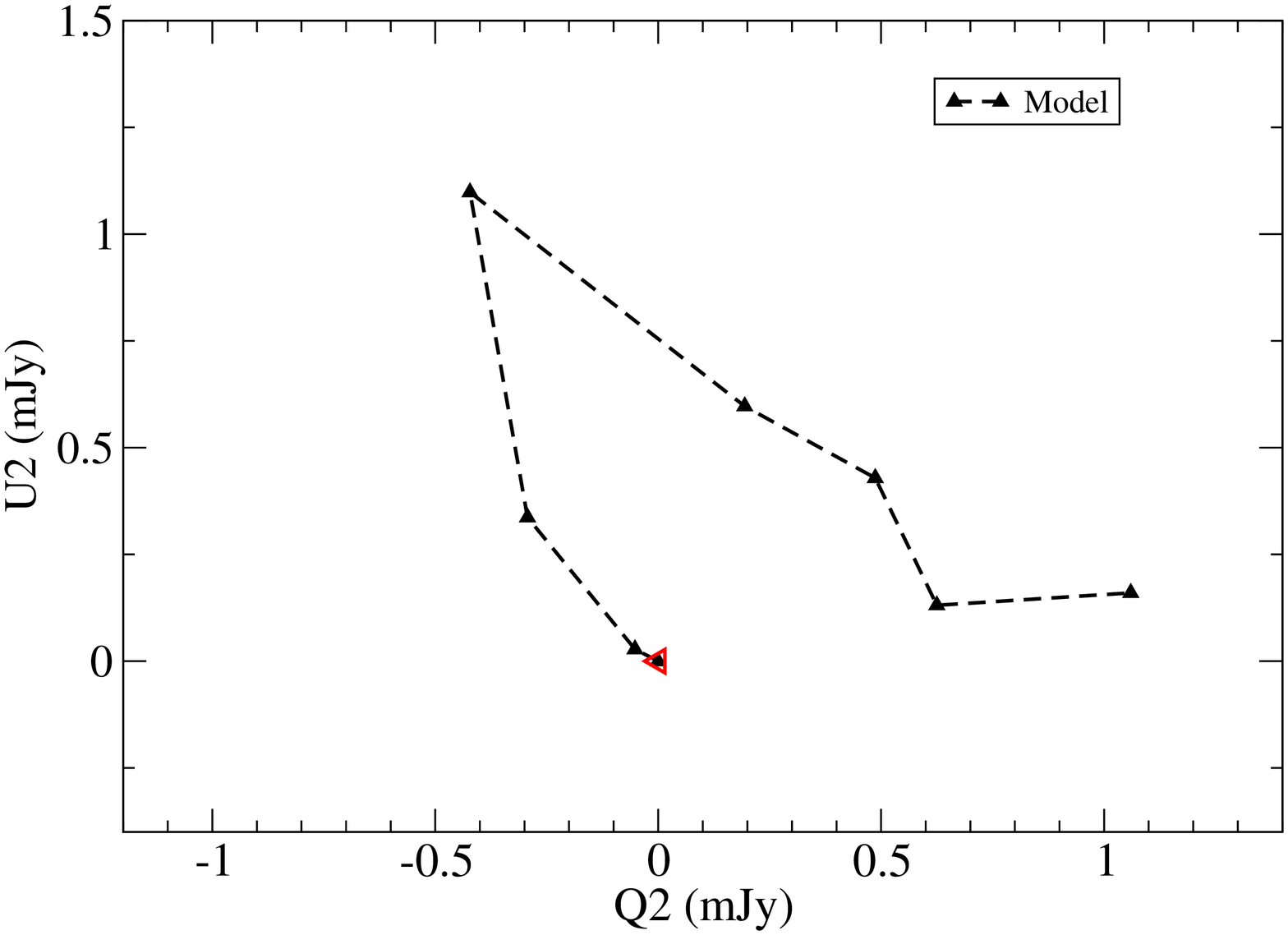}
     \caption{QU-plots for the integrated polarization (left panel) and
      for the flare component (right panel) of the 2007.8 outburst,
     both showing a large position angle swing. Triangles indicate the 
     beginning of the  QU tracks.}
      \end{figure*}
    \subsection{QU-plots}
        The QU-plot during the period of $\sim$JD2454354.5-360.5 for the 
     integrated outburst is shown in Figure 7 (left panel). The QU-track
      derived for the flare-component (right panel) is similar to that of
      the integrated  outburst.
      Both  QU-tracks reveal large position angle rotations, showing the nature
      of synchrotron emission of the 2007.8 outburst.
          \section{Model simulation of the light curves of periodic 
        optical outburst in December/2015}
      \subsection{Introduction}
     The first flare of the quasi-periodic optical outburst observed 
     in December/2015 (during the period of $\sim$2015.88-2015.96 with
     its peak at $\sim$2015.926) was identified as an unpolarized thermal
     flare predicted by the disk-impact model (Lehto \& Valtonen \cite{Le96}, 
     Valtonen et al. \cite{Va19}, Dey et al. \cite{De19}). 
     This interpretation is mainly based on the accurate timing of the 
     orbital motion of the putative black-hole binary in the nucleus of OJ287, 
     where the secondary black-hole penetrates the disk of the primary hole
     twice per one orbital cycle, causing the quasi-periodic optical outbursts
     with double structure through the bubble-production mechanism. 
     The low polarization degrees  observed in the first flares of
     a few periodic outbursts were regarded as firm evidence for
     their thermal origin.\\
      However, there are some distinct features observed in the December/2015
      outburst and in other periodic outbursts could not be explained 
      in terms of the impact-disk model. Relativistic jet models may be 
       required, suggesting that these  periodic outbursts could be  
       synchrotron in origin. For example,
       \begin{itemize}
       \item (1) For the December/2015 outburst,
       Kushwaha et al. (\cite{Ku18a}) observed that a GeV 
       $\gamma$-ray flare was simultaneous with the optical flares 
       at R- and V-bands, peaking concurrently at $\sim$2015.96 
       ($\sim$JD2457361.5). Obviously, the $\gamma$-ray flare could not be 
       co-spatial and associated with the thermal flare produced by the 
       bubble torn off the disk of the primary hole. Both the $\gamma$-ray
        and optical flares should be produced in the relativistic jet through
       synchrotron/inverse-Compton mechanism as in generic blazars.
       \item (2) The multi-wavelength (J, I, R, V, UV) light curves of 
       the December/2015 outburst are very similar to that of the strongly 
       polarized synchrotron flare occurred in March/2016
       (peaking at $\sim$JD2457450). This similarity might imply that the 
       December/2015 optical outburst is synchrotron in origin and this pair
       of outbursts could be interpreted in terms of the helical motion of 
       a superluminal optical knot through two helical cycles via lighthouse 
       effect, having a period of $\sim$90 days (see right/upper panel of
        Fig.1;  Qian \cite{Qi19a});
       \item (3) Generally, the light curves of the periodic optical
       outbursts consist of a number of subbursts or elementary flares,
       each having a symmetric profile, similar to the individual (isolated)
        non-periodic synchrotron flares occurred during the intervening periods
       (e.g., Valtaoja et al. \cite{Val20}, Qian \cite{Qi19b}). Symmetry in
       the outburst profiles  seems a significant feature (Sillanp\"a\"a et al.
       \cite{Si96a}, \cite{Si96b}) different from the standard  non-symmetric
       profiles of the thermal outbursts predicted by the disk-impact model 
       (Valtonen et al. \cite{Va11} ). Symmetric profiles are suggested to be
       explained in terms of the lighthouse model under the precessing 
       jet-nozzle scenario proposed by Qian et al.  (\cite{Qi19a}).
        \item (4) The variability behavior of optical polarization (especially
       polarization position angle) of the periodic optical outbursts
       seems particularly important for determining the origin of the
       emission from the outbursts. For example, a low-polarization degree
       can be due to the appearance of a thermal outburst, but can also be
       caused by the appearance of a synchrotron flare which has its 
       polarization perpendicular to that of the  preexisting steady 
       synchrotron component with a similar polarized flux. In this case
       large changes in polarization position angle should occur.
       The studies of the polarization behavior for the 
       optical outbursts in 1983.0 and 2007.8  (Holmes et al. \cite{Holm84},
       D'Arcangelo et al. \cite{Da09}, Qian \cite{Qi19b}, and this paper
       (Sections 4 and 5) indicate that these outbursts may be synchrotron 
       flares. In section 6 we will investigate the polarization behavior of
       the first flare of the  December/2015 outburst in detail.
      \item (5) Some periodic optical outbursts have been observed to exhibit
       simultaneous radio variations. The 1995.9 optical flare is the best
       example: Valtaoja et al. (\cite{Val20}) observed the simultaneous
       optical and radio (at 22 and 37GHz) flares, having similar substructures 
       and envelopes. Obviously, at least this periodic optical outburst must
       be synchrotron in origin and related to the relativistic jet 
       (Qian \cite{Qi19b}). In general, both the connection between the 
       optical and radio variations and the close correlation between the
       optical outbursts and the ejection of superluminal radio components 
       from the core  may indicate that the optical flares (both periodic 
       and non-periodic) are synchrotron flares (Tateyama et al. \cite{Ta99},
       Kikuchi et al. \cite{Ki88}, Britzen et al. \cite{Br18}, 
       Qian \cite{Qi18b}).
     \item (6) The multi-wavelength optical observations of the 
        December/2015 outburst show its color stability (Gupta et al.
        \cite{Gu16}), which is consistent with the monitoring results of
        OJ-94 project during the period 1993.8-1996.1 (Takalo \cite{Tak96a}, 
        Takalo et al. \cite{Tak96b}, Sillanp\"a\"a et al. \cite{Si96a}, 
        \cite{Si96b}). During the OJ-94 project of $\sim$2.5yr time-range
        both periodic outbursts (the pair of flares in 1994.59 and 1995.81)
        and a large number of non-periodic synchrotron flares were
        observed, indicating that the periodic and non-periodic outbursts 
        may originate from similar emission process. Otherwise how could we
        explain the periodic and non-periodic outbursts having similar 
        optical spectrum (Sillanp\"a\"a \cite{Si96a}, \cite{Si96b}).
        Obviously, the color stability is difficult to be explained in the 
        impact-disk model.
        \end{itemize}     
        In order to clarify the nature of the quasi-periodic optical outburst
        observed in December/2015, we have collected some polarization data 
        from the literature (Myserlis et al. \cite{My18}, Kushwaha et al. 
        \cite{Ku18a}, Valtonen et al. \cite{Va16}, \cite{Va17}, \cite{Va19}) 
       for investigating its 
       polarization behavior (especially the rotation in polarization position 
       angle of the flaring component) and showing that the 
       helical-motion model proposed by Qian (\cite{Qi19a})
        may be appropriate to interpret its light curves of flux density, 
        polarization degree and polarization position angle as a whole.
     \begin{figure*}
     \centering
     \includegraphics[width=5.5cm,angle=-90]{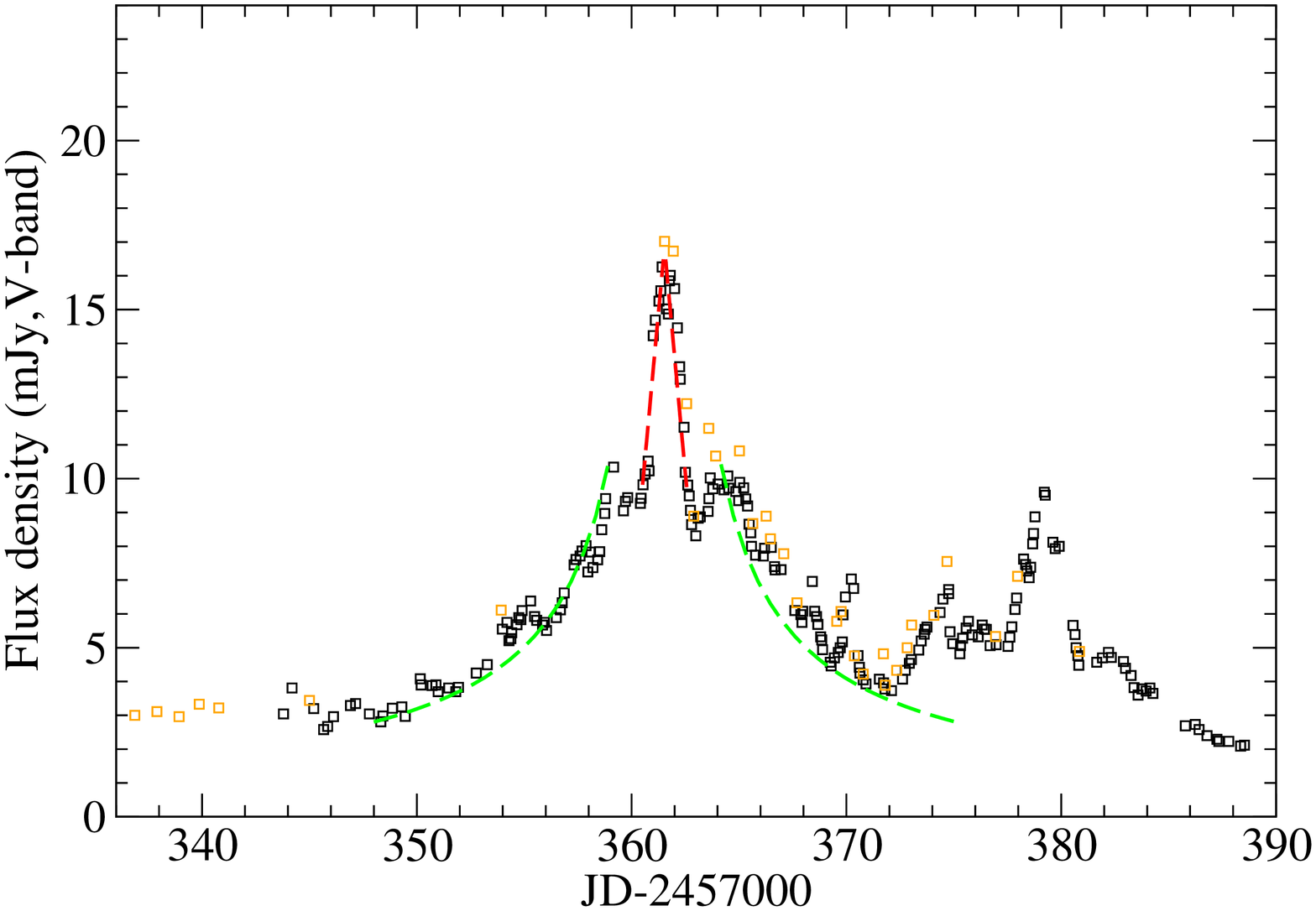}
     \includegraphics[width=5.5cm,angle=-90]{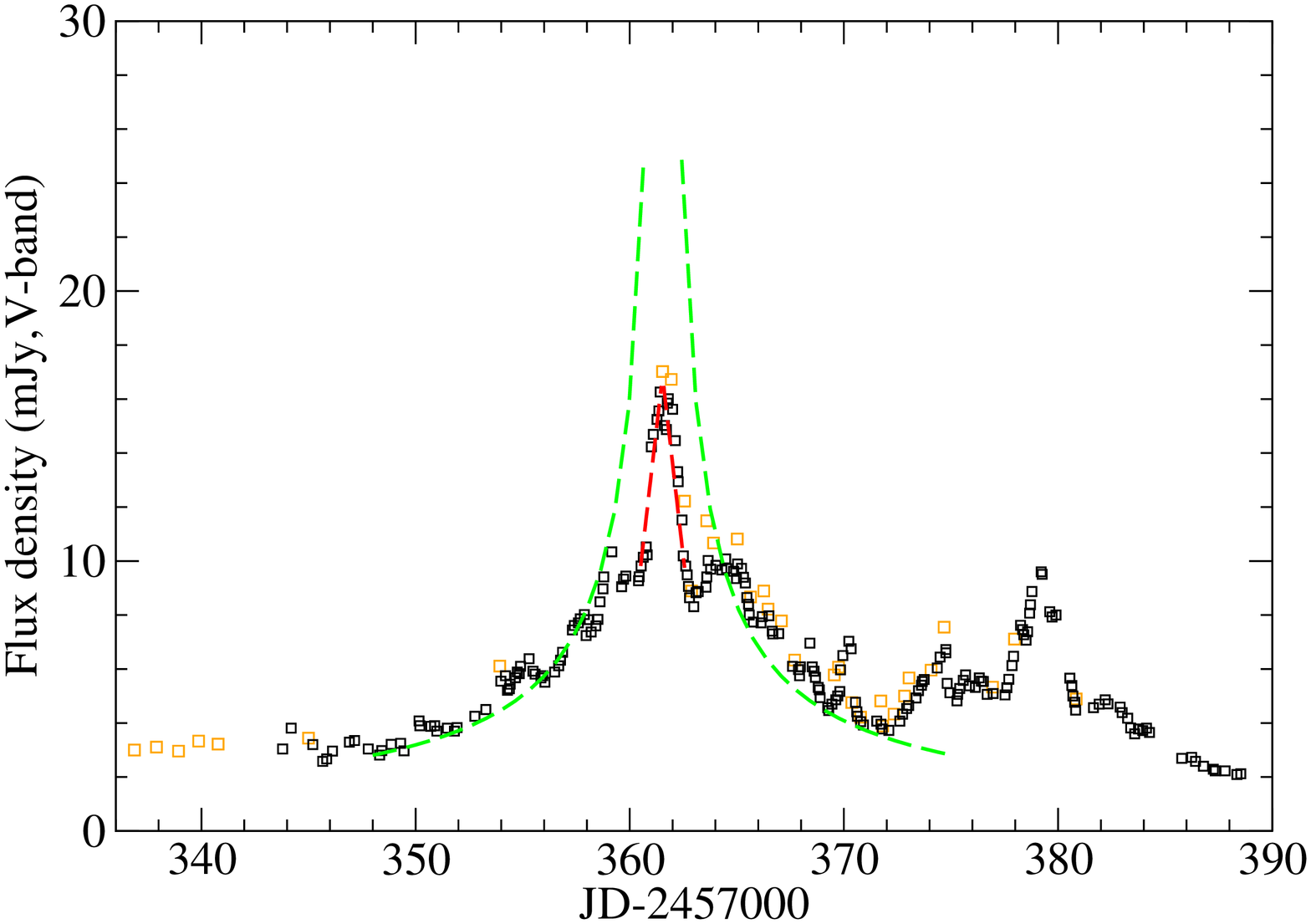}
     \caption{Model simulation of the light curve of the first flare of 
      December/2015 outburst (V-band, during the period $\sim$JD2457350-375),
      which is assumed to have a structure consisting of a central strong 
      spike-like flare (peaking at $\sim$JD2457361.5) and the wing flares 
      on its  either side. In the right panel the modeled light curves for the
      wing flares are extended toward the region of the central spike,
      showing the different rising and declining slopes for the spike and
      wing flares. The symmetry of the central spike (red lines) and 
      wing flares (green lines) relative to the peaking epoch $\sim$JD2457361.5
      are clearly revealed. Data points are collected from Valtonen et al.
      (\cite{Va19}; black squares) and from Kushwaha et al. (\cite{Ku18a};
      orange squares).}
     \end{figure*}
     \begin{figure*}
     \centering
     \includegraphics[width=5.5cm,angle=-90]{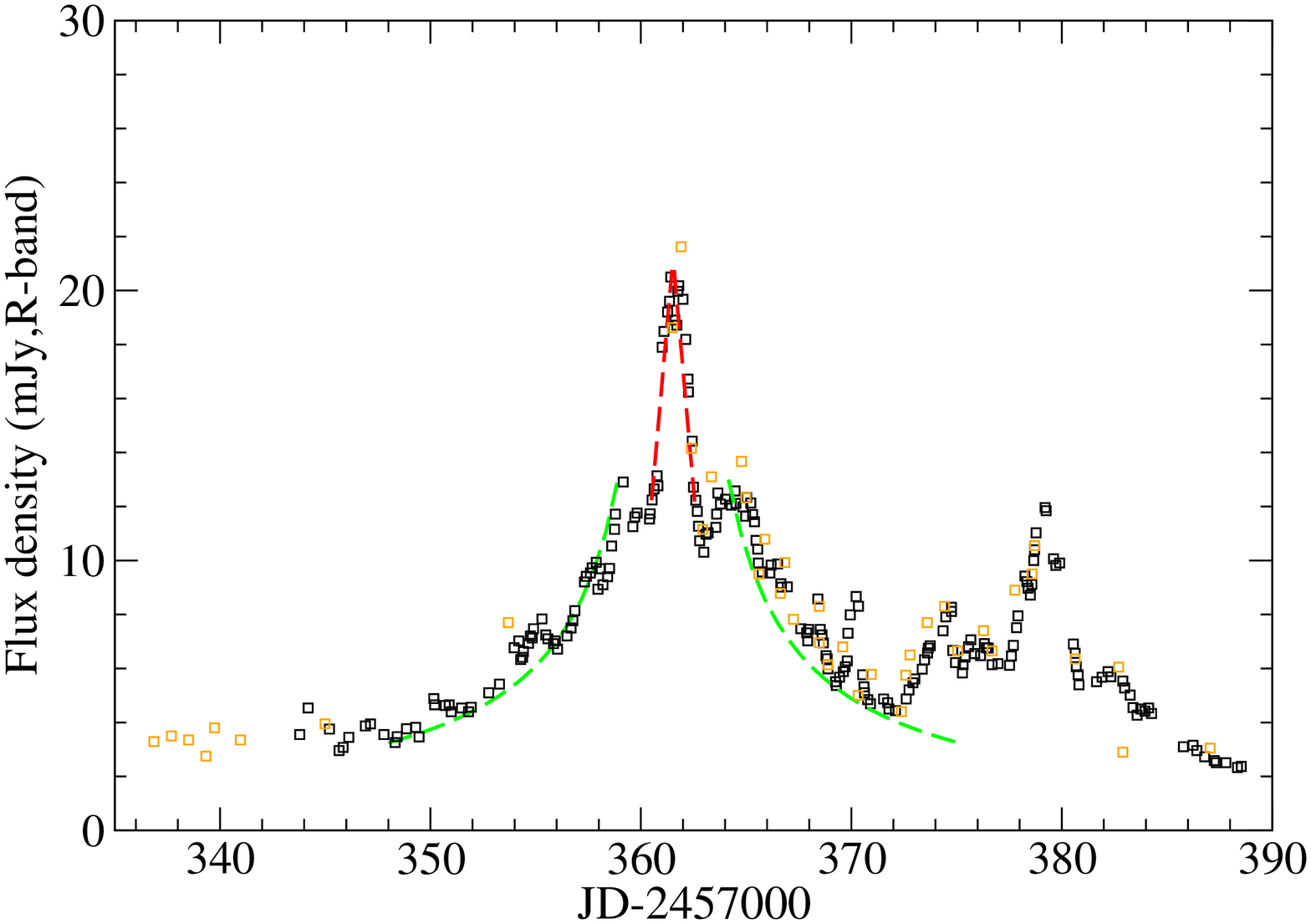}
     \includegraphics[width=5.5cm,angle=-90]{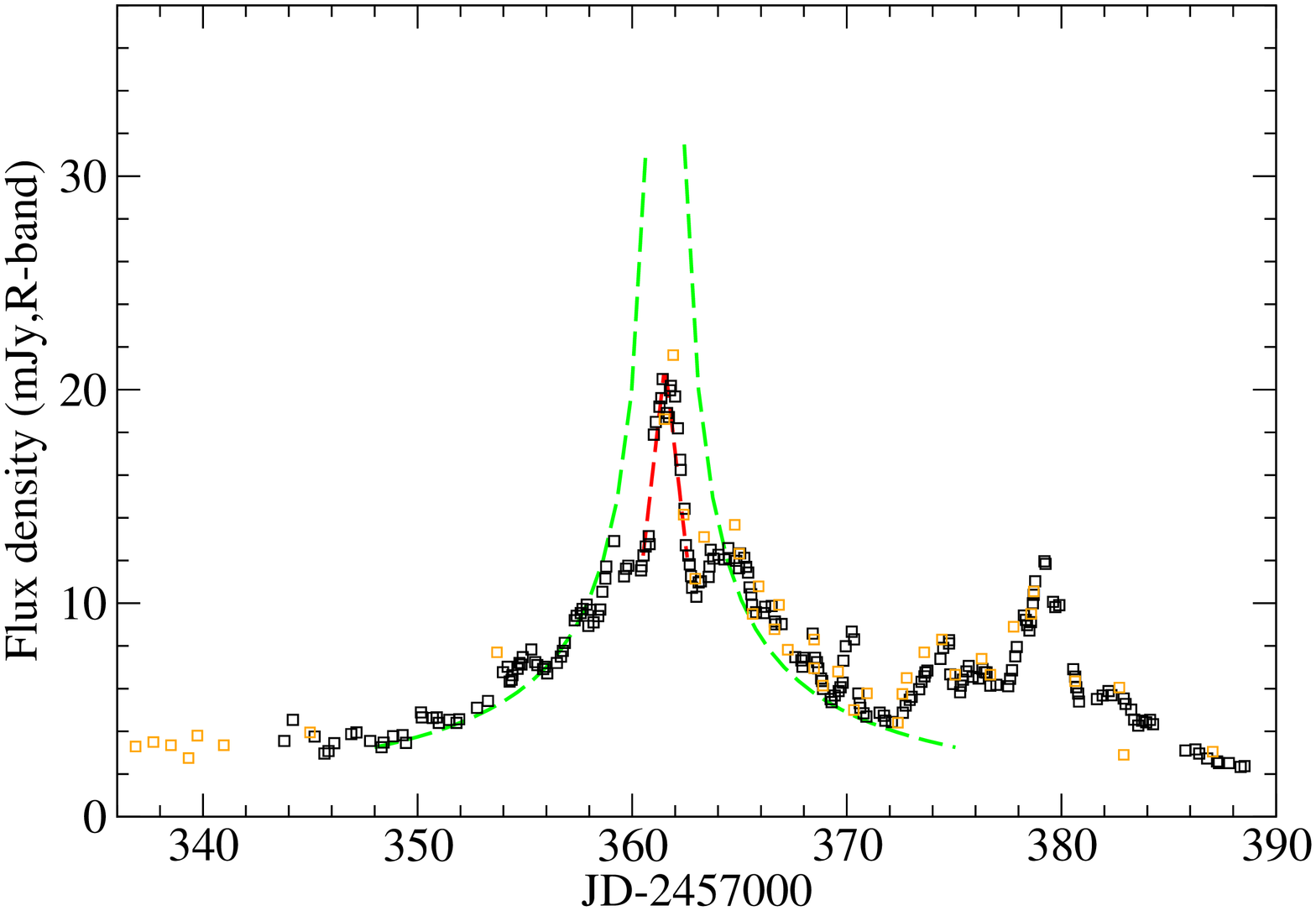}
     \caption{Same as in Figure 8, but for R-band.}
     \end{figure*}
     \subsection{Internal structure and symmetry in flare-profiles}
      It has been proposed that the optical outbursts observed in blazar 
      OJ287 may consist of a number of  elementary-flares with symmetric
      profiles (Qian \cite{Qi19a}). The light curves of the
      December/2015 outburst observed at V- and  R-bands are shown in Figures
      8 and 9.\\ 
       By visual inspection of the light curves some characteristic features 
      can be seen:
      \begin{itemize}
       \item (1) The December/2015 outburst  consists of two 
      flares: the first flare (during $\sim$JD2457350-372) 
      and the second flare (during $\sim$JD2457372-385). Both
      have internal substructures. The first flare has a central strong spike
      between $\sim$JD2457360.5 and $\sim$JD2457362.5 (peaking at 
      $\sim$JD2457361.5) with weaker wing bursts on its either side.
       Although the lower portion of the spike blends with the wing bursts,
      the smooth and almost completely-recorded light curve of its 
      upper portion seems demonstrating that this spike flare is an individual
        elementary flare with a timescale of $\sim$6 days (see below).
       Moreover, this strong spike-flare clearly  has a 
      symmetric profile relative to its peaking time (see the red lines in 
      Figures 8 and 9: flux density $\propto$ $\mid{t-t_{peak}}\mid$). 
       \item (2) In Figures 8 and 9 the modeled curves for the wing 
     flares are shown  in green (for both the rising and declining
      portions), which are also symmetric relative to the peaking time of 
      the central spike: flux density $\propto$ ${\mid{t-t_{peak}}}\mid^{-0.8}$.
      It is quite clear that the wing flares are individual flares 
        independent of the spike flare, because their slopes of the respective
      rising and declining portions (during $\sim$JD2457348.0-360.5 and during 
       $\sim$JD2457363.9-375.1, respectively) are much smaller than 
      the rising and declining slopes of the spike flare 
      (during $\sim$JD2457360.5-362.5). 
       \item (3) The second flare peaking at 
        $\sim$JD2457379.2  has an internal substructures similar to that of 
        the first flare: a strong central spike with weaker wing flares on its
       either side, but having a longer time scale. The sipke flare also 
       has a symmetric profile relative to the peaking epoch. The smooth 
       and almost fully-observed light curve pattern also demonstrates
       that this central relatively strong spike is an individual
       elementary flare with a timescale of $\sim$7 days, independent from 
       the wing flares. The model-simulation results for the second
      flare and its characteristic features can be seen in Figure 10.
      \end{itemize}
       The detailed analysis of the structure of the observed light curves
       for the December/2015 outburst given above may have provided 
       more evidence than before that this optical outburst actually 
       comprise a number of elementary flares.
       It is particularly important that these elementary flares have 
       symmetric profiles as clearly seen in the strong central spike flares
       (peaking at $\sim$JD2457361.5 and JD2457379.2) which have been 
       fully recorded at both R- and V-bands (Valtonen et al. \cite{Va19}, 
        Kushwaha et al. \cite{Ku18a}). It is known that some single 
       (or individual) non-periodic (synchrotron) flares also exhibit
       symmetric profiles (Qian \cite{Qi19a}). These investigations lead us to
       the conclusion that symmetry in the elementary flare profiles 
       is a general property of the outbursts (both periodic and non-periodic) 
       observed in OJ287.\\
       In the precessing jet-nozzle scenario proposed  by Qian (\cite{Qi19a})
       to model-simulate the light curves of the December/2015 optical outburst,
        its structure consisting of elementary flares and 
       their symmetric profiles are the two basic ingredients, which ensure
       the observed flux-density light curves to be explained in terms of the
        helical motion of discrete superluminal
       optical knots (shocks or blobs) in the relativistic jet via 
       lighthouse effect. In Subsection 6.4 below, we will further investigate 
       the optical polarization behavior  of the first flare (during
       $\sim$JD2457350-370) of the December/2015 outburst for trying to find 
       out the true nature of its optical emission.\\
        \begin{figure*}
     \centering
     \includegraphics[width=8cm,angle=-90]{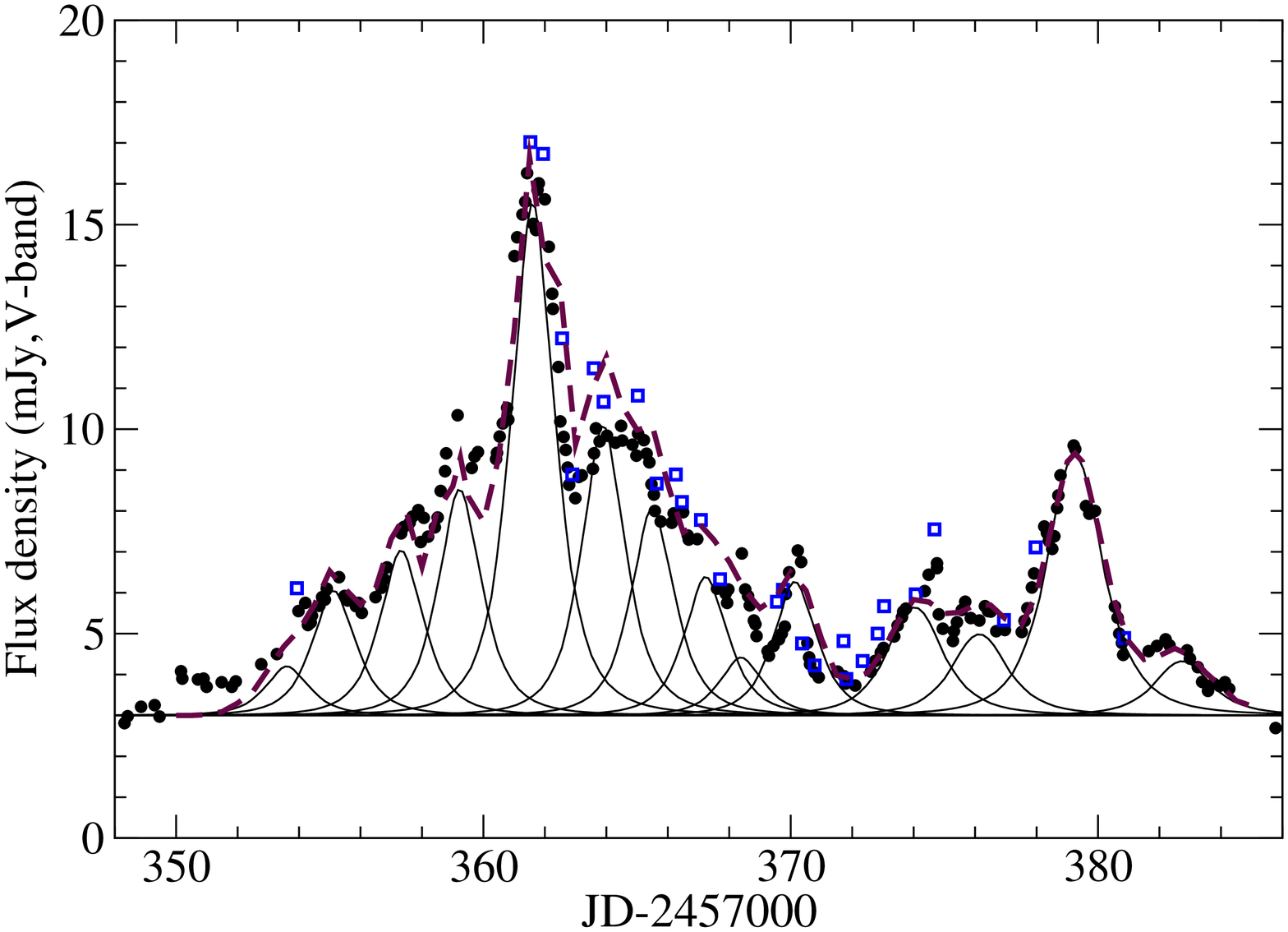}
     \includegraphics[width=8cm, angle=-90]{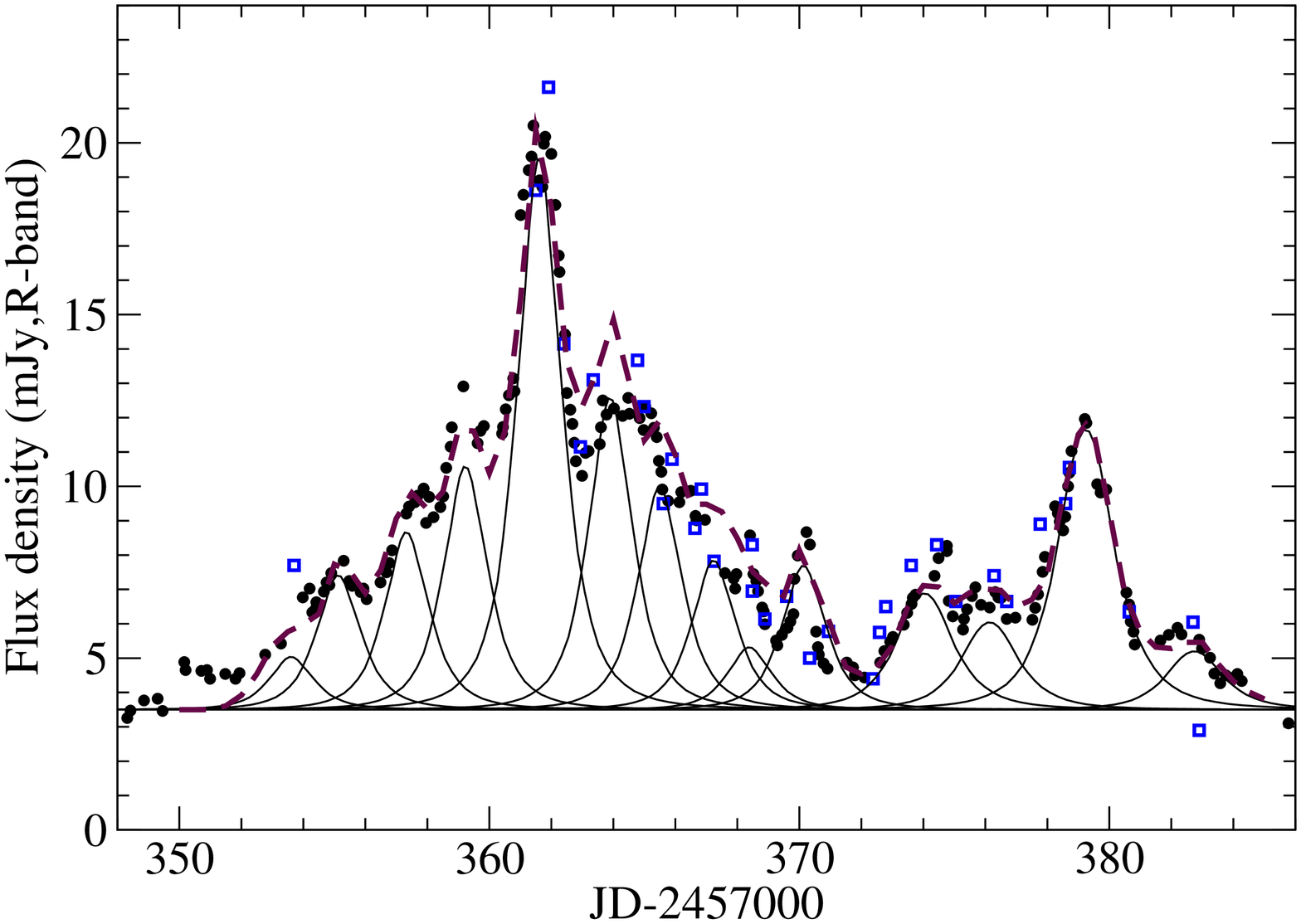}
     \caption{Modeling of the V- (upper panel) and R-band light curves
      in terms of the precessing nozzle model proposed in Qian (\cite{Qi19a}):
     the December/2015 optical outburst are model-simulated by 14 
     symmetric elementary flares: 10 for the first flare (during 
     $\sim$JD2457350-372; $\Gamma$=14) and 4 for the second
      flare (during $\sim$JD2457372-386; $\Gamma$=13).  
     They are all assumed to be produced
      by the helical motion of a succession of discrete optical superluminal
     knots via  lighthouse  effect. Note that the two central spike-like 
     flares peaking at $\sim$JD2457361.5 and $\sim$JD2457379.2 are 
     extremely well model-simulated by symmetric profiles, which
      convincingly confirms the 
     nature of their synchrotron radiation. Bold lines represent the summed 
     flux density.}
     \end{figure*}
      Moreover,  the first and secondary flares of the 
     December/2015 outburst have similar internal structures:
     a central strong spike-like flare with a symmetric profile and 
     weaker wing flares on either side. Two kind of explanations might be
     proposed: (1) If the first and second flares are independent, then they
    could be discrete relativistic shocks  produced in the jet consisting of 
    a shock-front component (causing the strong spike-like flare)
    and a weaker  wake component (causing the wing flares); (2) If the 
   first and  second flares are physically related, they could be a pair 
    of relativistic shocks (both forward and reverse) formed in the collision
    of relativistic flows in the jet, leading to concurrent outbursts.
    In this case there would be four emitting regions naturally formed:
    two shock fronts and two (weaker) wake regions
    (Bell \cite{Bel78a}, \cite{Bel78b}, Rees \cite{Re78}, Kong et al.
    \cite{Kon82}, Hughes et al. \cite{Hu85}, \cite{Hu11}, Lind \& Blandford 
    \cite{Lin85}, Carilli et al. \cite{Car88}, Cawthorne \& Wardle \cite{Caw88},
    G${\rm\acute{o}}$mez et al. \cite{Go94a}, \cite{Go94b}, Cohen et al. 
    \cite{Co18}). This structure of a pair
   of shocks might be just sufficient to explain the structure of the
    December/2015: the two shock fronts produce the two strong spike-like 
    flares, while the two wakes produce the wing
    flares between the two spikes (during $\sim$JD2457363-377). \\
     In both the cases suggested above, the December/2015 outburst would be
     produced via lighthouse effect, when the multiple emitting regions 
     tend to form a stable structure  moving along a helical trajectory in the 
    acceleration-collimation zone of the  relativistic jet in OJ287
     (Camenzind \cite{Cam93}, Camenzind \& Krockenberger \cite{Cam92},
     Schramm et al. \cite{Sc93}, Wagner et al. \cite{Wa95}, Dreissigacker
     \cite{Dr96a}, Dreissigacker \& Camenzind \cite{Dr96b}). In addition,
     there might be also emitting  regions formed in front of
    the forward shocks and/or the reverse shocks due to  some kind of 
    flow instabilities, which might be required to help explain the formation
    of the  wing flares at the beginning and at the end of the December/2015
    outburst.\\
     The proposed interpretation is only a speculative scenario, detailed
     theoretical models are required to specifically investigate the model
     parameters involved for the formation of a pair of 
    shock-wake structures in the
   collision of relativistic flows in OJ287 (e.g., Kong et al. \cite{Kon82}).\\
     We would like to note that
    the Lorentz factors $\Gamma$=14 and 13 adopted in the simulation 
    of the flux light curves respectively for the first and second flares
     are quite close to those derived from centimeter/millimeter variability
     and VLBI-imaging studies for OJ287: $\Gamma$=15.3 (Aller et al. 
    \cite{Al14}) , 15.1 (Lister et al. \cite{Lis13}), 
    16.3 (Jorstad et al. \cite{Jo05}).
     \begin{table*}
     \caption{Model parameters of the 14 elementary flares used to simulate
     the light curve of the December/2015 
     optical outburst:  epoch of the peak (JD-2457000), Lorentz factor,
     peak flux density $S_{V,p}$ and  $S_{R,p}$ (mJy) at V- and R-bands, the 
     corresponding co-moving flux density $S_{V,co}$ and $S_{R,co}$. The numbers
     given in the parentheses represent the power index of ten. For the 
     first and second flares $\Gamma$=14.0 and 13.0, respectively.
     Spectral index is assumed to be ${\alpha}_{RV}$=1.5.}
     \begin{flushleft}
     \centering
     \begin{tabular}{llllll}
     \hline
     epoch & $\Gamma$ & $S_{V,p}$ & $S_{R,p}$ & $S_{V,co}$ & $S_{R,co}$ \\
     \hline
     353.55 & 14.0 & 1.20 & 1.53 & 3.84(-7) & 4.92(-7) \\   
     355.07 & 14.0 & 3.04 & 3.90 & 9.78(-7) & 1.25(-6) \\
     357.27 & 14.0 & 4.02 & 5.16 & 1.29(-6) & 1.66(-6) \\
     359.17 & 14.0 & 5.51 & 7.06 & 1.77(-6) & 2.27(-6) \\
     361.55 & 14.0 & 12.5 & 16.0 & 4.02(-6) & 5.15(-6) \\
     363.84 & 14.0 & 7.07 & 9.06 & 2.27(-6) & 2.91(-6) \\
     365.44 & 14.9 & 5.06 & 6.48 & 1.63(-6) & 2.09(-6) \\
     367.18 & 14.0 & 3.38 & 4.33 & 1.09(-6) & 1.39(-6) \\
     368.33 & 14.0 & 1.41 & 1.81 & 4.54(-7) & 5.81(-7) \\
     370.08 & 14.0 & 3.26 & 4.18 & 1.05(-6) & 1.34(-6) \\
     \hline
     373.97 & 13.0 & 2.64 & 3.38 & 1.18(-6) & 1.51(-6) \\
     376.07 & 13.0 & 1.98 & 2.53 & 8.84(-7) & 1.13(-6) \\
     379.17 & 13.0 & 6.35 & 8.14 & 2.84(-6) & 3.63(-6) \\     
     382.67 & 13.0 &  1.32 & 1.69 & 5.91(-7) & 7.56(-7) \\
     \hline
     \end{tabular}
     \end{flushleft}
    \end{table*}
    \subsection{Model simulation of flux density light curves}
     We first discuss the  model-simulation results of the flux light curves
     observed  at V- and R-bands in terms of the precessing nozzle model.
     In comparison with the previous simulation (Qian\cite{Qi19a}), here
     we have taken into account of the internal structure of the
      December/2015 outburst.
     The modeling  results are shown in Figures 10 and 11, and Table 1.\\
     The December/2015 optical outburst has been assumed to comprise
     a number of elementary flares, each of which is produced by a superluminal
    optical knot moving along a helical trajectory via lighthouse effect. \\
      14 elementary flares are used to simulate  the observed light curves
     at V- and R-bands (see Figure 10 and Table 1). The modeled flux density
     profile, Lorentz/Doppler factor, apparent velocity and viewing angle 
     versus time for the central strong spike-like flares are given in 
     Figure 11. We emphasize that the two spike flares (during 
     $\sim$JD2457359-364 and $\sim$JD2457376-384) are very well fitted by 
     the model-simulated symmetric profiles (see Figure 10). \\
      The similarity in structure (strong spike with wing bursts) of 
      the first flare  and the second flare for the 
     December/2015 outburst may be also in favor of the suggestion that 
     the first flare is also a synchrotron flare as like the second flare, 
     which was observed to have a very high polarization degree ($\sim$40\%).
     The low polarization degree of $\sim$5-6\% during the first flare 
     may be a result due to composition of two or more polarized components
     (see below).  
     \begin{figure*}
     \centering
     \includegraphics[width=5cm,angle=-90]{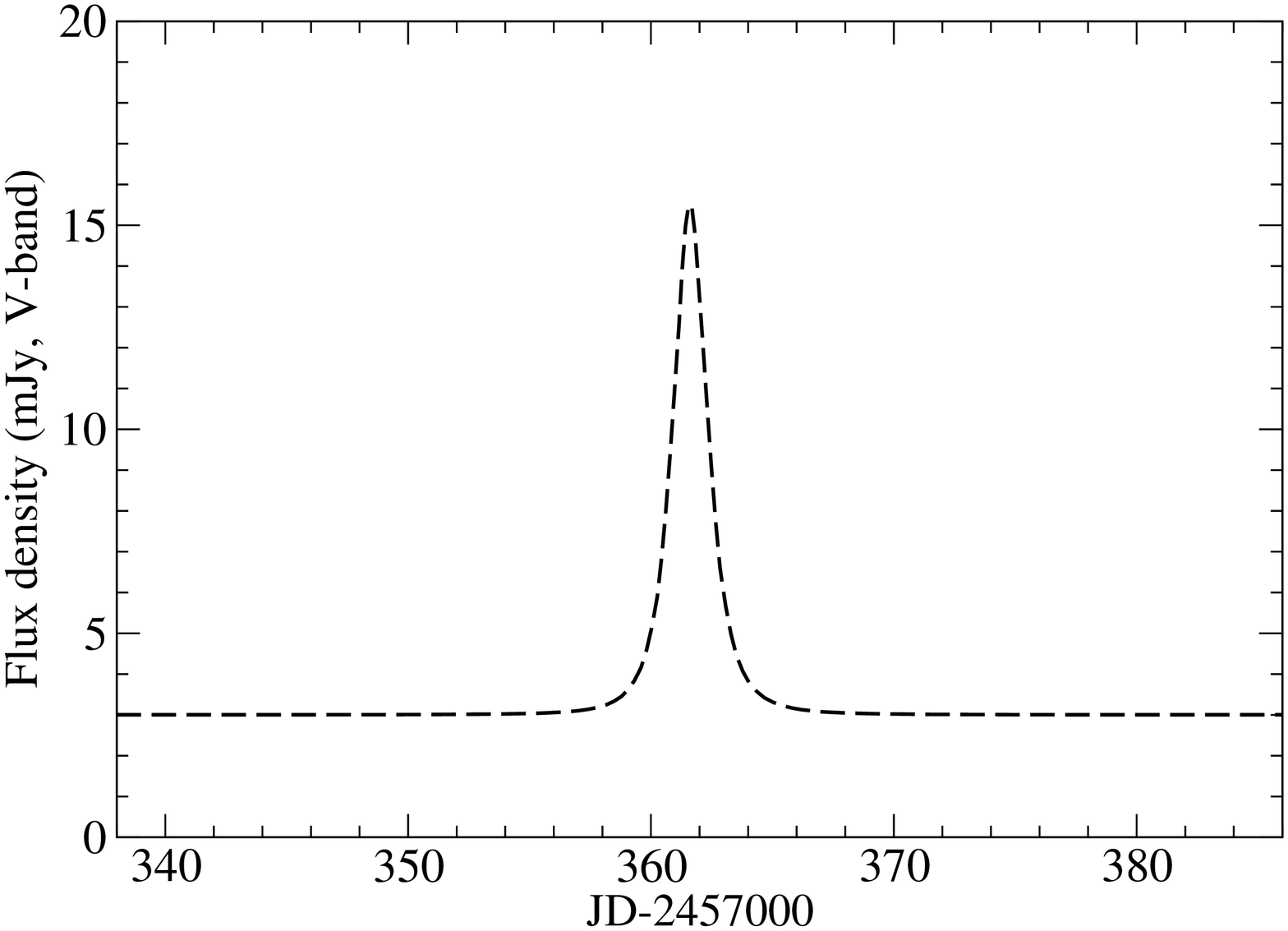}
     \includegraphics[width=5cm,angle=-90]{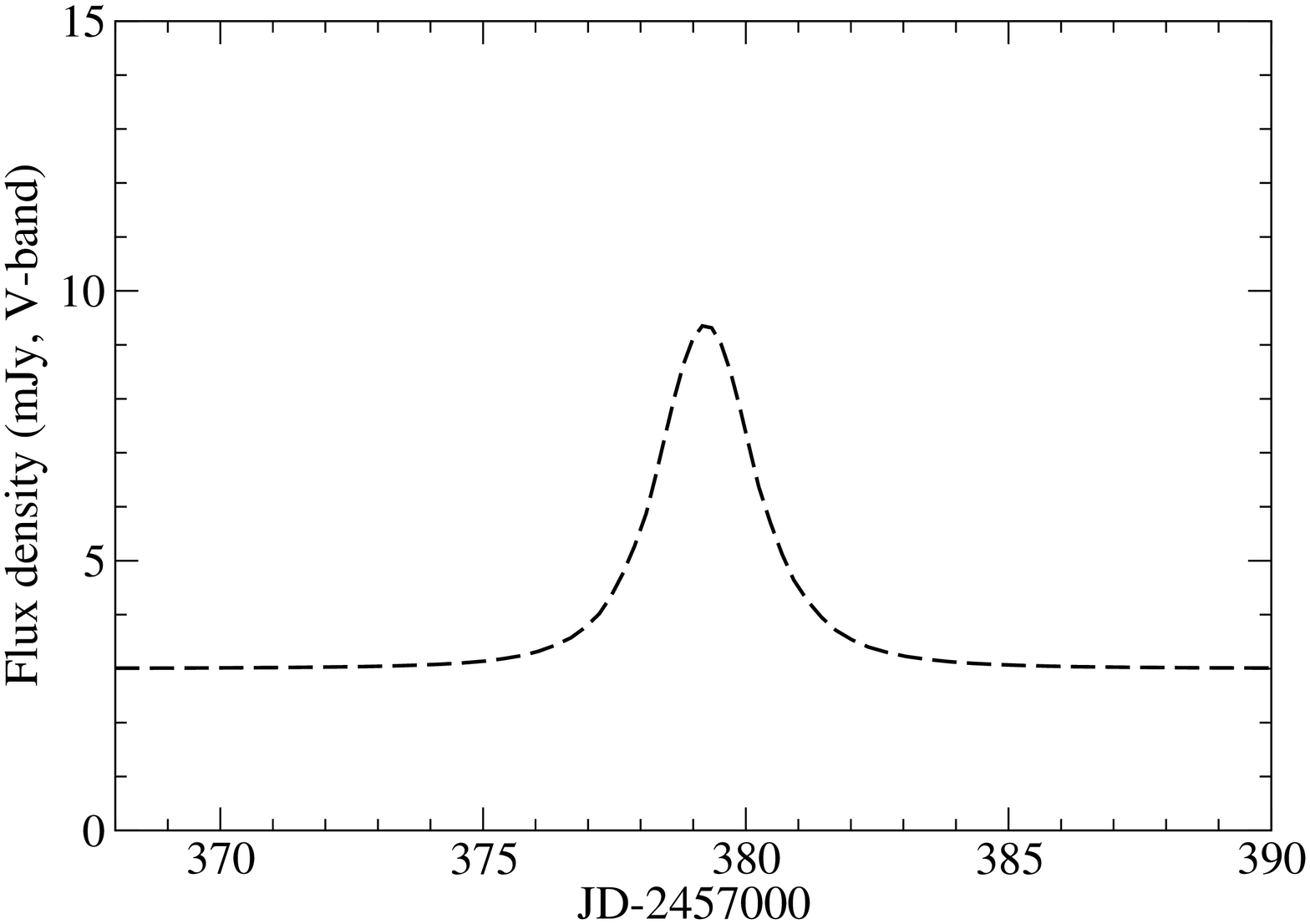}
     \includegraphics[width=5cm,angle=-90]{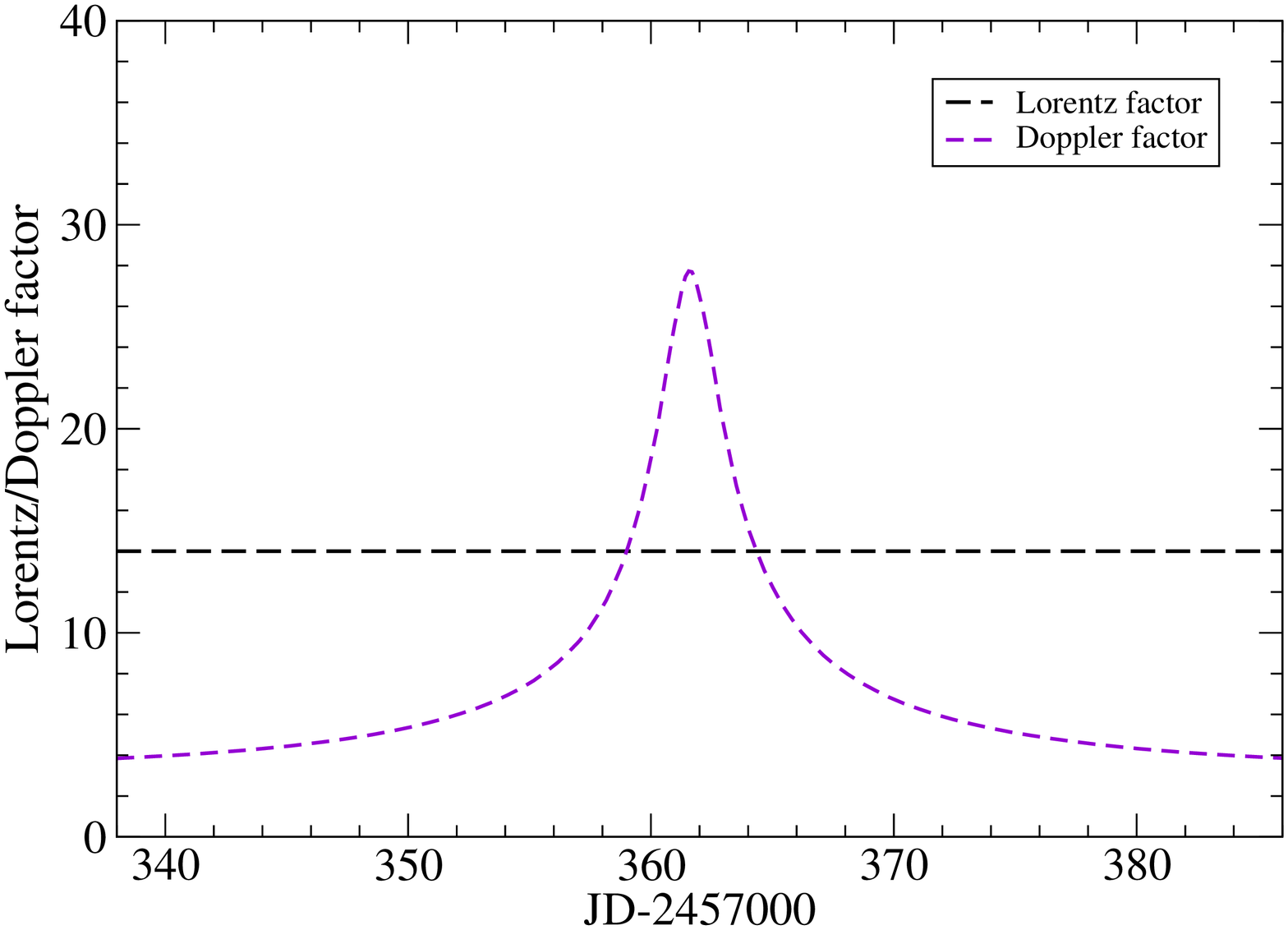}
     \includegraphics[width=5cm,angle=-90]{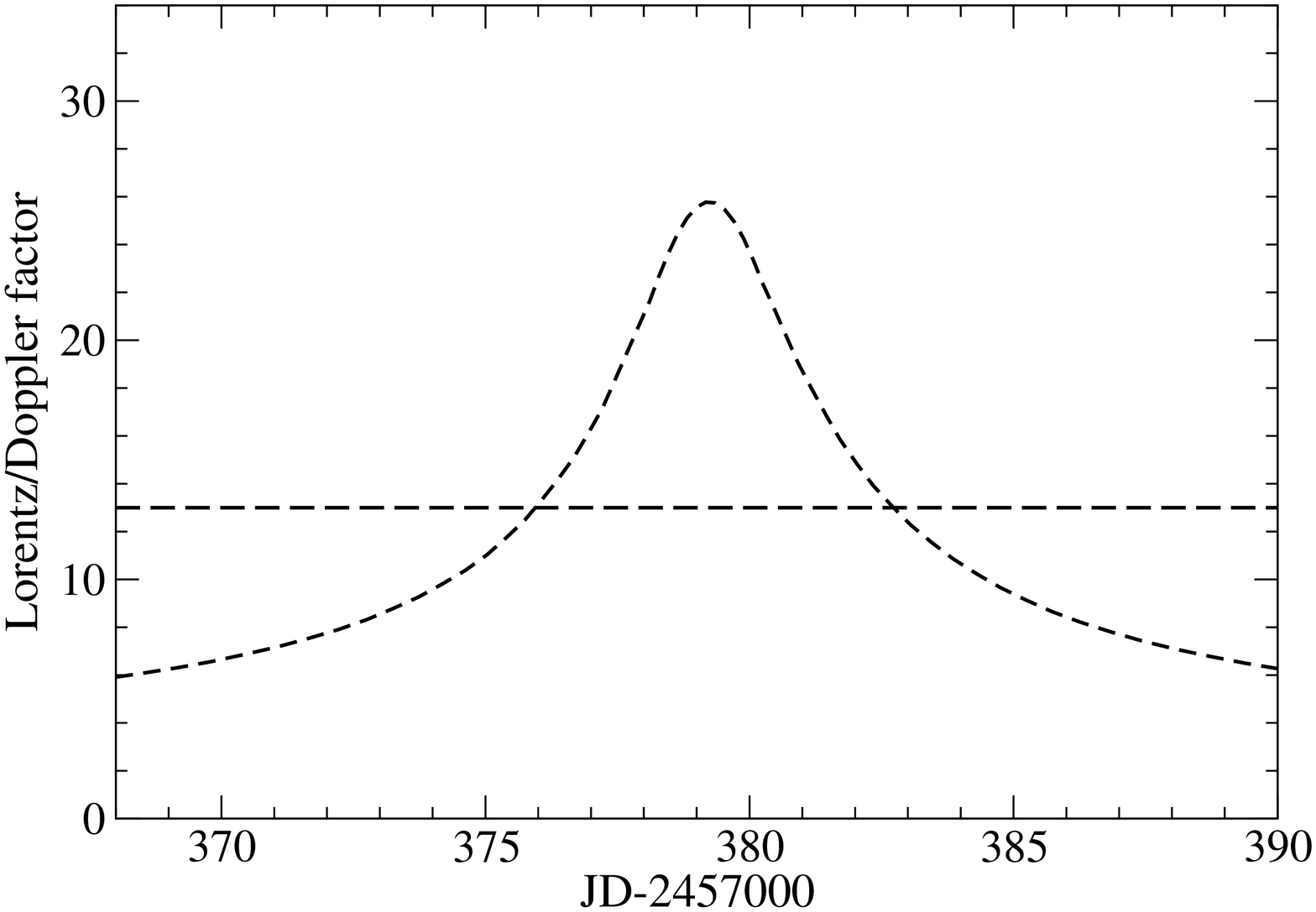}
     \includegraphics[width=5cm,angle=-90]{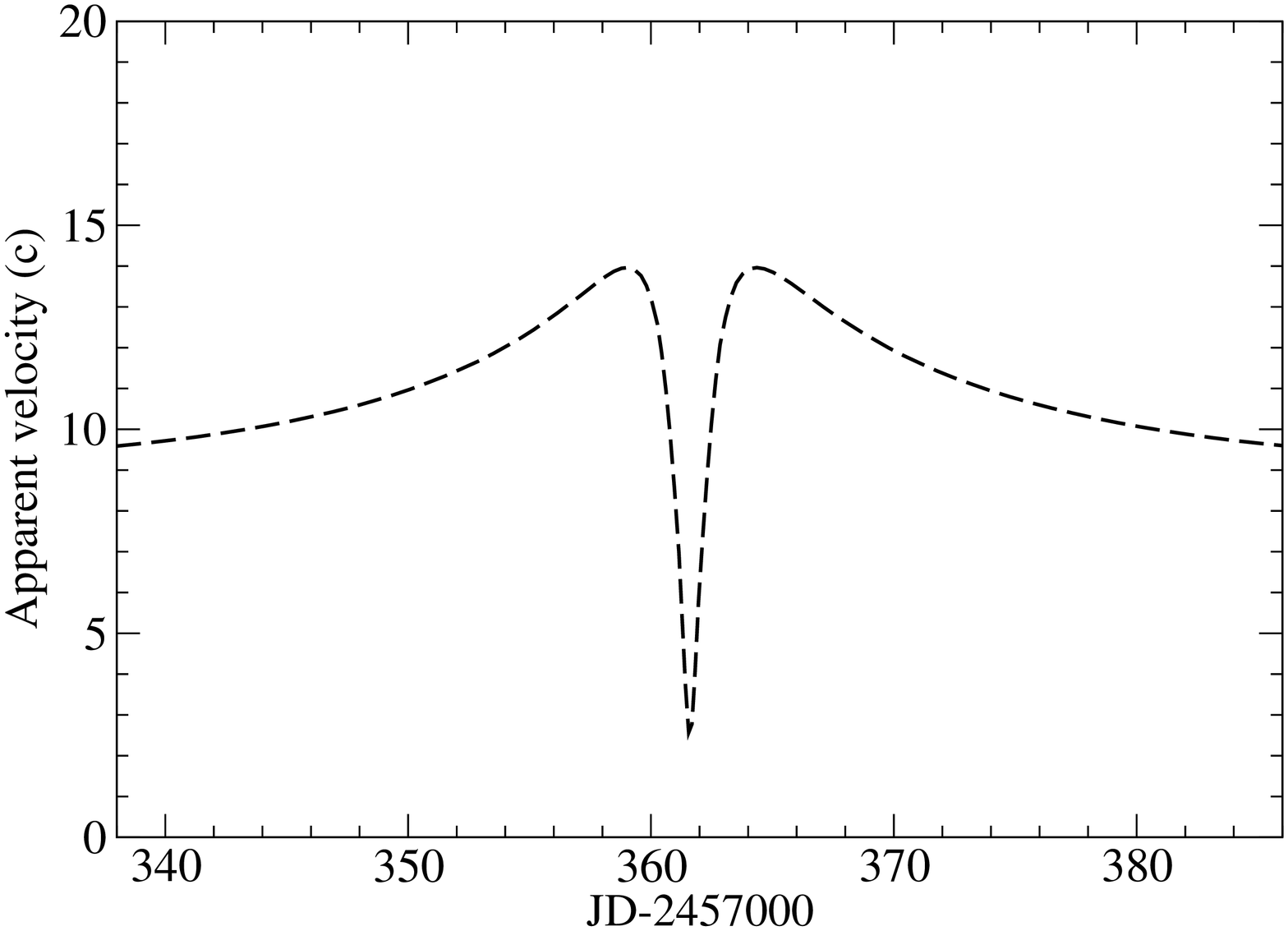}
     \includegraphics[width=5cm,angle=-90]{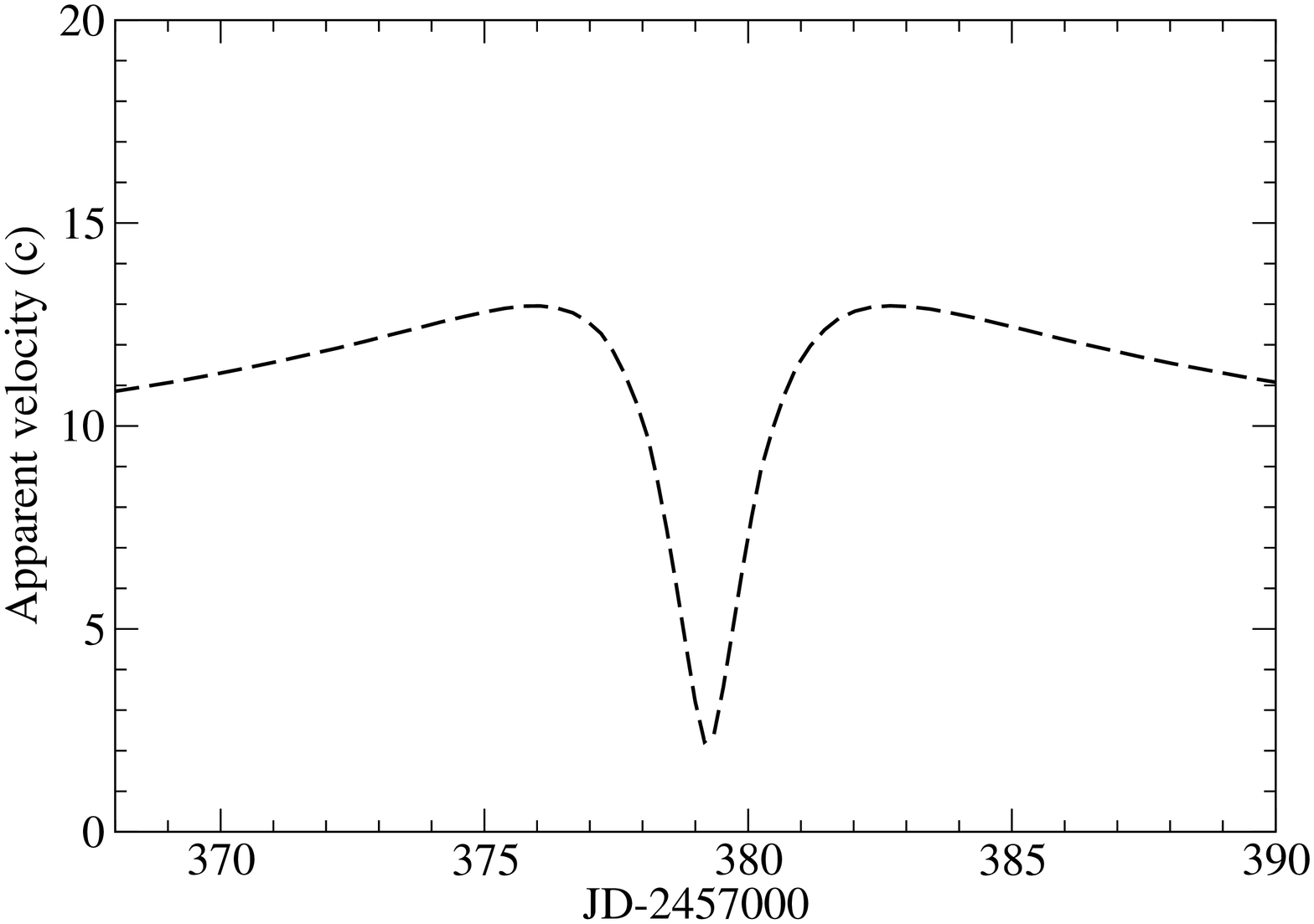}
     \includegraphics[width=5cm,angle=-90]{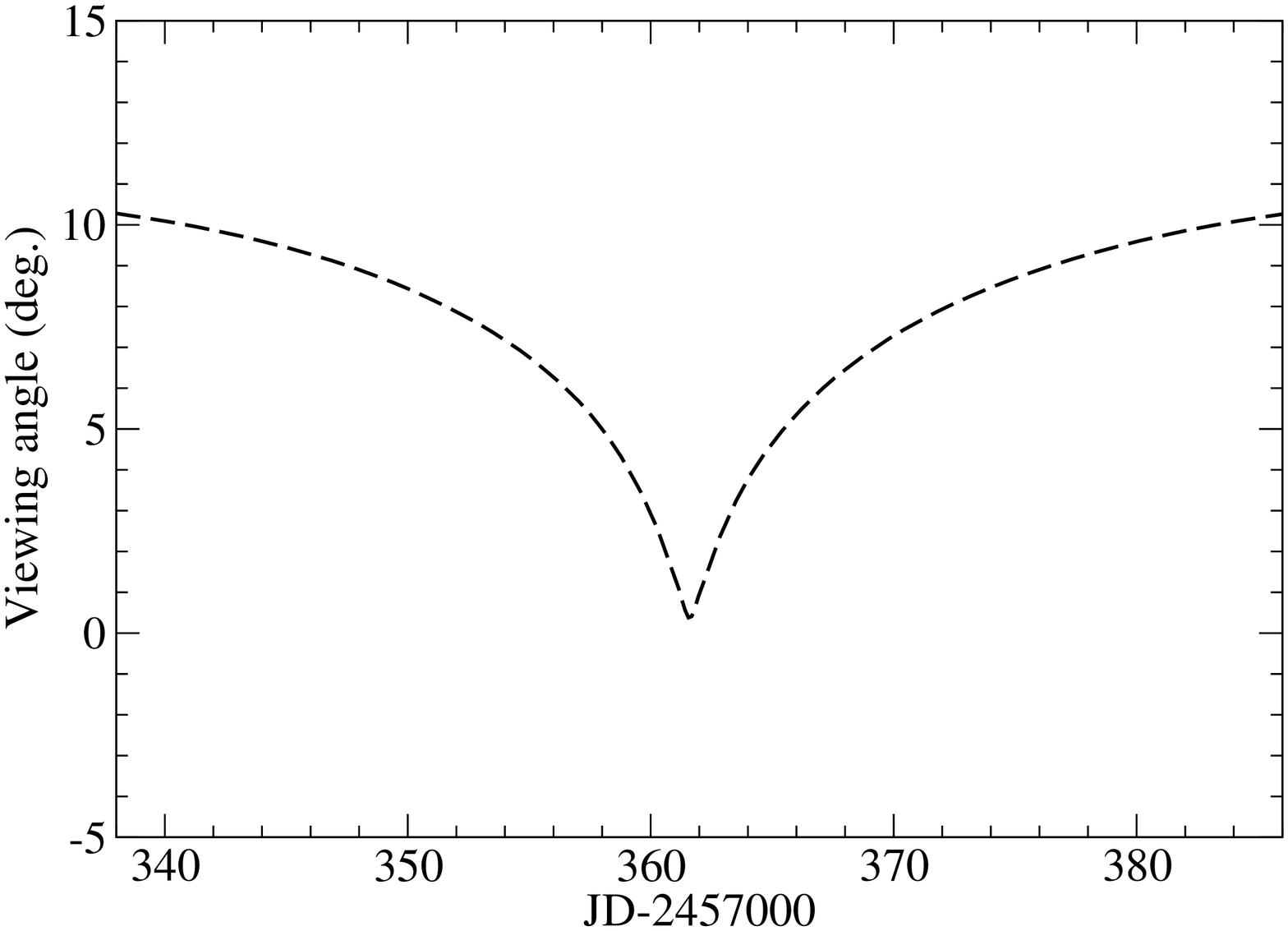}
     \includegraphics[width=5cm,angle=-90]{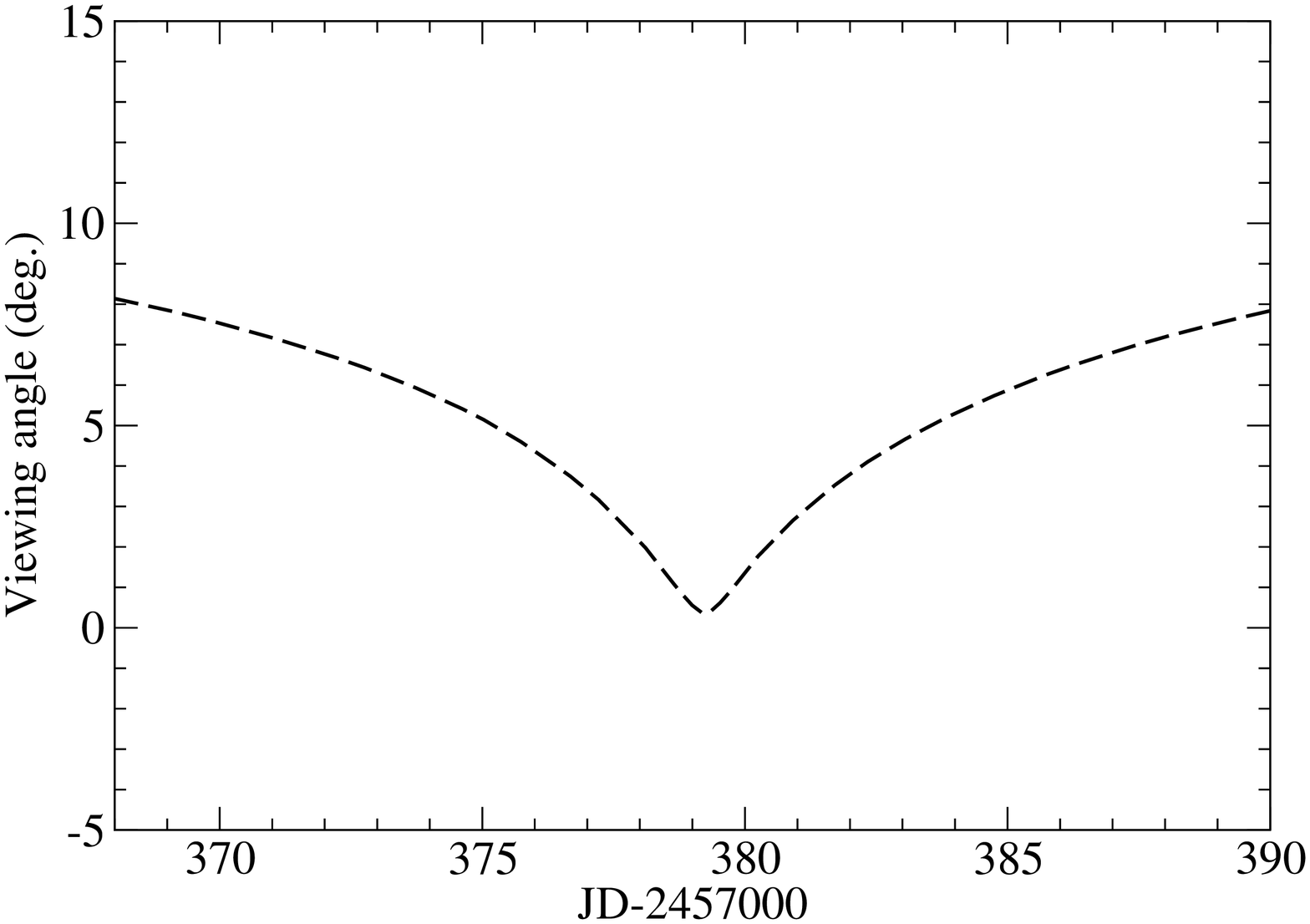}
     \caption{Modeled flux density profile, Lorentz/Doppler factor, apparent 
     velocity and viewing angle versus time for the central strong
     spike-like flares. Left column: for the first spike-flare (during 
     $\sim$JD2457359-364, $\Gamma$=14). The length of the helical trajectory
     along the beam-axis responsible for the first spike flare is 
      $\sim$0.42\,mas, corresponding to $\sim$1.9\,pc and a helical 
     period of $\sim$71\,day. Right column: for the
     second spike-flare (during $\sim$JD2457376-384, $\Gamma$=13).
     Note: the rising and declining portions of the two spike flares
     are very well fitted by symmetric profiles (also see Figure 10).}
     \end{figure*}
     \subsection{Polarization behavior of the December/2015 outburst}
      Investigating the polarization behavior of the periodic optical outbursts
      in 1983.0 and 2007.8 in Sections 4 and 5, we argued that the 
    polarization behavior (including the light curves of flux density, 
    polarization degree and position angle) of the
     periodic optical outbursts may be most important for identifying 
     the nature of their optical emission (thermal or nonthermal), and 
     the rapid large position angle swings during the outbursts
     may become the decisive factor.\\
    In order to clarify the nature of optical emission of the December/2015
     outburst, we shall use a two-component model to study the light curves of  
     flux, polarization degree and position angle of its first flare 
     (during $\sim$JD2457350-374) as a whole.
      It will be  shown that its polarization behavior (especially the
      rotation of the position angle) demonstrates its synchrotron in origin. 
    \subsubsection{Measurements of polarization position angle from 
         different authors}
     We have collected some observational data on the polarization position 
     angle (during $\sim$JD2457336-374; the first
     flare of the December/2015 outburst), which are 
     shown in Figure 12. The upper panel shows the comparison between the 
     measurements by Kushwaha et al. (\cite{Ku18a}; R-band) and Valtonen
     et al. (\cite{Va17}; R-band). The lower panel shows the comparison 
    between the measurements by Myserlis et al. (\cite{My18}; V-band) and
    Valtonen et al. (\cite{Va17}; R-band). These measurements at  R- 
    and V-bands are well consistent: especially the large position angle swing
    during the period $\sim$JD2457370-374. Additionally, there is a 
    large position angle swing clockwise ($\sim{80^{\circ}}$) first (during
    $\sim$JD2457358-359) and then counter-clockwise ($\sim{90^{\circ}}$)
     during $\sim$JD2457359-362. Note that this CCW position angle
     rotation appeared near the peak of the first spike-flare.
    \footnote{The pair of clockwise and counter-clockwise rotations will be
     alternatively interpreted as a continuous clockwise rotation by introducing
     an ambiguity of $-180^{\circ}$ (see Figure 14: left column/bottom panel). 
     The measured values of position angle at $\sim$JD2457371.8, JD2457372.8
     and 2457373.0 from Valtonen et al. also have been 
     added by $-180^{\circ}$ for 
    matching with the measurements by Kushwaha et al. and Myserlis et al.}
      \begin{figure*}
     \centering
     \includegraphics[width=10cm,angle=-90]{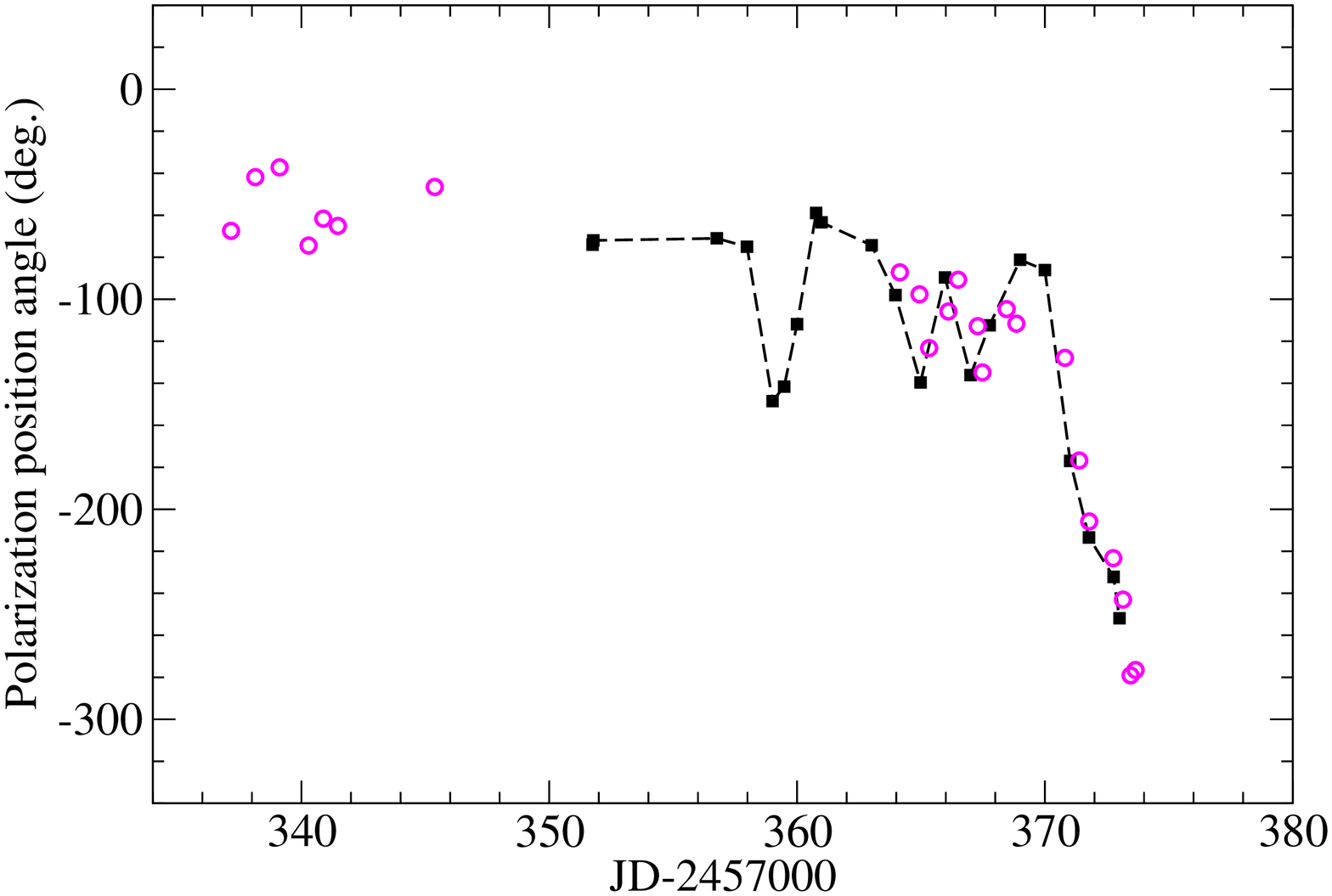}
     \includegraphics[width=10cm,angle=-90]{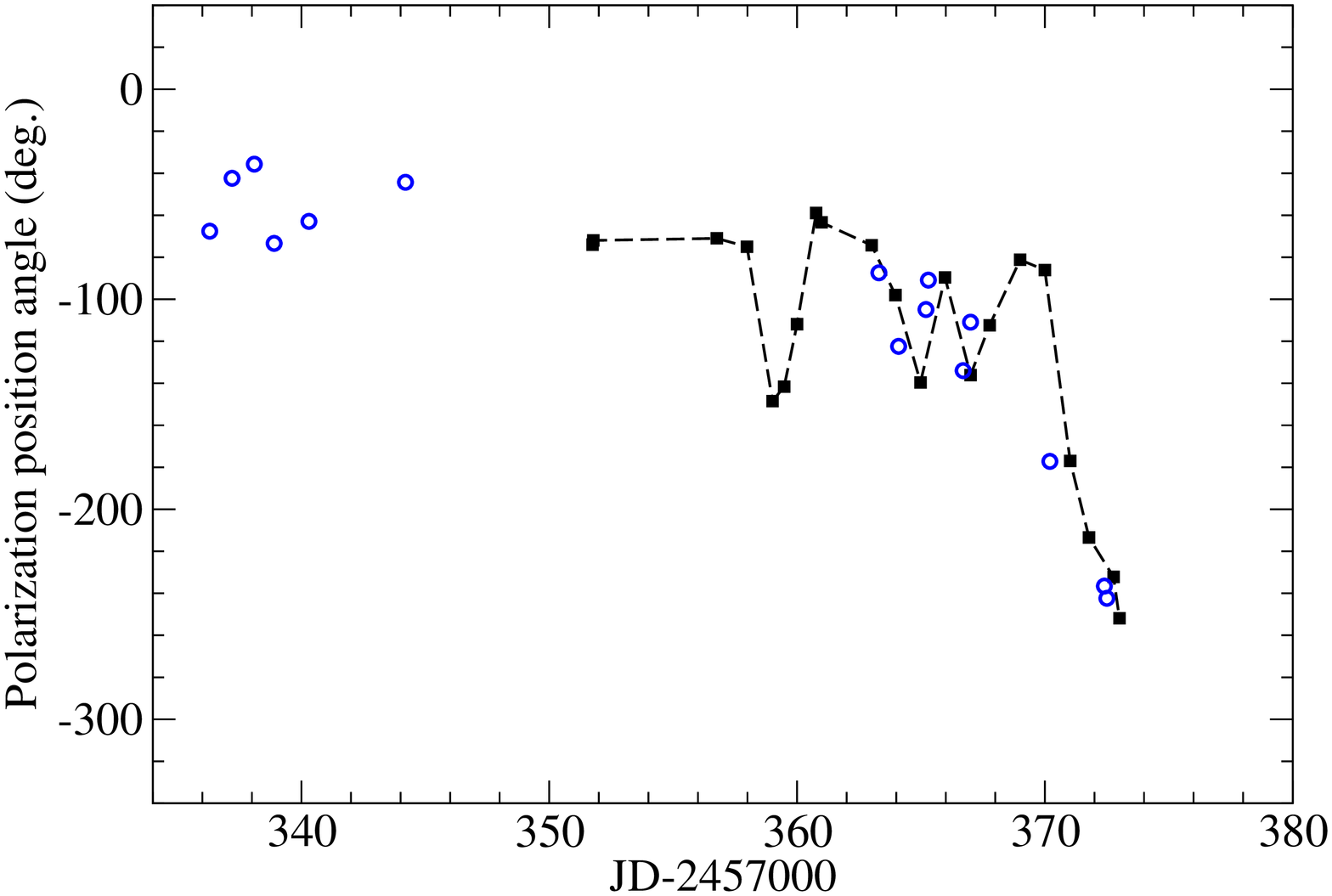}
     \caption{Comparison between the polarization position angle measurements 
      by different authors for the first flare of the December/2015 
      quasi-periodic optical outburst.
     Upper panel: solid black squares from Valtonen et al. (\cite{Va17}),
     open magenta circles from Kushwaha et al. \cite{Ku18a}. Lower panel:
     solid black squares from Valtonen et al. (\cite{Va17}), open blue
     circles from Myserlis et al. (\cite{My18}). Prominent position 
     angle swings were observed during $\sim$JD2457358-362 and during 
     $\sim$JD2457370-374.}
     \end{figure*}
     \begin{figure*}
     \centering
     \includegraphics[width=5.5cm,angle=-90]{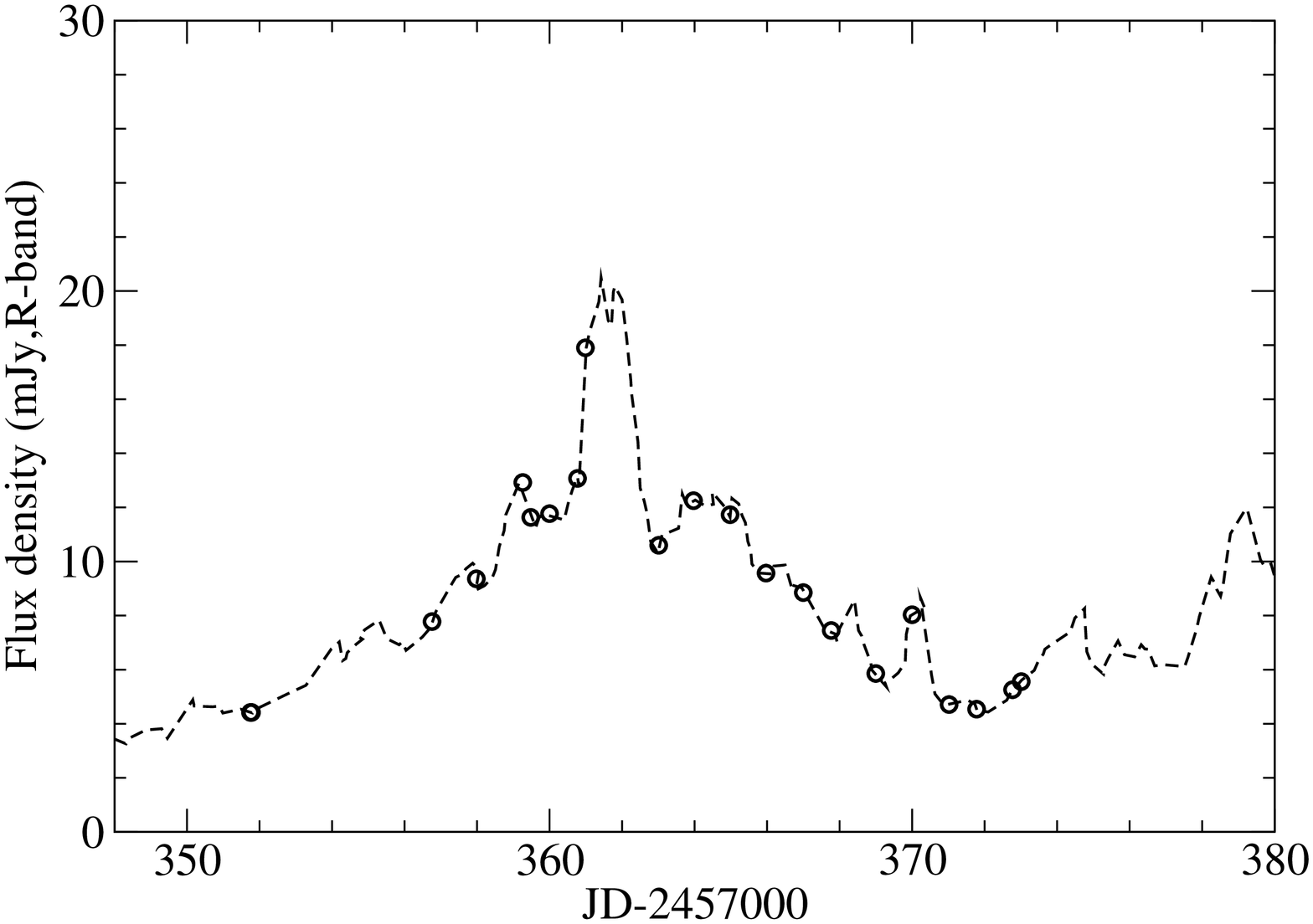}
     \includegraphics[width=5.5cm,angle=-90]{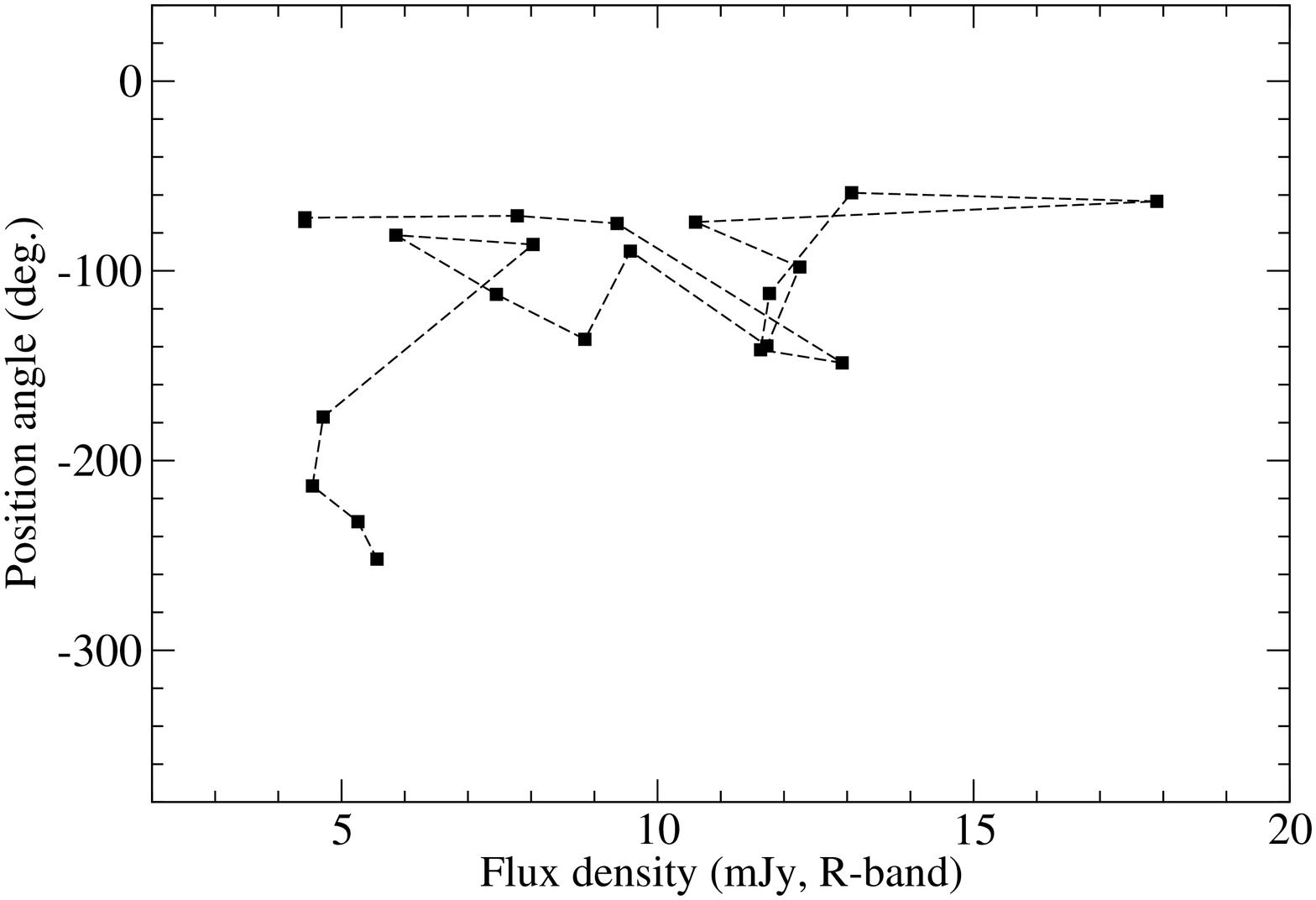}
     \includegraphics[width=5.5cm,angle=-90]{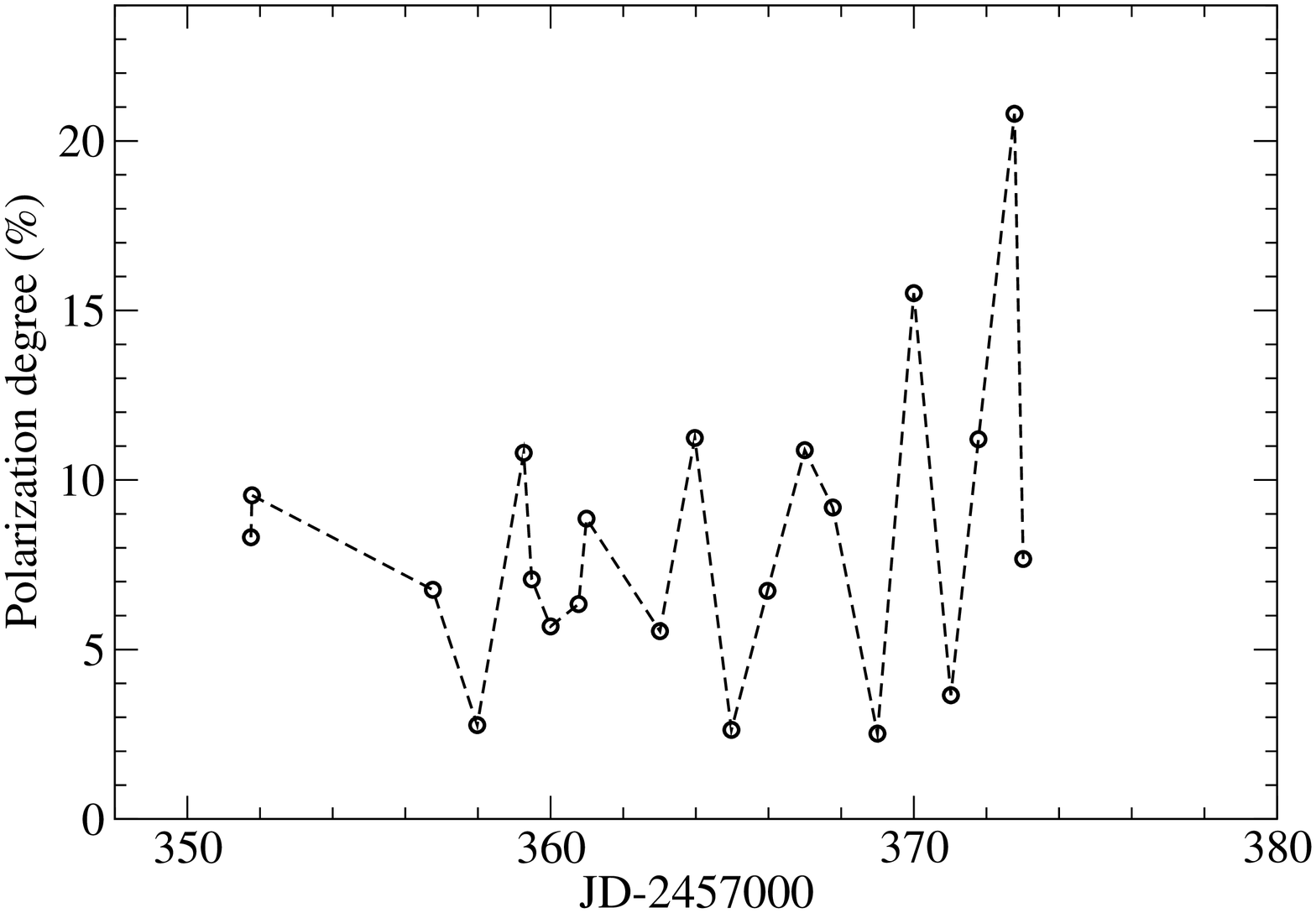}
     \includegraphics[width=5.5cm,angle=-90]{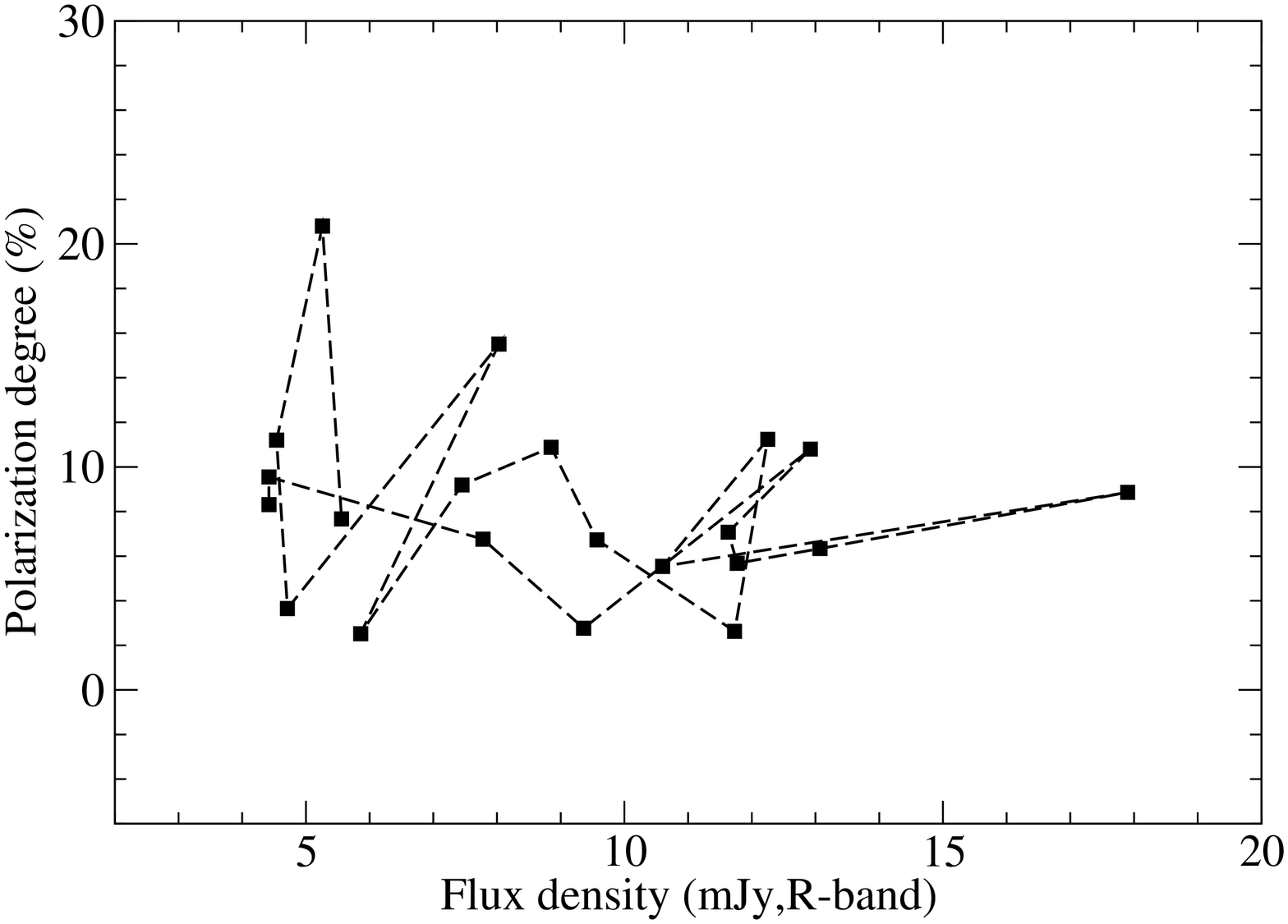}
     \includegraphics[width=5.5cm,angle=-90]{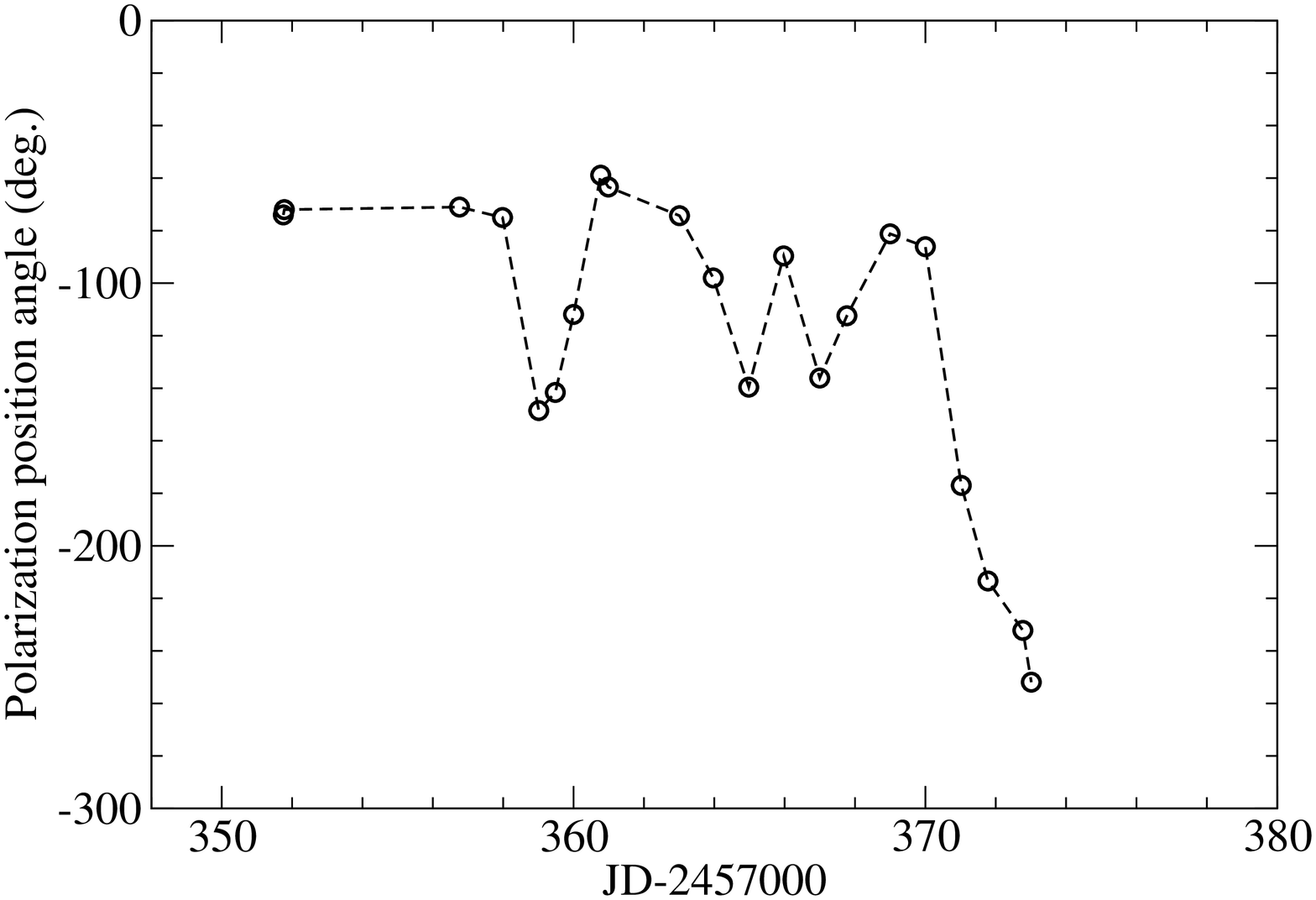}
     \includegraphics[width=5.5cm,angle=-90]{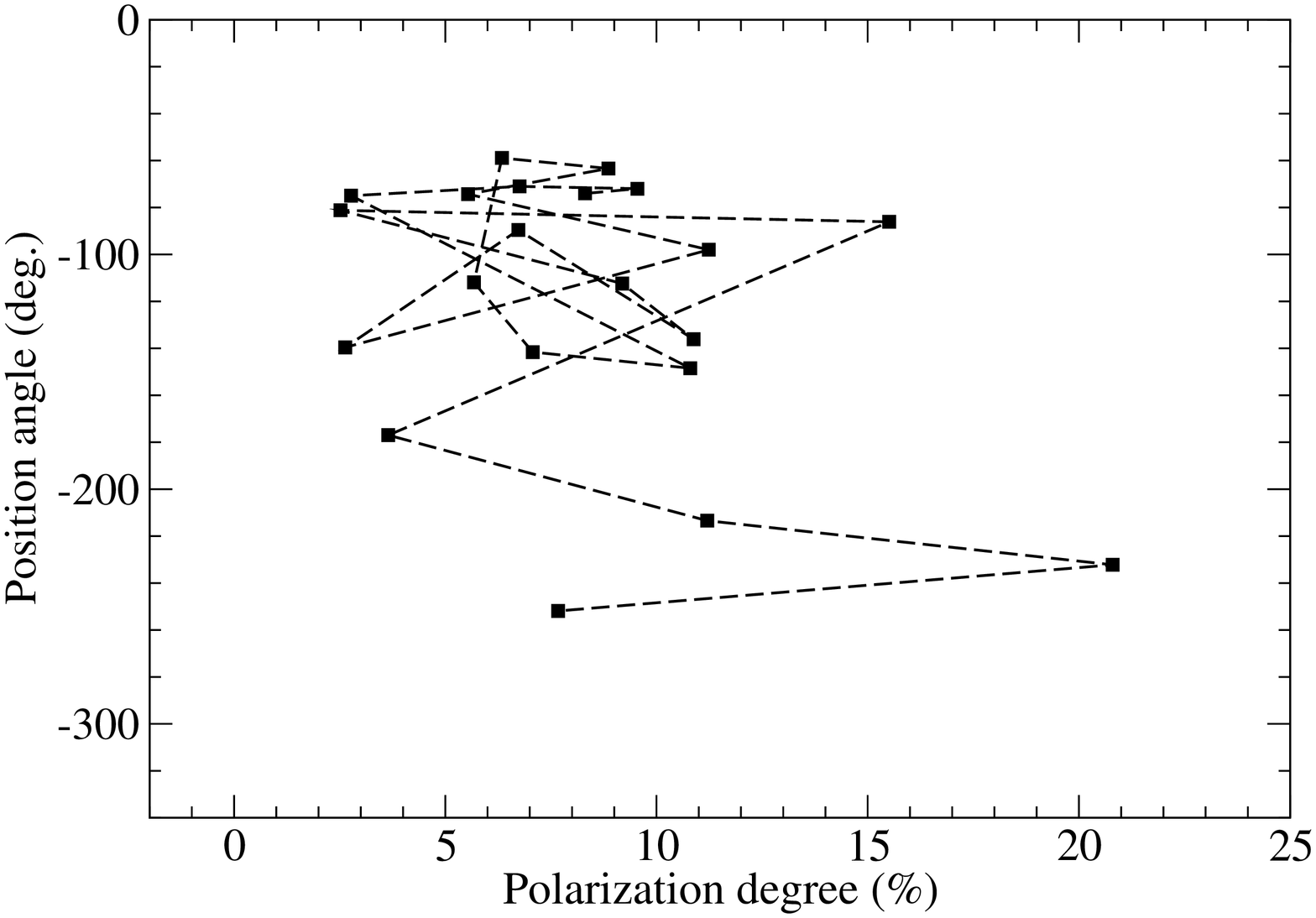}
     \caption{Left column: The observed (integrated) light curves of flux 
     density, polarization degree and position angle. Right column: relation
     between the integrated flux density and the  position angle (top 
     panel); relation between the polarization degree and the position angle 
     (middle); relation between the integrated polarization degree with the
      position angle. Swings in polarization position angle are clearly
      revealed. No sign shows the inverse-proportion relation between the
      polarization degree and the flux density, as required by the impact-disk
      model.}
     \end{figure*}
     \begin{figure*}
     \centering
     \includegraphics[width=5.5cm,angle=-90]{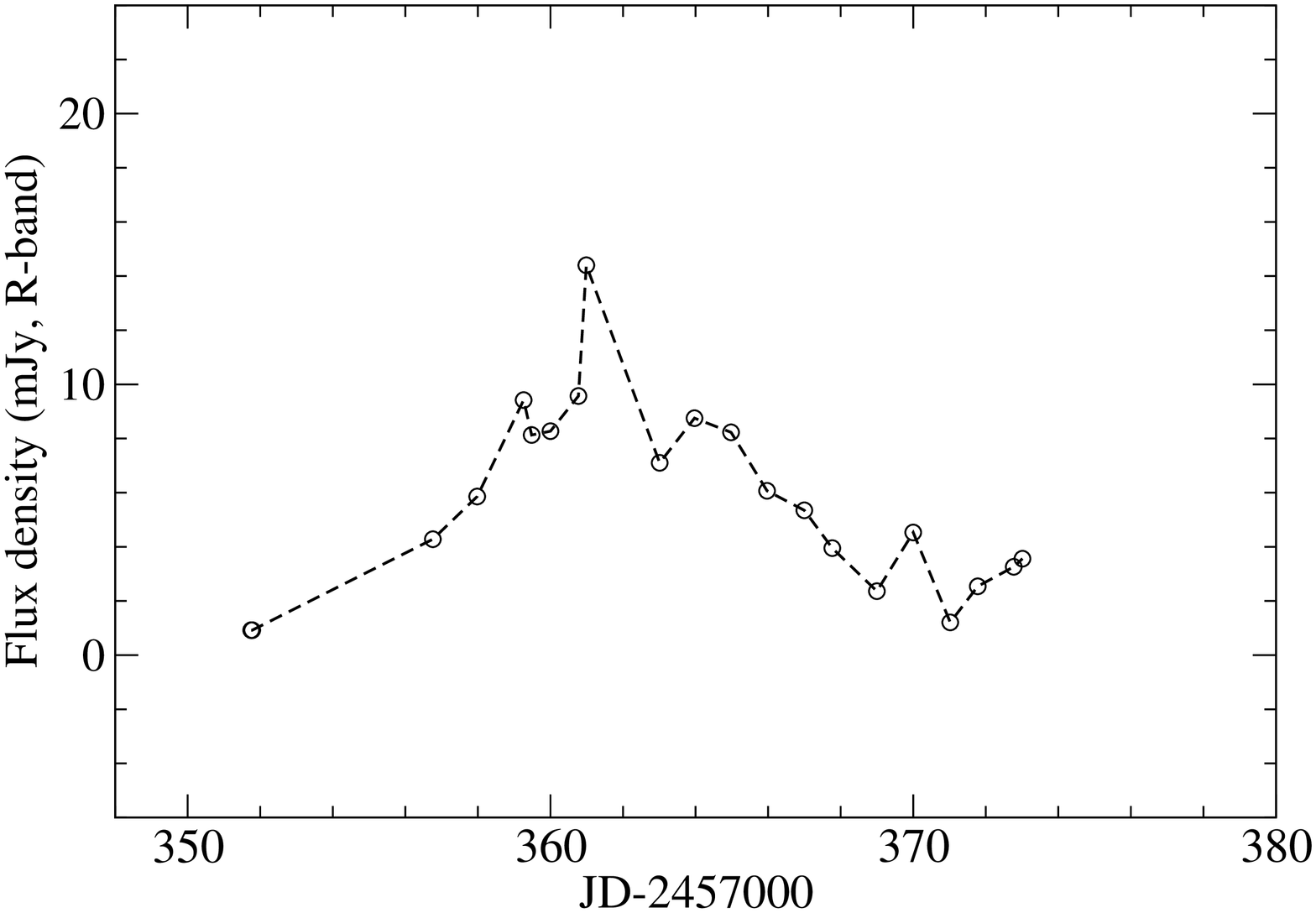}
     \includegraphics[width=5.5cm,angle=-90]{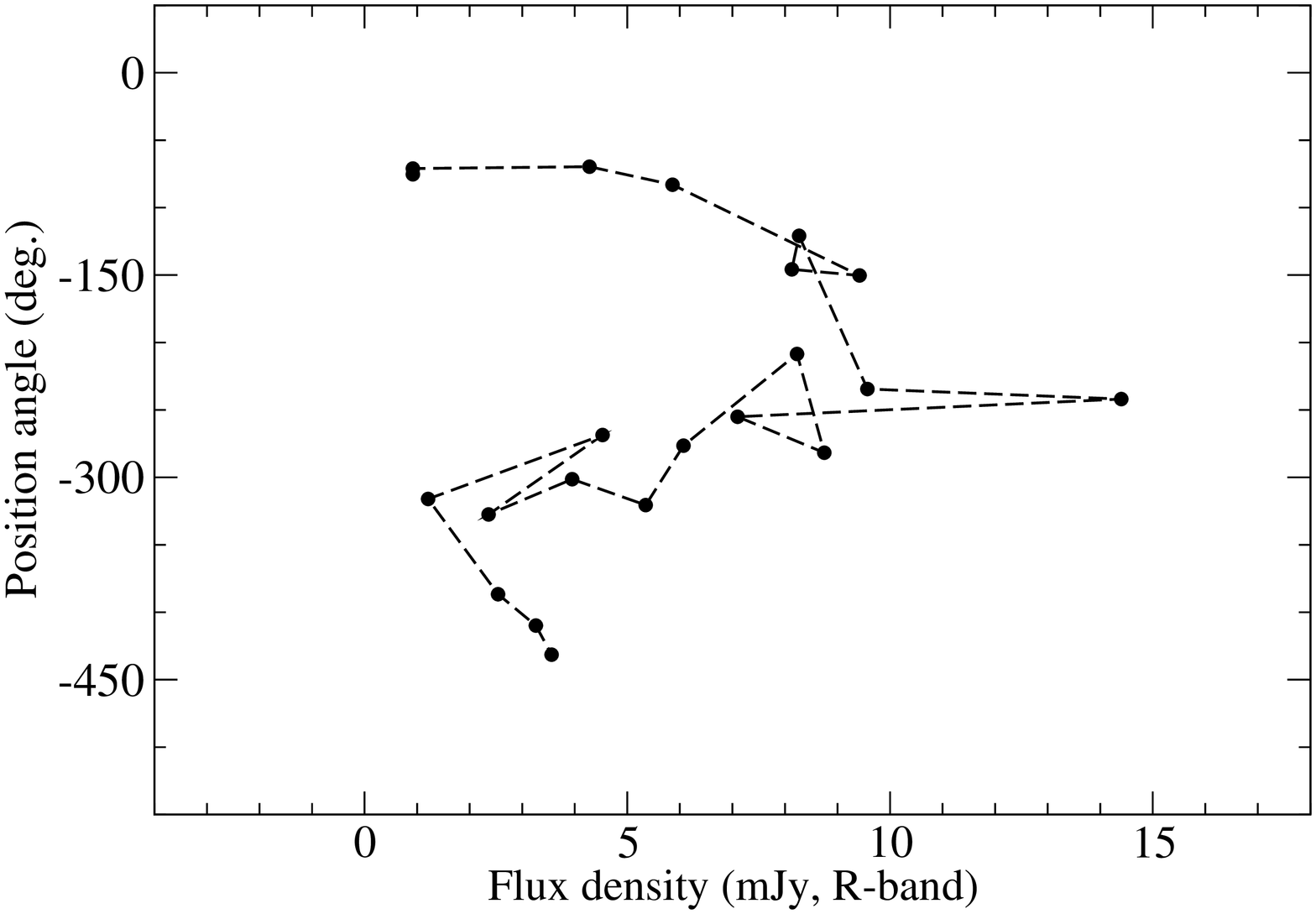}
     \includegraphics[width=5.5cm,angle=-90]{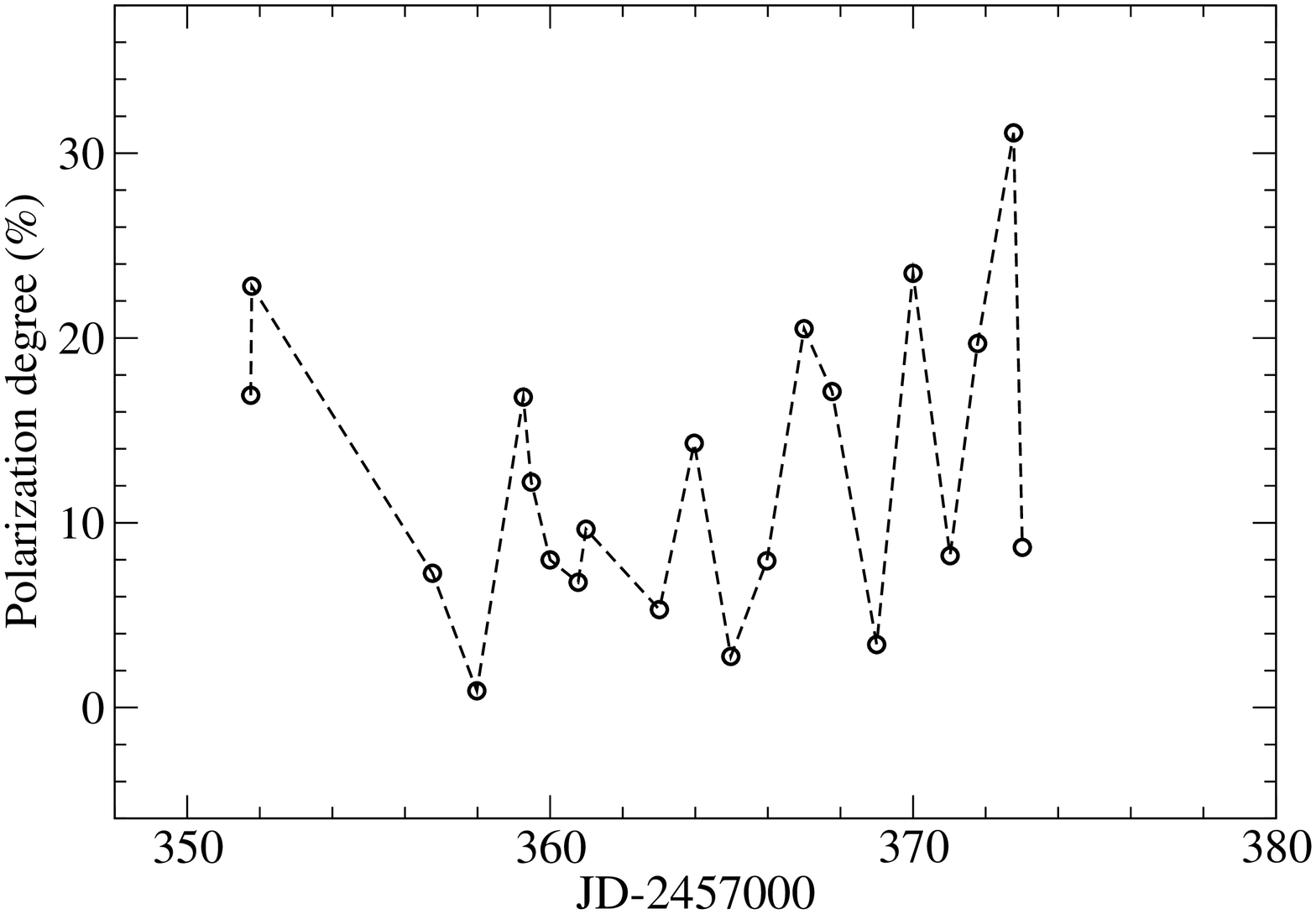}
     \includegraphics[width=5.5cm,angle=-90]{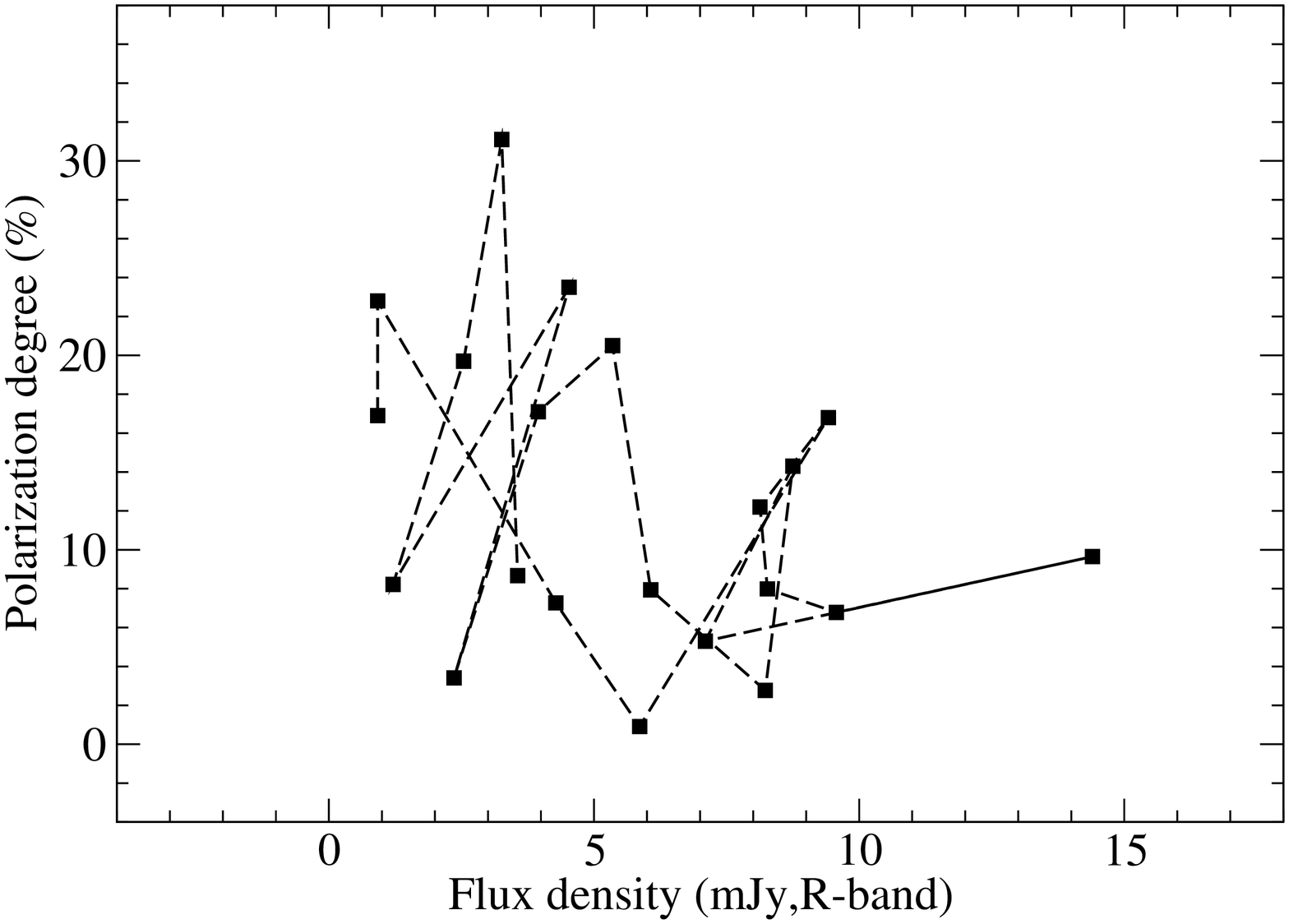}
     \includegraphics[width=5.5cm,angle=-90]{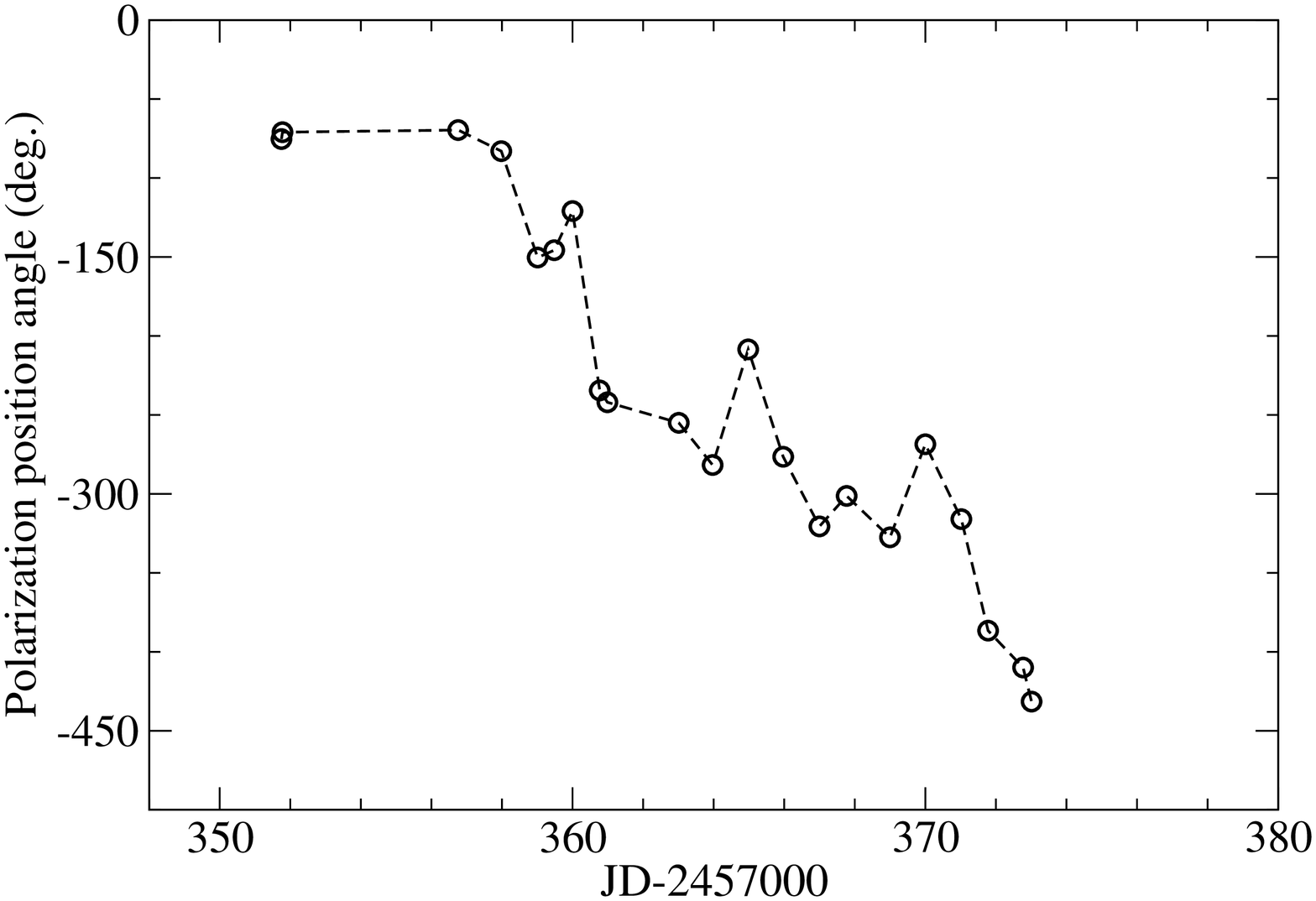}
     \includegraphics[width=5.5cm,angle=-90]{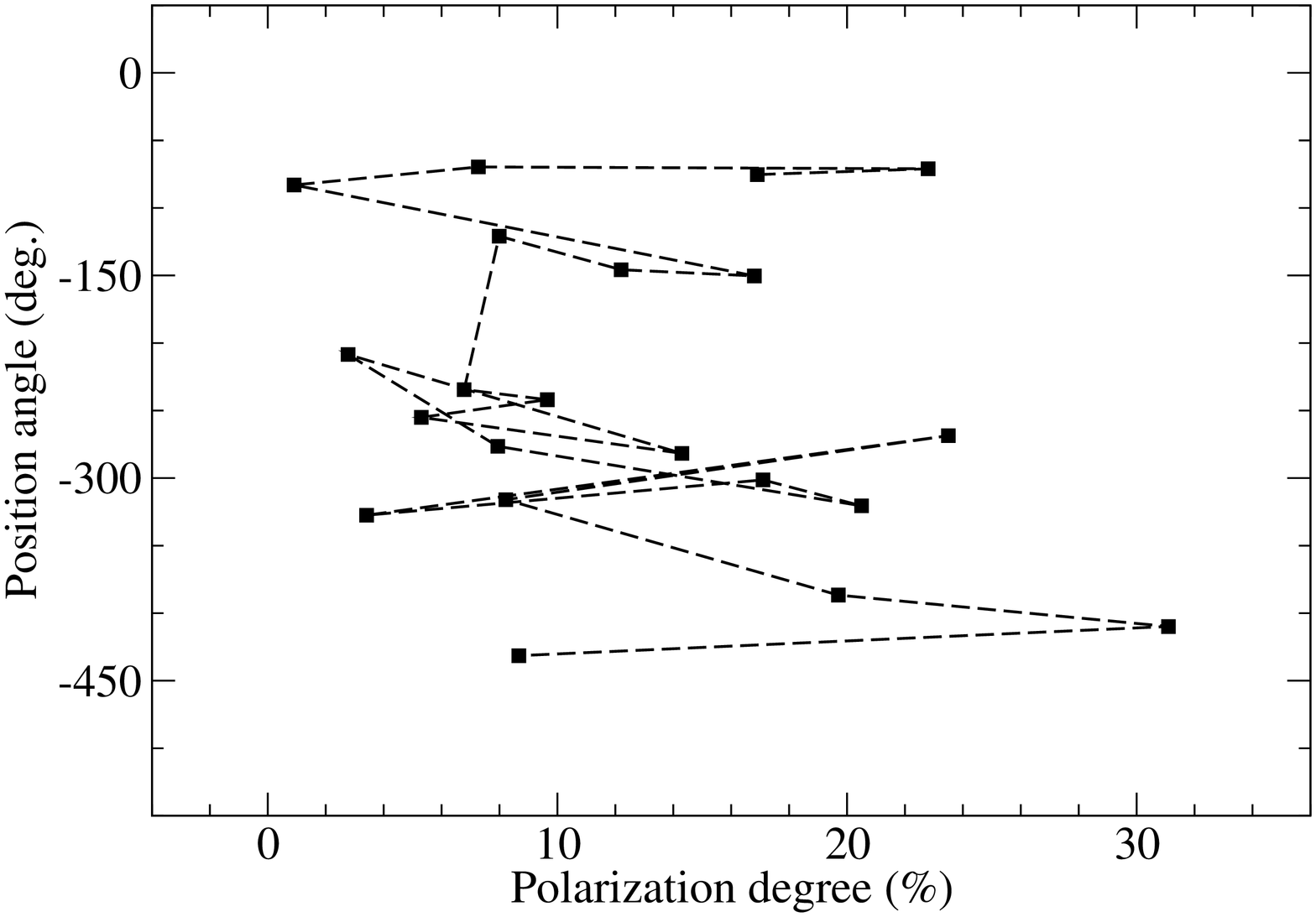}
     \caption{Modeling results for the flare component (Component-2).
      Left column: the modeled light curves of flux density, polarization
       degree and position angle. Right column: relation between the modeled 
       flux density and the  position angle (top panel); relation between
       the polarization degree and the position angle (middle); relation 
       between the modeled polarization degree and the position angle.
       Rotations in  polarization position angle are clearly revealed. 
       No sign shows the inverse-proportion relation between the polarization
       degree and the flux density, as required by the impact-disk model.}
     \end{figure*}
     \begin{figure*}
     \centering
     \includegraphics[width=5.5cm,angle=-90]{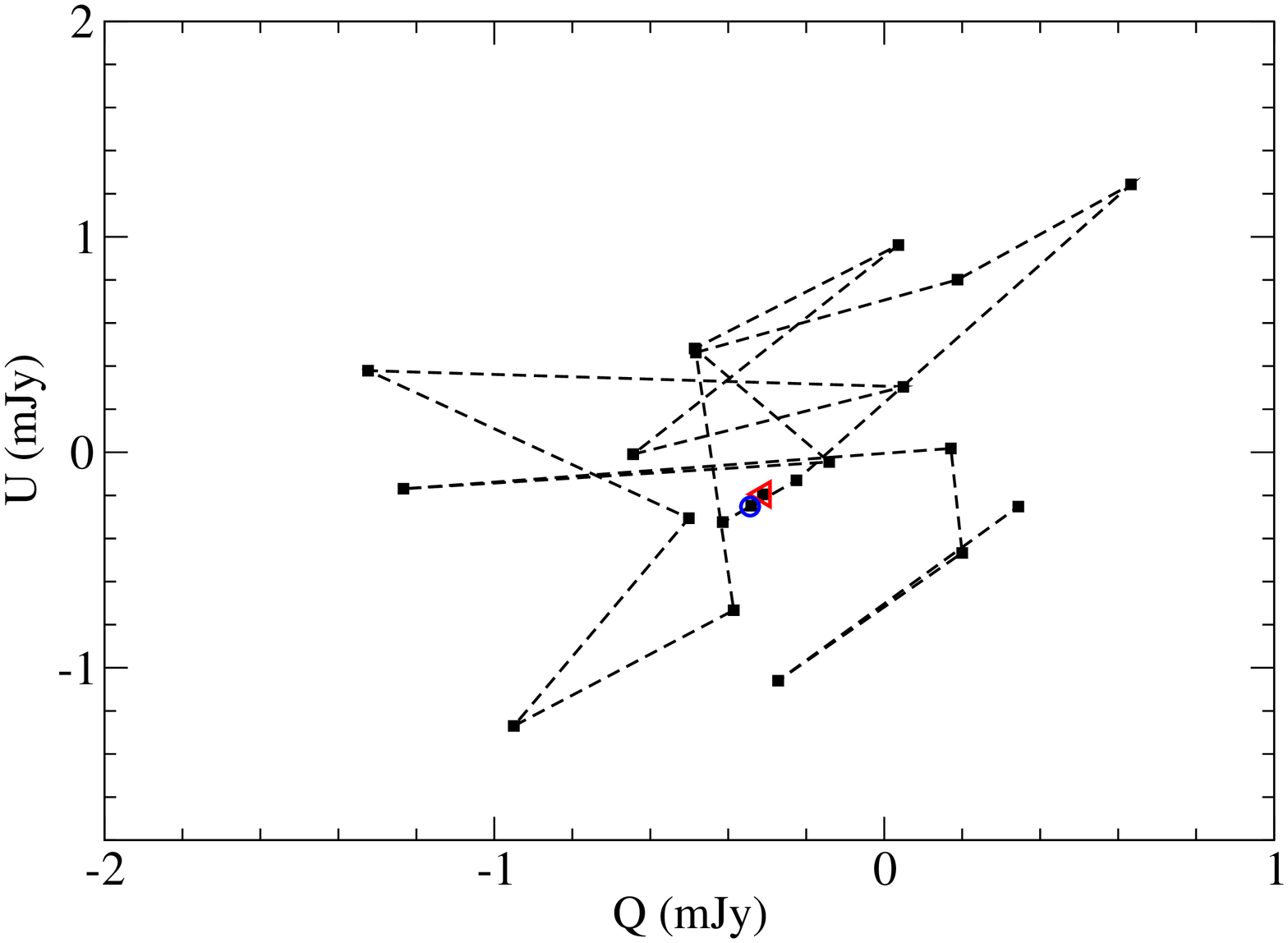}
     \includegraphics[width=5.5cm,angle=-90]{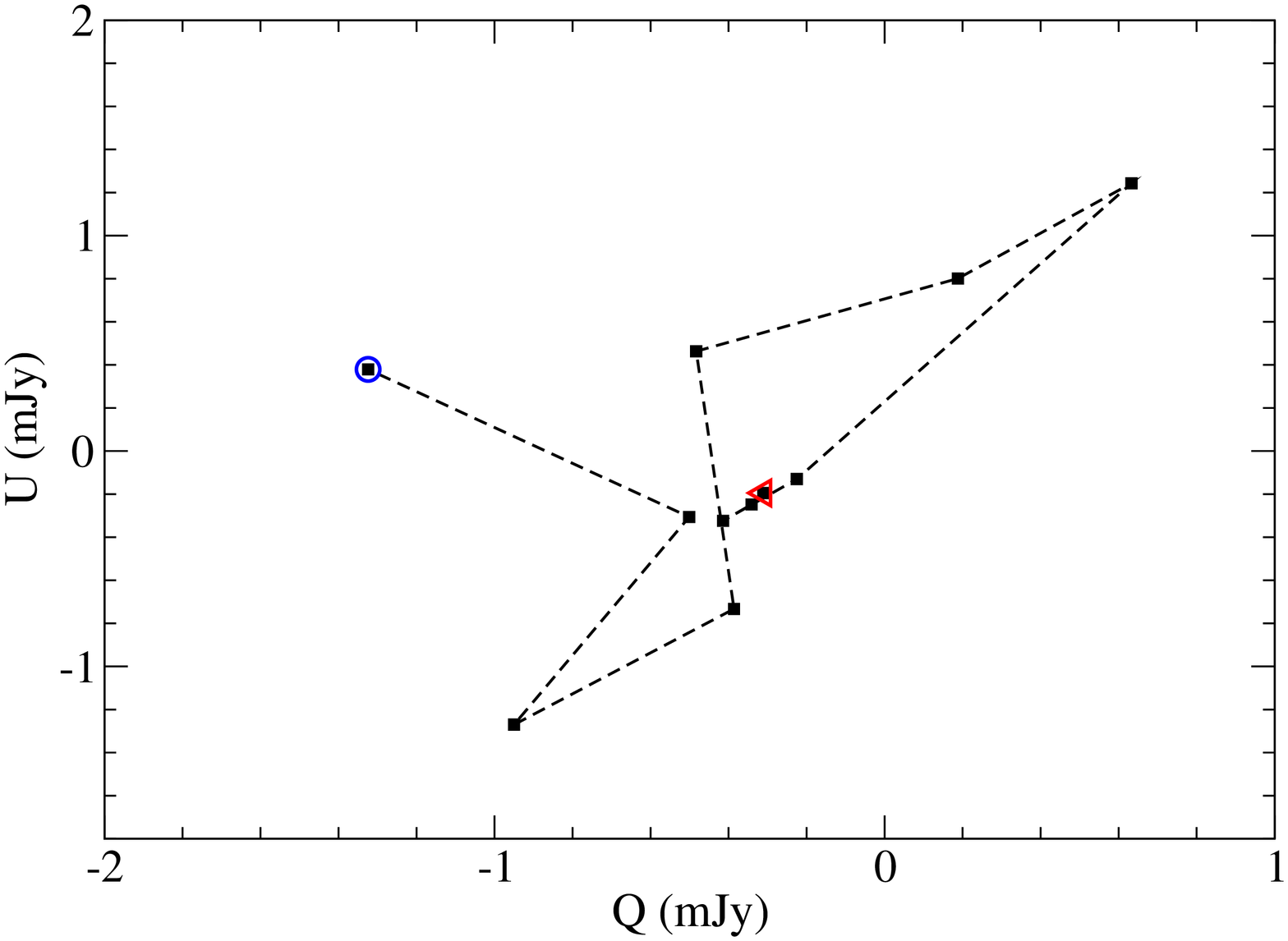}
     \caption{Left: Stokes Q-U plot for the integrated outburst (during 
     $\sim$ JD2457351.8-371.0). Right: Stokes Q-U plot for part of 
     the outburst during $\sim$JD2457357-364, which clearly reveals the
      polarization position angle swing. Black triangles represent the start
      of the tracks. }
     \end{figure*}
      \begin{figure*}
      \centering
      \includegraphics[width=5.5cm,angle=-90]{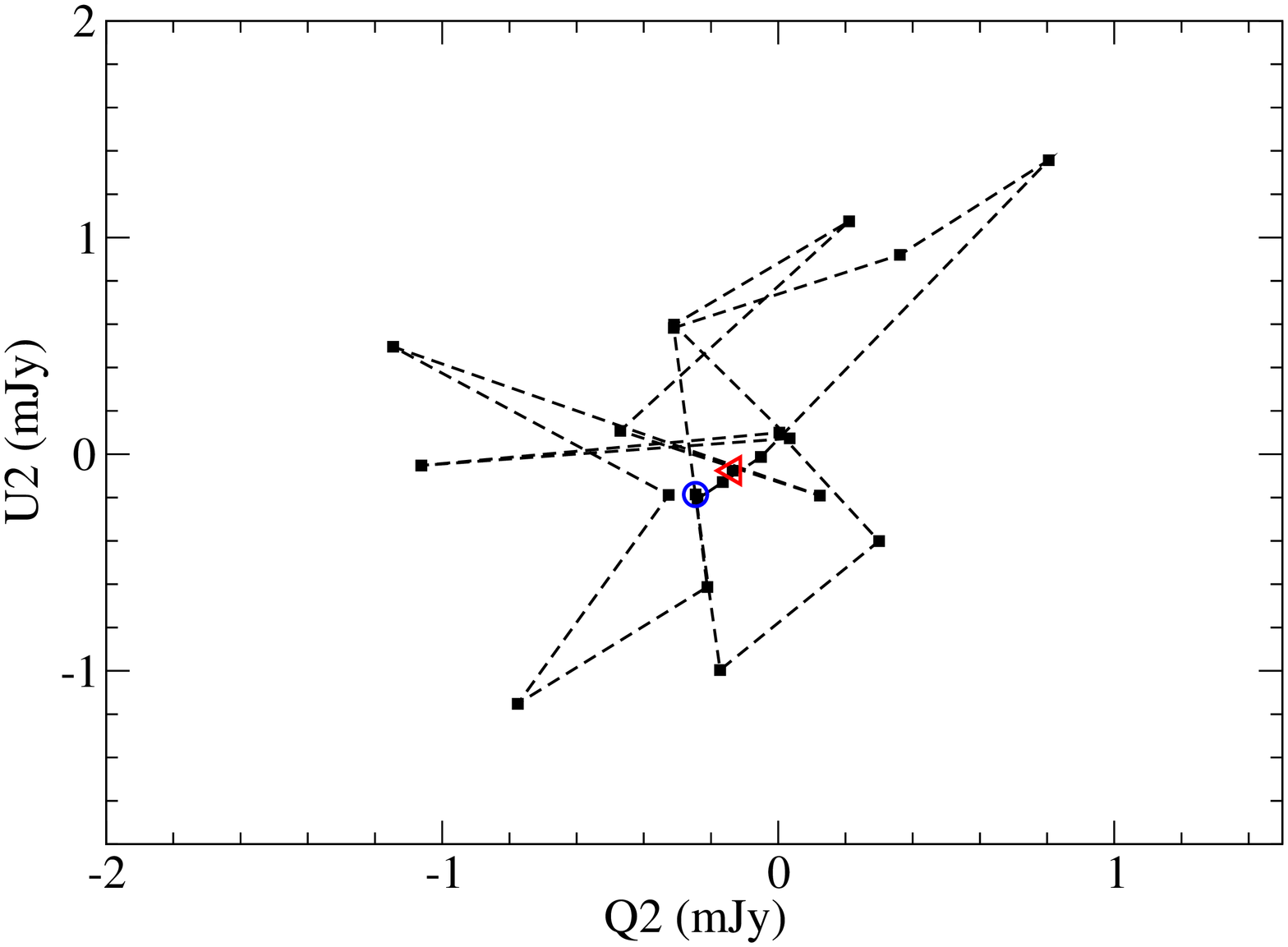}
      \includegraphics[width=5.5cm,angle=-90]{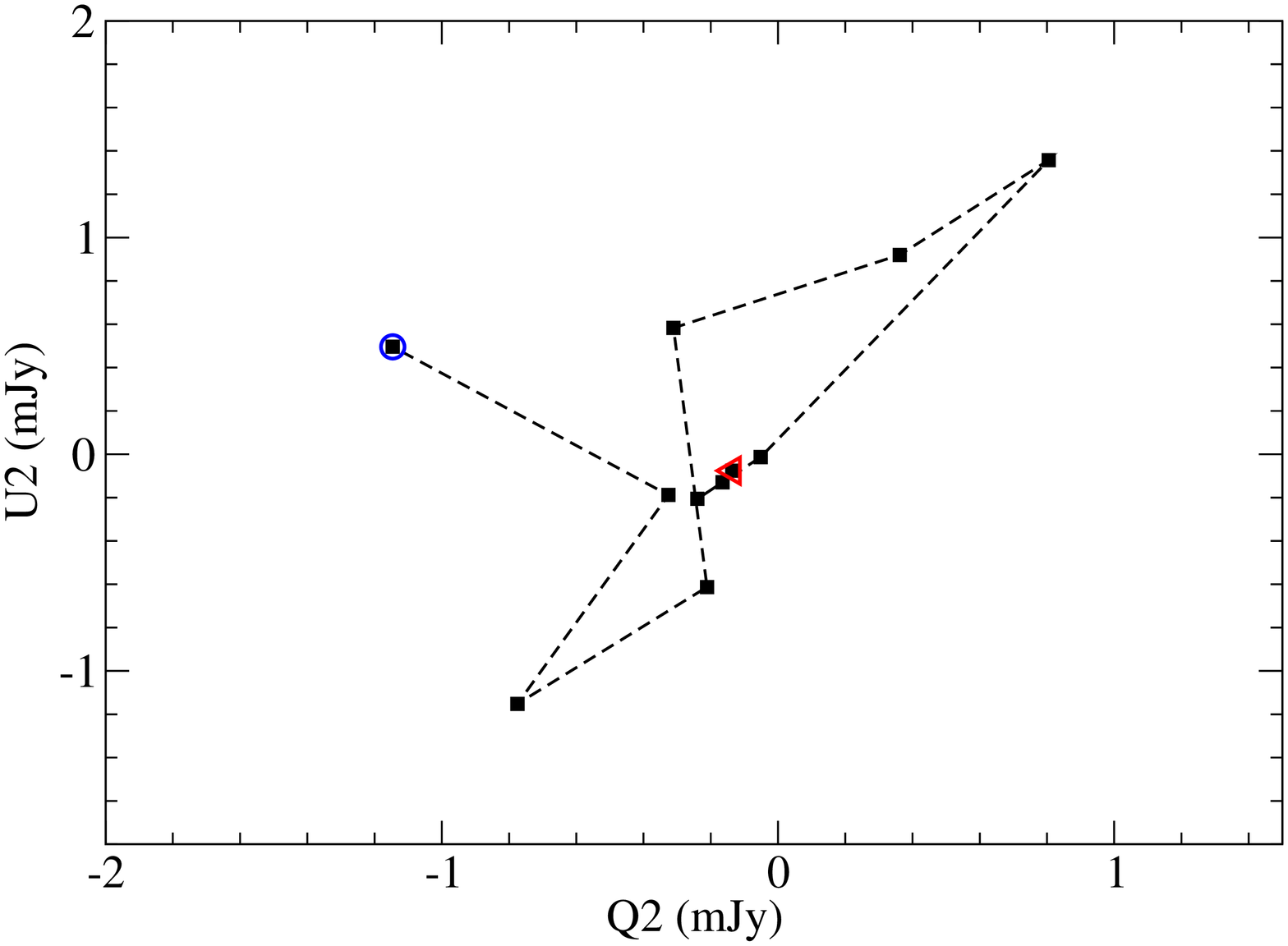}
      \caption{Left: Stokes Q-U plot of the modeled flare-component
     (Component-2) during $\sim$JD2457352-2457371. Right: Stokes Q-U plot 
     for part of the flare-component during $\sim$JD2457351-2457364, including
     the central strong spike. Black triangles represent the star of the 
     tracks.}
      \end{figure*}
     \subsubsection{A two component model}
     Now we turn to make model simulation of the light-curves for the first
      flare of the December/2015 outburst.\\
     We have already seen in the last subsection that large swings in 
     position angle were observed in this flare. Its integrated polarization
     degree was also highly variable as shown in Figure 13 (left column/middle
     panel). During the period $\sim$JD2457350-372 it varied  between $\sim$3\%
     and $\sim$15\%. The distinct features are:
     (1) The flux peak (at $\sim$JD2457361.5) of the strong central spike flare 
        is not concurrent with the minimal polarization degree  ($\sim3\%$
        at $\sim$JD2457359).  There are other three epochs when the 
       polarization degree was observed  to be between $\sim$2\% and $\sim$4\%,
       but all occurred during the declining stage; 
       (2) the light curve of polarization degree does not reveal an 
       inverse-proportion relation with the flux density 
       (right column/middle panel in Figure 13),as expected if the outburst
        is purely thermal, where the polarization degree should decrease when 
        its flux density  increases and vice versa. \\     
      As for the  1983.0 and 2007.8 outbursts discussed in Sections 4 and 5, 
      we have applied a two-component model to simulate the light curves of
      the observed (integrated) flux density, polarization degree and position 
     angle as a whole. We assume that the  first flare of the December/2015 
    outburst consists of two polarized components: one is a stable polarized
    component (or the steady underlying jet component before the 
    outburst; component-1) and the other is a flaring  component with variable
     polarization (component-2), respectively  defined by ($I_1,p_1,\theta_1$)
    and ($I_2,p_2,\theta_2$). We choose the following values for component-1:\\
       $I_1$=3.5\,mJy (R-band), $p_1$=6.0\%, $\theta_1=-73^{\circ}$ \\
     The value for $p_1$ is constrained by the drop of polarization degree from
     $\sim$6.8\% to $\sim$2.8\% during the period $\sim$2457356.5-2457358.5
     with slight change in position angle (Fig. 13; left column: middle and 
     bottom panels).  \\
     With the values of ($I_1,p_1,\theta_1$) chosen, the values 
     of $(I_2,p_2,\theta_2)$ can then be uniquely determined from solving 
     the combined equations (1) and (2). The modeling results of the 
     flaring component (component-2) are shown in 
     the left column of Figure 14. It can be seen that during the first
     flare ($\sim$JD2457358.0-369.0) the derived polarization degree 
     changes between $\sim$0.9\% and $\sim$20\%. The minimum degree occur 
     at $\sim$JD2457358, not coincided with the flux peaking epoch 
     ($\sim$JD2457361.5).\\
      The modeling results in terms of the two-component model clearly exhibit
      a rapid rotation of polarization position angle of the first flare
      during the period ($\sim$JD2457358-373), as shown in Figure 14 
     (left column/bottom panel). The position angle of the flaring component 
     changes by $\sim{350^{\circ}}$: from $\sim{-80^{\circ}}$
     to $\sim{-430^{\circ}}$. Its average clockwise rotation rate is
       approximately $\sim{-25^{\circ}}$/day. Interestingly, the fastest 
      rotation rate of  $\sim{-110^{\circ}}$/day derived 
      during $\sim$JD2457360-361 is
      coincident with the peak of the strong spike flare. The derived 
      polarization position angle rotates clockwise with rapid fluctuations,
      which could be due to the superposition of more than two polarized 
      components (e.g., a spike flare plus a wing flare and the stable
      component), as the modeled flux density light curve for the flaring
      component demonstrates.\\
       The position angle rotations in optical wavebands are much faster
      (by a factor of $\sim$20)  than that observed at radio wavelengths
      (Myserlis et al. \cite{My18}) and  the optical PA rotations precede
      the radio PA rotations. This phenomenon could well be explained in terms 
      of the precessing nozzle scenario with helical motion: (1) both optical
      and radio PA rotations are produced by  superluminal optical and radio
      knots moving along helical trajectory via lighthouse effect;(2)  
      radio knots may evolve from optical knots, and due to opacity effects in 
     radio-bands radio PA rotations should appear in  regions further out 
      in the jet, thus having a time delay  relative to the optical PA 
      rotations; (3) the pitch angle of the helical magnetic field may increase
      further out along the jet, causing the PA rotation rate in radio-bands 
      much smaller than the  rotation rate in optical bands; (4) combination 
      of polarized synchrotron radio/optical
      flares can create various types of polarization behavior as observed
      in generic blazars.\\
       It seems that the rotation of polarization position angle derived
       for the first flare of the December/2015 outburst could not be 
      interpreted in terms of the impact-disk model, where the first flare
      of the December/2015 outburst was interpreted to be purely thermal,
      because  a thermal flare with zero polarization alone can not 
      cause rapid changes in position angle. Our model-simulation results 
      favor the suggestion that the first flare of the  December/2015 outburst
      is synchrotron flare according to its polarization behavior. \\
      Polarization position angle swings have been observed in blazars for a 
      quite long time and interpreted by various authors, mostly invoking
      superluminal knots moving along helical trajectory in helical magnetic
      fields of jets or two-component model (Aller et al. \cite{Al81}, 
      Blandford \& K\"onigl 
      \cite{Bl79}, Holmes et al. \cite{Holm84}, 
      Kikuchi et al. \cite{Ki88}, K\"onigl \& Choudhuri 
      \cite{Ko85a}, \cite{Ko85b}, D'Arcangelo et al. \cite{Da09}, Marscher
       et al. \cite{Ma08}, Myserlis et al. \cite{My18}, Qian \cite{Qi92},
      \cite{Qi93}, \cite{Qi03}).\\
      For interpreting the PA rotations observed in optical and radio regimes 
     in OJ287 the relativistic jet
      models of superluminal knots moving along helical trajectory in magnetic 
      fields seem the most appropriate and consistent with 
      the processing nozzle scenario (Myserlis et al. \cite{My18},
      Qian \cite{Qi18b}, \cite{Qi19a}).
      \subsubsection{QU-plots}
     The Stokes QU-plots for the integrated outburst  is shown in
     Figure 15. And the QU-plots for the flaring-component (component-2)
     is shown in Figure 16. In the left panels are shown the entire Q-U tracks
     (during $\sim$20 days interval, $\sim$JD2457352-371)  which are very
     erratic, like a drunkards walk. This is because of the very rapid
     variations in its polarization and the considered time-interval is too
     long. However, for a shorter time-interval of $\sim$7\,day during
      $\sim$JD2457357-364 for the central
      strong spike flare, the QU-track  clearly reveals its position angle
      rotation. Same behavior is for the flaring component (component-2) 
     shown in Figure 16. 
   \section{Discussion and conclusion} 
     Based on  the precessing jet nozzle scenario 
    previously proposed by Qian (\cite{Qi19a}, \cite{Qi19b}, \cite{Qi18b},
     \cite{Qi16}) and Qian et al. (\cite{Qi91a}, \cite{Qi14},
    \cite{Qi17}, \cite{Qi18a}, \cite{Qi19c}), we have model-simulated
    the flux light curve of the December/2015 periodic optical outburst
    (during $\sim$JD2457358-390; Figure 10). We have also analyzed the
     polarization behavior of the periodic optical outbursts in 1983.0, 
    in 2007.8 and the  first flare of the December/2015 outburst
     (during period $\sim$JD2457358-372)  and showed that the
     rapid and large rotations in their polarization position angles 
     (Figures 3-4, 6-7 and 13-15) associated with the appearance of 
     low polarization degrees. \\
     A few conclusions can be made as follows.
     \begin{itemize}
     \item In order to determine the nature of emission (thermal or nonthermal)
      from the periodic optical outbursts, the light curves of flux density,
      polarization degree and polarization position angle 
      ($I(t), p(t), {\theta}(t)$) should be investigated and consistently interpreted
      as a whole. Low polarization degrees alone seem not appropriate to 
      be used as a unique  factor to recognize the emission from the 
      outbursts being thermal;
     \item The model-simulations for the three outbursts have revealed that they
     all show  large position angle rotations during the outbursts, implying
     that the three outbursts should be all synchrotron in origin and
      produced in the relativistic jets;
     \item The precessing jet-nozzle scenario may be helpful for understanding 
     the phenomena observed in OJ287. The simulation of the flux density 
     light curve for the first flare of the December/2015 shows that 
     this flare may comprise a number of elementary synchrotron flares, which
     are produced through a succession of superluminal optical knots moving 
    along helical trajectory via lighthouse effect. Most of the optical/radio
    emission features observed in OJ287 can be understood in terms of the
     precessing nozzle model. However, this scenario has been suggested only 
     deal with the emission features. The solution to the  
     mechanism(s) for the quasi-periodicity and double-peaked structure in the 
     optical  light curve requires different approaches to work out 
    (e.g., the disk-impact mechanism suggested in Lehto \& Valtonen \cite{Le96},
     Sundelius et al. \cite{Su97}).
    \end{itemize}
     The relativistic jet models under the precessing nozzle scenario suggested 
   for OJ287 by Qian (\cite{Qi19a}, \cite{Qi19b}, \cite{Qi18b}) may
   be useful to interpret the basic phenomena of its electromagnetic
    radiation and its nature of emission, including the following observational
    aspects.
    \begin{itemize}
    \item (1) The simultaneous $\gamma$-ray and optical outbursts observed in 
       the December/2015  flaring event (Kushwaha et al. \cite{Ku18a}) can be 
      interpreted, because
      both $\gamma$-ray and optical outbursts are suggested to be originated
       within the relativistic jet;
    \item (2) The quasi-periodic optical outbursts are composed of 
     elementary flares with timescales of $\sim$5\,-10\,days. Each of the 
     elementary flares has a symmetric profile and they blend together to
     form the very complex light curves of the flux, polarization and position
     angle. We have well model-simulated the flux light curves for the 
      periodic  outbursts in 1983.0, 1984.12, 1994.59 ,1995.84, 
     2005.74, 2007.69 and the 2015.87 optical 
     outburst (Qian \cite{Qi19a}). The symmetric profiles of elementary flares
     may be caused by the helical motion of the individual superluminal 
     optical knots via lighthouse effect. This mechanism is applicable to
     both periodic and non-periodic outbursts.
    \item (3) The simultaneous variations in radio/mm and optical bands (Qian
     \cite{Qi19b}) can be interpreted: we have suggested  that the superluminal 
     optical and radio knots might
     have a core-envelope structure with its synchrotron radiation distributed
     in the direction perpendicular to the direction of the helical motion. 
     In this case the 
     core-region dominates the optical radiation and its envelope dominates
      the radio/mm radiation, and
    the motion of this core-envelope structure can produce instantaneous
    optical and radio/mm flux variations (Qian \cite{Qi19b}). 
    This optical-radio/mm radiation 
    pattern is consistent with the stratification of the magnetic surface
    predicted by MHD theories for the magnetospheres produced by
    the black-hole/accretion disk systems ( e.g. Camenzind \cite{Cam90}).
    \item (4) The connection between the optical outbursts and the  delayed
    radio outbursts and ejection of superluminal radio knots can be understood
    , because the superluminal optical knots  evolve into time-delayed
     radio knots when they move outward to  large distances from the core.
     \item (5) The large-amplitude rotations in polarization position 
    angle of  the outbursts can be explained. 
     The helical motion of the superluminal optical knots through the 
     surrounding helical magnetic fields would result
    in large-amplitude rotations of polarization position angle of the 
    outbursts. This phenomenon  has been discovered and studied: e.g, for 
    BL Lacertae and OJ287 (Sillanp\"a\"a \cite{Si93}, Marscher \cite{Ma08},
     Holmes et al. \cite{Holm84}). In this paper we have also  demonstrated
    the large-amplitude position angle rotations for the 1983.0, 2007.8 
    and 2015.8 outbursts, providing
    strong evidence for the helical motion of superluminal optical knots in 
    blazar OJ287. \\
    Recently, Myserlis et al. (\cite{My18}) found the large-amplitude
    polarization position angle rotations in OJ287  at radio wavelengths (10.5,
    8.4 and 4.8GHz), which are delayed with respect to the position angle
    rotations at optical wavelengths. This is fully consistent with the 
    predictions of the precessing nozzle scenario (Qian \cite{Qi19a}, 
   \cite{Qi19c}): during 
    the optical outbursts the radiation at radio wavelengths can not escape
    due to opacity effects and radio emission can be observed only when
     the radio-emitting regions become transparent. At the same time the
     timescales of the
    PA rotations at radio wavelengths would be much longer than that at optical
    wavelengths, because the radio PA rotations  occurred in outer jet regions 
    where the pitch angle of the helical magnetic field may be much larger than
    those in the optical-emitting regions (with coiled magnetic fields ). 
     This is just the case as observed in
    Myserlis et al. during  December 2015 to January 2017.
     \item (6) The  analysis of the kinematics of the superluminal 
    radio knots on VLBI-scale at 15GHz in OJ287 (Qian \cite{Qi18b}) 
    has shown that the precessing nozzle model can be used to explain
    the VLBI-kinematics of the radio knots. The precession of the
    jet-nozzle is a key ingredient to understand the phenomena in OJ287.
    A tentative study  indicates that OJ287 might have a double-jet structure,
     because only in this case the kinematics of its superluminal components 
     C11 and C12 
     (having similar position angles but four-year separation in ejection 
     epochs) can be well fitted (Qian \cite{Qi18b}).
    \end{itemize}
   Although the precessing nozzle model OJ287 (Qian \cite{Qi18b}, 
    \cite{Qi19a}) can be used to 
    understand most of the emission properties of the optical/radio outbursts
    observed in OJ287, its basic assumptions are still to be tested 
   and confirmed.\\ 
    In addition, the precessing nozzle model (as a relativistic jet model)
    is only
    applied to interpret the emission properties of the outbursts in OJ287.
    It does not deal with the mechanism of the quasi-periodicity and
    double-peaked structure, because this subject mainly involves the physical
    processes occurred in the course of binary orbital motion, e.g. as
    suggested  by the impact-disk scenario. These may include the
    penetration of secondary hole into the primary disk, 
    interaction between the secondary hole and the magnetosphere of the 
    spinning primary hole, and the consequential effects from the 
    impact-disturbances
     in mass accretion on the ejection of superluminal optical knots, etc.\\
      Some HD and MHD simulations (e.g., Artymowicz \& Lubow 
    \cite{Ar96}, Artymowicz \cite{Ar98}, Hayasaki et al. \cite{Ha08},
    Cuadra et al. \cite{Cu09}, Farris et al. \cite{Fa14}, Shi et al. 
   \cite{Sh12}, \cite{Sh15}, D'Orazio et al.\cite{Dor13}) have suggested that
    cavity-accretion models with two-stream accretion flows toward the binary 
    holes could interpret the production of the quasi-periodic pair-flares.
    \footnote{Note that there is some evidence for double-jets in OJ287 from
     the analysis of the VLBI-kinematics of superluminal radio knots (Qian
    \cite{Qi18b}).} However,  cavity-accretion models, e.g., as proposed 
     for OJ287 by Tanaka et al. (\cite{Tan13}), 
     can only produce a pair of thermal flares which is contradictory
    to the polarization behavior observed in the quasi-periodic outbursts
     (Qian, this paper; Myserlis et al.\cite{My18}; Homles et al.\cite{Holm84}).
     Moreover, this model is not able to provide an interpretation for
    the quasi-periodicity and prediction of flaring times of the 
    impact-flares.\\ 
    At present, only the impact-disk model (Lehto \& Valtonen \cite{Le96}, 
    Sundelius et al. \cite{Su97}, Valtonen et al. \cite{Va19}) has been 
    proposed to explain the quasi-periodicity and double-peaked 
    structure in the 
    optical light curve. According to Laine et al. (\cite{La20}) this model
    has successfully predicted the pair of quasi-periodic outbursts 
    in 2015/2019.
    Based on the calculations of the precessing orbital motion under the 
    impact-disk scenario, the accurate timing of the quasi-periodic flares 
    can be applied to test general relativity (Einstein \cite{Ei16},
    \cite{Ei18}; e.g., gravitational waves, precession of binary orbit,
    no-hair theorem, etc.).\\
      Based on the model simulation of the light curves of flux density,
     polarization degree and polarization position angle as a whole and the 
     investigation on the nature of emission from the 
      optical outbursts in 1983.0, 2007.8 and 2015.8, we find that these 
     periodic outbursts may be all synchrotron in origin, inconsistent 
     with the predictions from the disk-impact scenario. This issue
     might be helpful for understanding the entire phenomena observed in
     blazar OJ287. More multi-wavelength observations (in $\gamma$-rays and 
     in optical/radio bands) and  theoretical works are required  
     to find  some solutions. As a
     conjecture, for example, we would have to consider the possibility:
     if the impact-disk scenario for explaining the quasi-periodicity with
     double-peaked structure is unique
      and if the suggestion of the periodic outbursts being synchrotron in 
     origin is correct, then there should exist some mechanism(s) 
     $\it{directly}$ connecting the ejection of superluminal optical knots 
     with the disk-impacts without producing strong thermal optical outbursts.


\begin{thebibliography}{}

   \bibitem[2011]{Ac11}
   Ackermann M., Ajello M., Allafort A., et al., 2011, ApJ 743, 171
   \bibitem[1981]{Al81}
   Aller, H.D., Hodge, P.E., Aller, M.F., 1981, ApJ,248, L5 
   \bibitem[2010]{Al10}
   Aller, M.F., Hughes, P.A., Aller, H.D. 2010, in ``Fermi Meets Jansky - AGN
    in Radio and Gamma-rays'', Eds.: Savolainen, T., Ros, E., Porcas, R.W. and
    Zensus, J.A., p65
   \bibitem[2014]{Al14}
   Aller, M.F., Hughes, P.A., Aller, H.D., et al. 2014, ApJ, 791, 53  
   \bibitem[2016]{Al16}
   Aller, M.F., Hughes, P.A., Aller, H.D., et al. 2016, Galaxies,
   arXiv: 1609.06332v1 
   \bibitem[1980]{An80}
   Angel, J.R.P, \& Stockman, H.S. 1980, Ann.Rev.Astr.Ap., 18,321
   \bibitem[1996]{Ar96}
   Artymowicz, P., Lubow, S.H. 1996, ApJ, 467, L77
   \bibitem [1998]{Ar98}
   Artymowicz, P. 1998, in: Theory of Black Hole Accretion Disks, 
    ed. M.A.~Abramowicz, G.~Bj\"ornsson,  J.E.~Pringle, p202
    \bibitem[1978a]{Bel78a}
  Bell, A.R. 1978a, MNRAS, 182, 147
  \bibitem[1978b]{Bel78b}
   Bell, A.R., 1978b, MNRAS, 182, 443
   \bibitem[2010]{Be10}
   Beskin, V.C. 2010, Physics-Uspekhi, 53, 1199 
  \bibitem[1982]{Bj82}
   Bj\"ornsson, C.I. 1982, ApJ, 260, 855
  \bibitem[1977]{Bl77}
    Blandford, R.D., \& Znajek, R.L. 1977, MNRAS, 179, 433
   \bibitem[1979]{Bl79}
   Blandford, R.D., K\"{o}nigl A., 1979, ApJ 232, 34   
    \bibitem[1982]{Bl82}
    Blandford, R.D., \& Payne, D.G. 1982, MNRAS, 199, 883   
     \bibitem[2018]{Br18}
   Britzen, S., Fendt, C., Witzel, G., Qian, S.J., et al. 2018, 478, 3199
   \bibitem[1990]{Cam90}
    Camenzind, M. 1990,  Reviews in Modern Astronomy, Vol.3, 234
   \bibitem[1992]{Cam92}
    Camenzind, M. \& Krockenberger, M. 1992, A\&A, 255, 59
   \bibitem[1993]{Cam93}
   Camenzind, M. 1993, in: Proc. 1st MEGAPHOT workshop: "The Need  for a
    Dedicated Optical Quasar Monitoring Telescope", eds., U. Borgeast et al.,
    p.12     
   \bibitem[1988]{Car88}
   Carilli, C.L., Perley, R.A., Dreher, J.H. 1988, ApJ, 334, L73
   \bibitem[1988]{Caw88}
   Cawthorne, T.V. \& Wardle , J.F.C., 1988, ApJ 332, 696
  \bibitem[2017]{Co17}
  Cohen, M.H. 2017, Galaxies, 5, 12
  \bibitem[2018]{Co18}
   Cohen, M.H., Aller, H.D., Aller, M.F., et al. 2018, ApJ,  862, 1
  \bibitem[2009]{Cu09}
  Cuadra,J., Armitage, P.J., Alexander, R.D., Begelman, M.C., 2009, MNRAS 393,
   1423
  \bibitem[2009]{Da09}
  D'Arcangelo, F.D., Marscher, A.P., Jorstad, S.D., et  al. 2009, ApJ 697, 
    985               
  \bibitem[2019]{De19}
  Dey, L., Gopakumar, A., Valtonen, M., et al. 2019, arXiv:1905.02689
   \bibitem[2018]{De18}
   Dey, L., Valtonen, M.J., Gopakumar, A., et al. 2018, ApJ, 866, article 
     id. 11D 
   \bibitem[2013]{Dor13}
   D'Orazio, D.J., Haiman, Z., \& Macfadyen A. 2013, MNRAS, 436, 2997
   \bibitem[1996a]{Dr96a}
    Dreissigacker, O. 1996a, in: Extragalactic Radio Sources, eds.,
                   R.~Ekers et al., p421       
   \bibitem[1996b]{Dr96b}
   Dreissigacker, O \& Camenzind, M. 1996b, in: Blazar Continuum Variability
    (ASP Conference Series, eds. H.R.~Miller, J.R.~Webb, and J.C.~Noble),
     Vol.110, p.377
   \bibitem[1916]{Ei16}
   Einstein, A. 1916, Sitzungberichte der K\"oniglich Presssichen Akademie der
    Wissenshafte (SPAW, Berlin), 688
   \bibitem[1918]{Ei18}
   Einstein, A. 1918, Sitzungberichte der K\"oniglich Presssichen Akademi der
    Wissenshafte (SPAW, Berlin), 154
  \bibitem[2014]{Fa14}
   Farris, B.D.,  Duffell,P., MacFadyen, A.I., Haiman, Z., 2014, ApJ 783, 134 
   \bibitem[1999]{Ga99}
   Gabuzda, D.C. et al., 1999, in: ``BL Lac Phenomenon'', ASP Conference
    Series,eds., L.O.Takalo and Sillanp\"a\"a, Vol.159, p.447 
   \bibitem[2001]{Ga01}
   Gabuzda, D.C., G\'omez, J.L., 2001, MNRAS, 320, L49 
  \bibitem[2004]{Ga04}
    Gabuzda, D.C.,Murray, \'E, Cronin, P. 2004, MNRAS, 351, L89
   \bibitem[1994a]{Go94a}
   G$\acute{o}$mez, J.L., Alberti, A., Marcaide, J.M., 1994a, A\&A, 284, 51
  \bibitem[1994b]{Go94b}
   G$\acute{o}$, J.L., Alberti, J.M., Marcaide, J.M., et al., 1994b, 
    A\&A, 292, 33
  \bibitem[2016]{Gu16}
  Gupta, A.C., Agarwal, A., Mishra, A., et al. 2016, MNRAS, 458, 1127  
  \bibitem[2008]{Ha08}
  Hayasaki, K., Mineshige, S. \& Ho, L.C. 2008, ApJ, 682, 1134
  \bibitem[2017]{Hod17}
   Hodgson, J.A., Krichbaum, T.P., Marscher, A.P., et al. 2017, A\&A, 597, 80
  \bibitem[1999]{Ho99}
  Hogg, D.W. 1999, astro-ph/9905116
   \bibitem[1984]{Holm84}
   Holmes, P.A., Brand, P.W.J.L., Impey, C.D.,et al. 1984, MNRAS, 211, 497
  \bibitem[1985]{Hu85}
  Hughes, P.A., Aller, H.D., Aller, M.F. 1985, ApJ, 298, 301 
  \bibitem[2011]{Hu11}
  Hughes, P.A., Aller,M.F., Aller, H.D.,2011, ApJ, 735, 81
   \bibitem[2005]{Jo05}
   Jorstad, S.G., Marscher, A.P., Lister, M.L., et al., 2005, AJ, 130, 1418
   \bibitem[1997]{Ka97}
   Katz, J.I., 1997, ApJ, 478, 527
   \bibitem[1988]{Ki88}
   Kikuchi, S., Inoue, M., Mikami, Y., et al. 1988, A\&A, 190, L8
   \bibitem[1982]{Kon82}
   Kong, X.Y., Xu, Y. H., Zhong Z.X. 1982, Acta Astrophys. Sinica, 2, 81
   \bibitem[1985a]{Ko85a}
   K\"onigl, A. \& Choudhuri, A.R. 1985a, ApJ, 289, 173
   \bibitem[1985b]{Ko85b}
   K\"onigl,A. \& Choudhuri, A.R., 1985b, ApJ 289,  188 
   \bibitem[2009]{Ko09}
   Komatsu, E., Dunkley, J., Nolta, M.R., et al. 2009, ApJS, 180, 330    
   \bibitem[2018a]{Ku18a}
  Kushwaha, P., Gupta, A.C., Wiita, P.J., et al. 2018a, MNRAS, 473, 1145
   \bibitem[2018b]{Ku18b} 
  Kushwaha, P., Gupta, A.C., Wiita, P.J., et al. 2018b, MNRAS, 479, 1672
   \bibitem[2020]{La20}
   Laine, S., Gopakumar, A., Valtonen, M., et al. 2020, AAS, 23530501L
   \bibitem[1980]{Lai80}
   Laing, R.P. 1980, MNRAS, 193, 439
   \bibitem[1996]{Le96}
   Lehto, H.J., \& Valtonen, M.J. 1996, ApJ,  460, 207
   \bibitem[1992]{Lizy92}
    Li, Z.Y., Chiueh, T.H., Begelman, M.C. 1992, ApJ, 394, 459
   \bibitem[1985]{Lin85}
   Lind, K.R., Blanford, R.D. 1985, ApJ, 295, 358
   \bibitem[2013]{Lis13}
   Lister, M.L., Aller, M.F., Aller, H.D., et al., 2013, AJ, 146, 120
   \bibitem[2008]{Ma08}
   Marscher A.P., Jorstad s.G., D'Acangelo F.D., et al., 2008, Nature,
   Vol.452,doi:10.1038/06895, p.966       
   \bibitem[2001]{Mei01} 
   Meier, D.L., Koide, S., Uchida, Y.  2001, Science, Vol.291, 84
    \bibitem[2013]{Mei13}
    Meier, D.L. 2013, EPJ Web of Conference 61, 01001
  \bibitem[2018]{My18}
   Myserlis I., Komossa E., Angelakis E., et al., 2018, A\&A 619, A88
  \bibitem[1999]{Pe99}
   Pen, U.L., ApJS, 120,49 
   \bibitem[2013]{Pi13}
   Pihajoki, p., Valtonen, M., \& Ciprini, S. 2013, MNRAS 434, 3122
    \bibitem[1991a]{Qi91a}
   Qian, S.J., Witzel, A., Krichbaum, T.P., et al. 1991a, Acta Astron. Sin.,
             32, 369 (english translation: in Chin. Astro. Astrophys., 16, 
             137 (1992)) 
    \bibitem[1991b]{Qi91b}
   Qian, S.J.,  Quirrenbach, A., Witzel, A., et al. 1991b, A\&A, 241, 15
   \bibitem[1992]{Qi92}
   Qian, S.J. 1992, Chin. Astron. Astrophys., Vol.16, 266 
   \bibitem[1993]{Qi93}
   Qian, S.J., 1993, Chin. Astron. Astrophys., 17, 229 
   \bibitem[1996]{Qi96}
   Qian, S.J., Krichbaum, T.P., Zensus, J.A. et al., 1996, A\&A,308,395
    \bibitem[2003]{Qi03}
   Qian, S.J. \& Zhang X.Z., 2003, Chin.J. Astron. Astrophys., Vol.3,  
   75
  \bibitem[2009]{Qi09}
   Qian, S.J., Witzel, A., Zensus, J.A., et al. 2009, RAA, Vol.9, No.2, 137
    \bibitem[2013]{Qi13}
   Qian, S.J. 2013, Res. Astron. Astrophys., 13, 783
   \bibitem[2014]{Qi14}
   Qian, S.J., Britzen, S., Witzel,A., et al., 2014, RAA, Vol.14, No.3, 249
   \bibitem[2015]{Qi15}
   Qian, S.J. 2015, Res. Astron. Astrophy., 15, 687
   \bibitem[2016]{Qi16}
   Qian, S.J., 2016, RAA, Vol.16, No.1, 20 (15pp)   
  \bibitem[2017]{Qi17}
   Qian, S.J., Britzen, S., Witzel, A., et al. 2017, A\&A, 604, A90 
  \bibitem[2018a]{Qi18a}
   Qian, S.J., Britzen, S., Witzel, A. et al., 2018a, A\&A, 615, A123 
  \bibitem[2019c]{Qi19c}
   Qian , S.J., Britzen, S.J., Krichbaum, T.P., Witzel, A., 2019c, 
    A\&A 621, A11
    \bibitem[2018b]{Qi18b}
  Qian, S.J. 2018b, arXiv-1811.11514 
  \bibitem[2019a]{Qi19a}
   Qian, S.J., 2019a, arXiv-1906.09782 
  \bibitem[2019b]{Qi19b}
  Qian, S.J. 2019b, arXiv-1904.03357 
  \bibitem[1978]{Re78}
   Rees, M.J., 1978, MNRAS, 184, 61p
  \bibitem[1993]{Sc93}
   Schramm K.-J., Borgeest U., Camenzind M.,et al., 1993, A\&A 278,391
  \bibitem[2012]{Sh12}
   Shi, J.M., Krolik, J.H., Loubow, S.H., Hawley J.F. 2012, ApJ, 749, 118
 \bibitem[2015]{Sh15}
   Shi, J.M., Krolik, J.H., 2015, ApJ 807 131 
  \bibitem[1988]{Si88}
   Sillanp\"a\"a, A., Haarala, S., Valtonen, M.J., et al. 1988,
                                ApJ, 325, 628 
  \bibitem[1993]{Si93}
   Sillanp\"a\"a, A., Takalo, L.O., Nilsson K., Kikuchi S., 1993, APSS, 206
   55
   \bibitem[1996a]{Si96a}
    Sillanp\"a\"a, A., Takalo, L.O., Pursimo, T., et al. 1996a,  A\&A, 
             315, L13
  \bibitem[1996b]{Si96b}
    Sillanp{\"a\"a}, A., Takalo, L.O., Pursimo, T., et al. 1996b, 
    ASP conference series, Vol. 110 (Blazar Continuum Variability), 
    eds. H.R.~Miller, J.R.~Webb and J.C.~Noble, p74
   \bibitem[1987]{Sm87}
  Smith, P.S., Balonek, T.J., Elston, R., \& Heckert, P.A. 1987, ApJS, 64, 495
  \bibitem[2003]{Sp03}
  Spergel, D.N., Verde, L., Peiris, H.V., et al. 2003, ApJS, 148, 175
  \bibitem[1997]{Su97}
   Sundelius, B., Wade, M., Lehto, H.J., et al. 1997, ApJ, 484, 180 
  \bibitem[1996a]{Tak96a}
    Takalo, L.O. 1996a, ASP Conference Series, Vol.110 (Blazar Continuum
     Variability), eds.,  H.R.~Miller, J.R.~Webb and J.C.~Noble, p70
  \bibitem[1996b]{Tak96b}
   Takalo, L.O., Sillanp\"a\"a, A., Lehto, H.J. 1996b, Mem.S.A.It., 67, 545
  \bibitem[2013]{Tan13}
   Tanaka, T.L. 2013, MNRAS, 434, 2275
  \bibitem[1999]{Ta99}
   Tateyama, C.E., Kingham, K.A., Kayfmann, P., et al. 1999, ApJ, 520, 627
  \bibitem[1979]{Us79}
   Usher, P.D. 1979, Astron. J., 84, 1253  
   \bibitem[2000]{Val20}
   Valtaoja, E., Terasranta, H., Tornikoski, M., et al. 2000, ApJ, 531, 744  
  \bibitem[2008]{Va08}
   Valtonen, M.J., Lehto, H.J., Nilsson, K., et al. 2008, Nature, 452, 851
  \bibitem[2011]{Va11}
   Valtonen M.J., Mikkola S., Lehto H.J., et al. 2011, ApJ, 742, 22
   \bibitem[2016]{Va16}
   Valtonen, M.J., Zola, S., Ciprini, S.,  et al.  2016, ApJ, 819, L37  
   \bibitem[2017]{Va17}
    Valtonen, M.J., Zola, S., Jermak, H., et al. 2017,  Galaxies, 5, 83
   \bibitem[2019]{Va19}
    Valtonen, M.J., Zola, S., Pihajoki, P. et al. 2019, ApJ,      
      (arXiv:1907.11011v1[Astro-ph.HE])
    \bibitem[1998]{Villa98}
   Villata, M., Raiteri, C.M., Sillanp{\"a\"a}, A., et al. 1998, MNRAS 293, L13
   \bibitem[2010]{Vil10} 
   Villforth, C., Nilsson, K., Heidt, J.,  et al. 2010, MNRAS 402, 2087
   \bibitem[2004]{Vl04}
   Vlahakis, N., \& K\"onigl, A. 2004, ApJ, 605, 656
   \bibitem[1995]{Wa95}
   Wagner S.J., Camenzind M., D reissigacker O., et al., 1995, A\&A 298, 688 
   \end{thebibliography}
  \end{document}